\documentclass[a4paper,11pt]{article}
\pdfoutput=1 % if your are submitting a pdflatex (i.e. if you have
\usepackage{jinstpub} % for details on the use of the package, please
\usepackage{lineno,hyperref}
\usepackage{amssymb}
\usepackage{amsmath}
\usepackage{url}
\usepackage[utf8]{inputenc}
\usepackage{ifthen}
\usepackage{ltxtable}
\usepackage{subfigure}

\title{\boldmath Calibration of the Logarithmic-Periodic Dipole Antenna (LPDA) Radio Stations at the Pierre Auger Observatory using an Octocopter}

\author[63]{A.~Aab,}
\author[70]{P.~Abreu,}
\author[48,47]{M.~Aglietta,}
\author[29]{I.~Al Samarai,}
\author[16]{I.F.M.~Albuquerque,}
\author[1]{I.~Allekotte,}
\author[8,11]{A.~Almela,}
\author[62]{J.~Alvarez Castillo,}
\author[78]{J.~Alvarez-Mu\~niz,}
\author[38]{G.A.~Anastasi,}
% ...

\affiliation[1]{Centro At\'omico Bariloche and Instituto Balseiro (CNEA-UNCuyo-CONICET), Argentina}
% ...
\affiliation[8]{Instituto de Tecnolog\'\i{}as en Detecci\'on y Astropart\'\i{}culas (CNEA, CONICET, UNSAM), Centro At\'omico Constituyentes, Comisi\'on Nacional de Energ\'\i{}a At\'omica, Argentina}
% ...
\affiliation[11]{Universidad Tecnol\'ogica Nacional -- Facultad Regional Buenos Aires, Argentina}
% ...
\affiliation[16]{Universidade de S\~ao Paulo, Inst.\ de F\'\i{}sica, S\~ao Paulo, Brazil}
% ...
\affiliation[29]{Laboratoire de Physique Nucl\'eaire et de Hautes Energies (LPNHE), Universit\'es Paris 6 et Paris 7, CNRS-IN2P3, France}
% ...
\affiliation[38]{Gran Sasso Science Institute (INFN), L'Aquila, Italy}
% ...
\affiliation[47]{INFN, Sezione di Torino, Italy}
\affiliation[48]{Osservatorio Astrofisico di Torino (INAF), Torino, Italy}
% ...
\affiliation[62]{Universidad Nacional Aut\'onoma de M\'exico, M\'exico}
\affiliation[63]{Institute for Mathematics, Astrophysics and Particle Physics (IMAPP), Radboud Universiteit, Nijmegen, Netherlands}
% ...
\affiliation[70]{Laborat\'orio de Instrumenta\c{c}\~ao e F\'\i{}sica Experimental de Part\'\i{}culas -- LIP and Instituto Superior T\'ecnico -- IST, Universidade de Lisboa -- UL, Portugal}
% ...
\affiliation[78]{Universidad de Santiago de Compostela, Spain}
% ...
\affiliation[c]{also at Vrije Universiteit Brussels, Brussels, Belgium}
% ...

\emailAdd{auger\_spokespersons@fnal.gov}

\abstract{An in-situ calibration of a logarithmic periodic dipole
antenna with a frequency coverage of $30\,\rm{MHz}$ to $80\,\rm{MHz}$ is performed. Such antennas are part of
a radio station system used for detection of cosmic ray induced air
showers at the Engineering Radio Array of the Pierre Auger Observatory,
the so-called Auger Engineering Radio Array (AERA). The directional and frequency
characteristics of the broadband antenna are investigated using a remotely piloted aircraft
carrying a small transmitting antenna. The antenna sensitivity is
described by the vector effective length relating the measured voltage
with the electric-field components perpendicular to the incoming signal
direction. The horizontal and meridional components are determined with
an overall uncertainty of $7.4^{+0.9}_{-0.3}\,\rm{\%}$ and $10.3^{+2.8}_{-1.7}\,\rm{\%}$
respectively. The measurement is used to correct a simulated response of
the frequency and directional response of the antenna. 
In addition, the influence of the ground conductivity and permittivity on
the antenna response is simulated. Both have a negligible influence given the ground
conditions measured at the detector site. The overall uncertainties 
of the vector effective length components result in an uncertainty of $8.8^{+2.1}_{-1.3}\,\rm{\%}$
in the square root of the energy fluence for incoming signal directions
with zenith angles smaller than $60\rm{^{\circ}}$.}


\begin{document}
\maketitle
\flushbottom

\section{Introduction}
When ultrahigh-energy cosmic rays (UHECRs) hit the Earth, they collide with air nuclei and create a particle cascade 
of millions of secondary particles, a so-called air shower. The atmosphere acts thereby as a giant calorimeter of
${\sim} 11$ hadronic interaction lengths. Instrumentation of such a 
giant detector volume is challenging in every respect, especially concerning readout, calibration and 
monitoring. Well-established solutions are stochastic measurements of the remaining secondary
particles at ground level and direct detection of fluorescence light emitted from air 
molecules excited by the particle cascade. Both techniques are successfully applied in the Pierre 
Auger Observatory in Argentina, covering $3000\,\rm{km^{2}}$ with $1660$ water-Cherenkov detectors and 
$27$ telescopes for detection of fluorescence light \cite{Auger}.\\
In recent years, measurement of radio emission from air showers in the 
megahertz (MHz) regime has become a 
complementary detection technique \cite{LOPES, Codalema, AERA, TRex, LOFAR, Tim, Frank}. For this, 
the Pierre Auger Observatory was extended by $153$ radio stations, the so-called Auger Engineering Radio Array (AERA). These antenna stations at 
ground level provide information on the radio signal and are used to reconstruct the electric field generated by an air shower. \\
Two mechanisms contribute to coherent radio emission from air showers, namely the geomagnetic 
effect induced by charged particle motion in the Earth's magnetic field \cite{LOPES, Geo-KahnLerche, Geo-Allan, Geo-HuegeFalcke, Geo-Codalema, Geo-Auger} and the time varying negative charge excess in the shower front.
The charge excess is due to the knock-out of electrons from air molecules and annihilation of positrons in the shower front \cite{CE-Ask, CE, CE-Codalema, CE-Lofar, CE-AERA}. 
The radio emission can be calculated from first principles using classical electrodynamics \cite{CORSIKA, ZHAires, CoREAS, Radio}.
The emission primarily originates from the well-understood electromagnetic part of the air shower.
Thus, the theoretical aspect of radio measurements is on solid grounds \cite{Tim}. \\
As the atmosphere is transparent to radio waves, the radio technique has a high potential for precision measurements in cosmic-ray physics.
Correlation of the strength of the radio signal with the primary cosmic-ray energy has
meanwhile been demonstrated by several observatories \cite{LOPES-Energy, LOFAR-Energy, TREX-Energy, AERA-Energy-PRL, AERA-Energy-PRD}.
Furthermore, the radiation energy, i.e., the energy contained in the radio signal has
been determined \cite{AERA-Energy-PRD}. It was shown that the radio energy resolution is competitive with the results of particle measurements at ground level.
Furthermore using above-mentioned first-principle calculations, a
novel way of a stand-alone absolute energy calibration of the atmospheric
calorimeter appears feasible \cite{AERA-Energy-PRL}.\\
In all these considerations, the antenna to detect the electric field and a thorough description of its characteristics is of central importance. 
Precise knowledge of the directional antenna characteristics is essential to reconstruct the electric field and therefore enables high quality measurements of the cosmic-ray properties.
For a complete description of the antenna characteristics an absolute antenna calibration needs to be performed. The uncertainties of the absolute calibration directly impact the energy scale
for air shower measurements from radio detectors. Therefore, a central challenge of the absolute antenna calibration
is to reduce the uncertainties of the antenna characteristics to the level of $10\,\rm{\%}$ which is a significant improvement in comparison with the
uncertainties obtained in calibration campaigns at other radio detectors \cite{LOPES-Calib, TRex-Calib, LOFAR-Calib}.\\
In this work, the reconstruction quality of 
the electric-field signal from the measured voltage trace, which includes the directional 
characteristics of the antenna and dispersion of the signal 
owing to the antenna size, is investigated. All information are described with the 
vector effective length $\vec{H}$, a complex measure that relates the measured voltage to the incoming electric field.
One antenna of the subset of $24$ radio stations equipped with 
logarithmic periodic dipole antennas (LPDAs) is investigated here exemplarily.
This antenna is representative of all the LPDAs which are mechanically and
electrically identical at the percent level \cite{KlausPhD}.
While the low-noise amplifier attached to the antenna was included in the signal chain during the calibration,
amplifiers and the subsequent electronics of all radio stations have been characterized individually.
The LPDA antennas have the advantage of low sensitivity to radio waves reflecting from the ground which makes 
them largely independent of potentially changing ground conditions.\\
The LPDA antennas have been studied before and a first absolute calibration of one signal polarization was performed in $2012$ 
giving an overall systematic uncertainty of $12.5\,\rm{\%}$ \cite{AERA-Antennas}.
In comparison to the first absolute calibration of AERA, in this paper a new absolute calibration is presented using
a new setup enabling a much more dense sampling of the arrival directions, more field polarization measurements, 
and an extended control of systematic effects including the full repetition of calibration series.
To ensure far-field measurements, instead of the previously used balloon,
a drone was employed, carrying a separate signal generator and a calibrated transmitting antenna. \\
This work is structured as follows.
Firstly, a calculation of the absolute value of the vector effective length $|\vec{H}|$ of the LPDA 
is presented. Then, the LPDA antenna and the calibration setup are specified.
In the next section the calibration strategy is presented using one example flight
where $|\vec{H}|$ is measured on site at the Pierre Auger Observatory at one of the radio stations. 
The main section contains detailed comparisons of all the measurements with the calculated vector 
effective length and the determination of the uncertainties in the current understanding of the antenna.
Finally, the influence of the calibration results are discussed in applications 
before presenting the conclusions.
\section{Antenna Response Pattern}
This section gives a theoretical overview of the antenna response pattern. The vector effective length (VEL) is introduced as a measure of the directional dependent antenna sensitivity. 
Furthermore, it is explained how the VEL is obtained for an uncalibrated antenna. For more details refer to \cite{AERA-Antennas}.
\subsection{The Vector Effective Length (VEL)}
Electromagnetic fields induce a voltage at the antenna output. The antenna signal depends on the incoming field $\vec{E}(t)$, the contributing frequencies $f$, as well as on the incoming direction 
with the azimuthal angle $\Phi$ and the zenith angle $\Theta$ to the antenna. The relation between the Fourier-transformed electric field $\vec{\mathcal{E}}(f)$ and the Fourier transformed observed voltage $\mathcal{U}$ for $\Phi, \Theta, f$ is referred to
as the antenna response pattern and is expressed in terms of the VEL $\vec{H}$:
\begin{linenomath}
\begin{equation}
    \mathcal{U}(\Phi, \Theta, f) = \vec{H} (\Phi, \Theta, f) \cdot \mathcal{\vec{E}} (f)
    \label{eq:AntResponse} 
\end{equation}
\end{linenomath}
The VEL $\vec{H}$ is oriented in the plane perpendicular to the arrival direction of the signal and can be expressed as a superposition of a horizontal component $H_{\phi}$ and a component $H_{\theta}$ oriented 
perpendicular to $H_{\phi}$ which is called meridional component:
\begin{linenomath}
\begin{equation}
    \vec{H} = H_{\phi} \vec{e}_{\phi} + H_{\theta} \vec{e}_{\theta}.
    \label{eq:VELSpherical} 
\end{equation}
\end{linenomath}
The VEL is a complex quantity $H_{k} = |H_{k}| e^{i\alpha_{k}}$ with $k=\phi,~\theta$
and accounts for the frequency-dependent electrical losses within the antenna as well as reflection effects which arise in the case of differences between the antenna and read-out system impedances.
Both effects lead to dispersion of the signal shape.\\
The antenna response pattern is often expressed in terms of the antenna gain based on the directional dependence of the received power. With the quadratic relation between voltage and power, the
antenna gain and the absolute value of the VEL are related by:
\begin{linenomath}
\begin{equation}
    |H_{k}(\Phi, \Theta, f)|^{2} = \frac{c^{2} Z_{R}}{f^{2} 4 \pi Z_{0}} G_{k}(\Phi, \Theta, f).
    \label{eq:VELGain} 
\end{equation}
\end{linenomath}
Here, $f$ is the signal frequency, $c$ is the vacuum speed of light, $Z_{R}=50\,\rm{\Omega}$ is the read-out impedance, $Z_{0} \approx 120 \, \pi \, \Omega$ is the impedance of free space,
the index $k=\phi$ or $\theta$ indicates the polarization, and $\Phi$ and $\Theta$ denote the azimuth and zenith angle of the arrival direction.
\subsection{Calculating the Absolute Value of the VEL from a Transmission Measurement}
The antenna characteristics of an uncalibrated antenna under test (AUT) is determined by measuring the antenna response of the AUT in a transmission setup using a calibrated transmission antenna.
The relation between transmitted and received power is described by the Friis 
equation \cite{Friis} considering the free-space path loss in vacuum as well as 
the signal frequency:
\begin{linenomath}
\begin{equation}
    \frac{P_{r}(\Phi, \Theta, f)}{P_{t}(f)}=G_{t}(f) G_{r}(\Phi, \Theta, f) \left( \frac{c}{f 4 \pi R} \right)^{2},
    \label{eq:Friis} 
\end{equation}
\end{linenomath}
with the received power $P_{r}$ at the AUT, the transmitted power $P_{t}$ induced on the transmission antenna, the known antenna gain $G_{t}$ of the calibrated transmission antenna, the unknown
antenna gain $G_{r}$ of the AUT, the distance $R$ between both antennas and the signal frequency $f$.\\
By considering Eq.~\eqref{eq:VELGain} and Eq.~\eqref{eq:Friis} the VEL of the AUT in a transmission setup is then determined by
\begin{linenomath}
\begin{equation}
    |H_{k}(\Phi, \Theta, f)| = \sqrt{\frac{4 \pi Z_{R}}{Z_{0}}} R \sqrt{\frac{P_{r,k}(\Phi, \Theta, f)}{P_{t}(f) G_{t}(f)}}
    \label{eq:VEL} 
\end{equation}
\end{linenomath}
\subsection{Calculating the Absolute Value of the Antenna VEL with separate Amplifier from a Transmission Simulation}
In this work, the NEC-2 \cite{NEC2} simulation code is used to simulate the response pattern of the passive part of the AUT.
With the passive part of the AUT the antenna without a then following low-noise amplifier stage is meant.
These simulations provide information about the received voltage directly at the antenna footpoint (AF)
which is the location where the signals of all dipoles are collected and converted to the then following $50\, \Omega$ system of the read-out system.
In the case of an amplifier (AMP) connected to the AF, the voltage at the output of the AMP is the parameter of interest.
The AMP is connected to the AF using a transmission line (TL). Both, the AMP and the TL, then constitute the active part of the AUT.
In the simulation, mismatch and reflection effects between the AF, the TL and the AMP,
which arise if the impedances $Z_{j}$ ($j=\rm{AF}, \rm{TL}, \rm{AMP}$) of two connected components differ from each other, have to be considered separately.
Moreover, the attenuation of the TL with a cable length $l_{TL}$ as well as the AMP itself described by the AMP S-parameters have to be taken into account.
The transformation of the received voltage at the AF to the received voltage at the AMP output is described by the transfer function $\rho$:
\begin{linenomath}
\begin{equation}
    \rho = \frac{1} {\sqrt{r}} \frac{Z_{TL}}{Z_{TL}+Z_{AF}/r} \left( 1 +  
\Gamma_{AMP} \right) \frac{e^{(\gamma + i\frac{2 \pi 
f}{c_{n}})l_{TL}}}{e^{2 (\gamma + i\frac{2 \pi f}{c_{n}})l_{TL}}-\Gamma_{AMP} 
\Gamma_{AF}} \frac{S21}{1+S11}
    \label{eq:SimVEL} 
\end{equation}
\end{linenomath}
with $\Gamma_{AMP} = \frac{Z_{AMP}-Z_{TL}}{Z_{AMP}+Z_{TL}}$ and $\Gamma_{AF} = 
\frac{Z_{AF}/r-Z_{TL}}{Z_{AF}/r+Z_{TL}}$. Furthermore, $\gamma$ denotes the 
attenuation loss along the transmission line, $f$ is the frequency 
of the signal, $c_{n}$ denotes the transfer rate inside the TL, and $r$ is the 
transfer factor from an impedance transformer at the AF which transforms the 
balanced signal of the antenna to an unbalanced signal of a TL. For more details 
refer to \cite{AERA-Antennas}.
\section{Logarithmic Periodic Dipole Antenna (LPDA)}
\begin{figure}
    \begin{center}
        \includegraphics[scale=0.33]{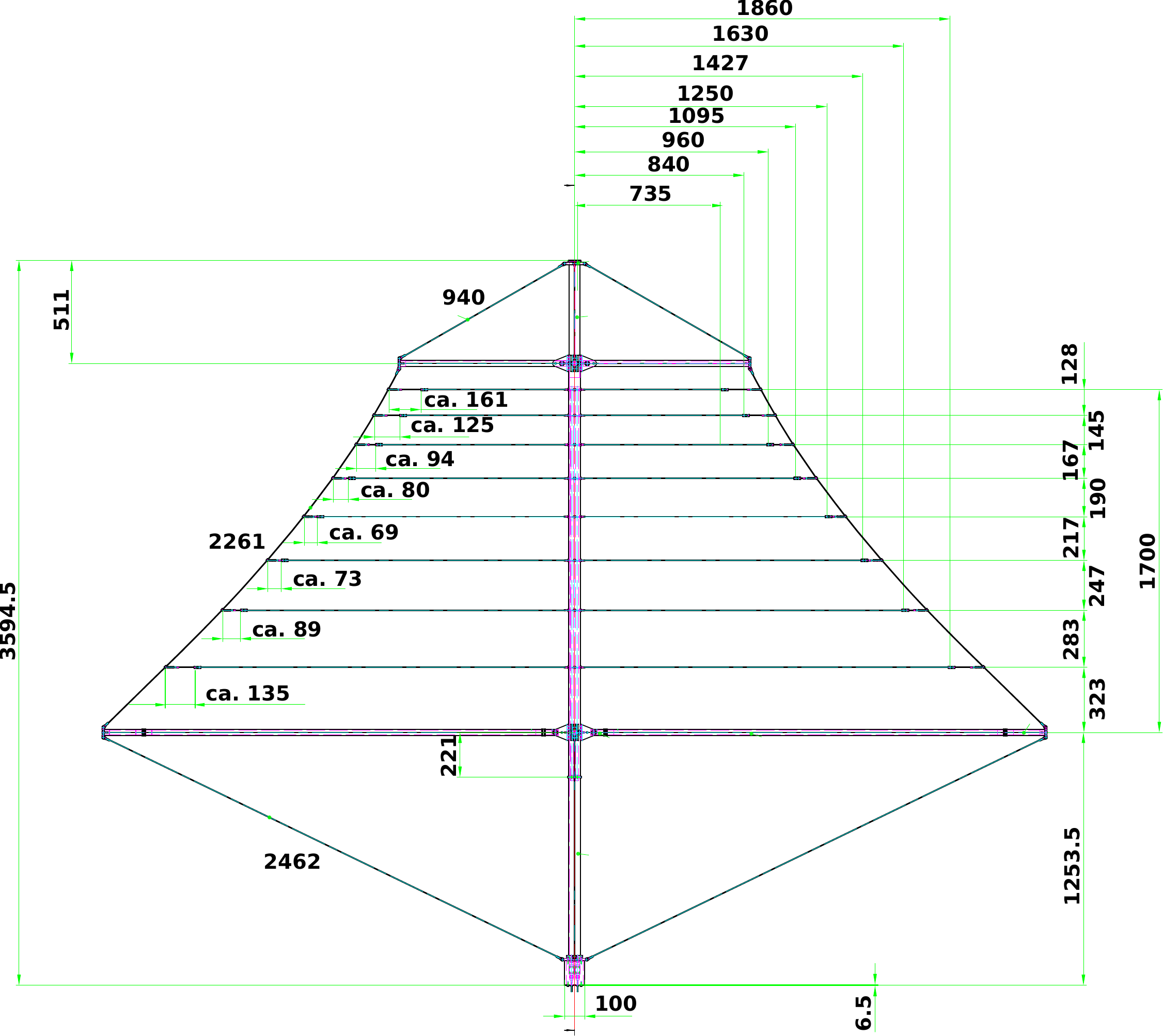}
        \caption{\it Drawing of the Logarithmic Periodic Dipole Antenna (LPDA), units are millimeter.}
    \label{fig:LPDA}
    \end{center}
\end{figure}

In this section, the Logarithmic Periodic Dipole Antenna (LPDA) which is used in a subset of the radio stations of AERA is presented. 
An LPDA consists of several $\lambda/2$-dipoles of different lengths which are combined to one single antenna with the largest dipole located at the bottom and the shortest dipole at the top of the LPDA.
The sensitive frequency range is defined by the length of the smallest $l_{min}$ and largest $l_{max}$ dipole. 
The ratio of the distance between two dipoles and their size is described by $\sigma$ and
the ratio of the dipole length between two neighboring dipoles is denoted by $\tau$. The four design parameters of the LPDAs used at AERA are $\tau=0.875$, $\sigma=0.038$, $l_{min}=1470\,\rm{mm}$ and $l_{max}=4250\,\rm{mm}$. 
These values were chosen to cover the frequency range from around $30\,\rm{MHz}$ to $80\,\rm{MHz}$ and to combine a high antenna sensitivity in a broad field of view using a
limited number of dipoles and reasonable dimensions. They lead to a LPDA with nine separate dipoles. For more details refer to \cite{AERA-Antennas}.
A full drawing of the LPDA used at AERA including all sizes is shown in Fig.~\ref{fig:LPDA}.
Each radio station at AERA consists of two perpendicular polarized antennas which are aligned to magnetic north with a precision better than $1\rm{^{\circ}}$. The dipoles are connected to a waveguide with the 
footpoint at the top of the antenna. The footpoint is connected by an RG58 
\cite{RG} coaxial transmission line to a low-noise amplifier (LNA) which 
amplifies the signal 
typically by $(18.1 \pm 0.2) \,\rm{dB}$. The LNA of the radio station and used in the calibration setup amplifies the signal by $18.1 \,\rm{dB}$.
The amplification is nearly constant in the frequency range $30\,\rm{MHz}$ to $80\,\rm{MHz}$ and variates at the level of $0.5 \,\rm{dB}$. 
For more technical details about the LNA refer to \cite{LNA}.
\section{Calibration Setup}
\label{sec:Calib}
\begin{figure}
    \begin{center}
        \includegraphics[scale=0.47]{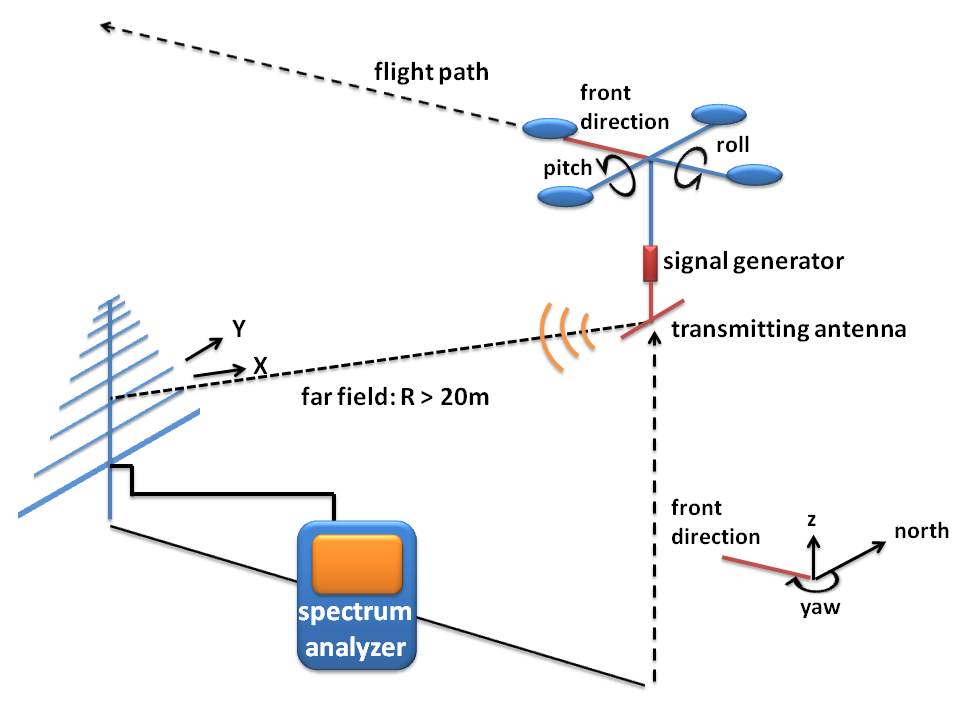}
%         \rulesep
	\line(1,0){250}\\
        \includegraphics[scale=0.42]{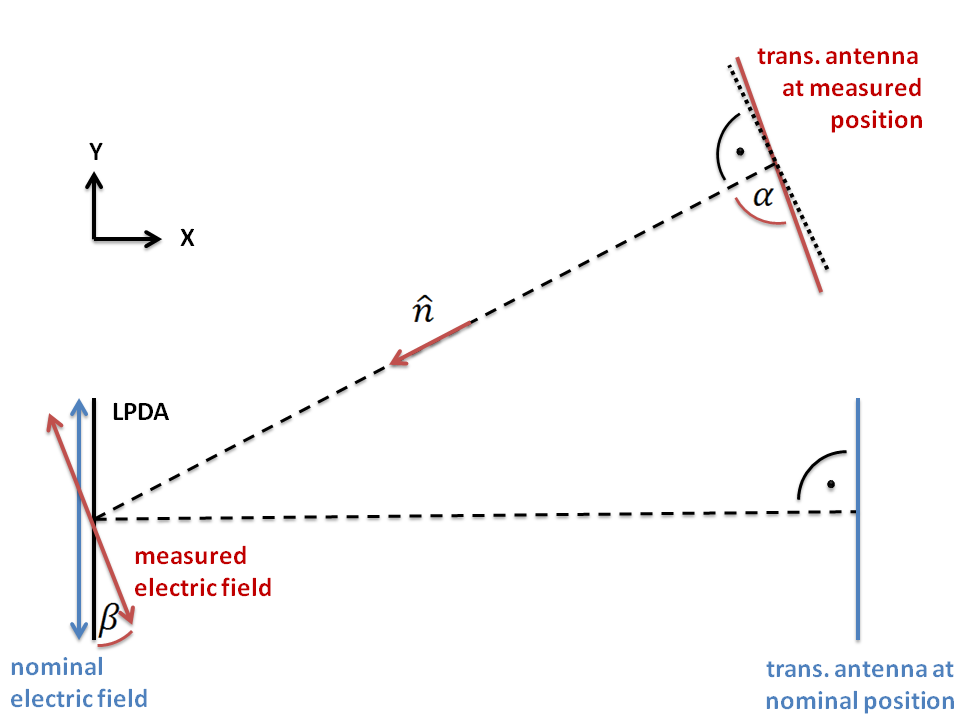}
        \caption{\it \textbf{(top)} LPDA calibration setup. The calibration signal is produced by
a signal generator and radiated by a transmitting antenna. Both the signal
generator and the transmitting antenna are attached underneath a flying
drone, a so-called RPA, to realize far-field conditions during the measurement. On
arrival of the signal at the LPDA, the antenna response is measured using a
spectrum analyzer. The orientation of the RPA is described by the yaw (twist
of front measured from north in the mathematically negative direction), and the 
tilt by the
pitch and the roll angles. \textbf{(bottom)} Sketch of the expected (blue arrow) and measured (red arrow) electric field polarization at the LPDA emitted by the transmitting antenna from the nominal (blue)
and measured (red) position. The real transmitting antenna position is shifted from the nominal position, e.g., due to GPS accuracy. This misplacement changes the 
electric-field strength and polarization measured at the LPDA and, therefore, influences the measurement.}
    \label{fig:Calib}
    \end{center}
\end{figure}

The antenna VEL of the LPDA is determined by transmitting a defined signal from a calibrated signal source from different arrival directions and measuring the LPDA response. The signal source consists
of a signal generator, producing known signals, and a calibrating transmitting antenna with known emission characteristics. The transmission measurement needs to be done in the far-field region, which is
fulfilled to a reasonable approximation at a distance of $R > 2\lambda = 20\,\rm{m}$ for the LPDA frequency range of $30\,\rm{MHz}$ to $80\,\rm{MHz}$. \\
In a first calibration campaign \cite{AERA-Antennas} a large weather balloon was used to lift 
the transmitting antenna and a cable to the signal source placed on ground.
As a vector network analyzer was used to provide the source and to measure the AUT output this
transmission measurement allowed to determine both, the VEL magnitude and phase. This setup has the disadvantages that it 
requires calm weather conditions and the cost per flight including the balloon and gas are high. 
Moreover, the cable potentially impacts the measurements if not properly shielded. In this first calibration campaign only the horizontal VEL was investigated. A new calibration campaign was necessary and a new setup was developed.\\
Now, a signal generator as well as a transmission antenna were both mounted beneath a flying drone, a so-called remotely piloted aircraft (RPA), to position the calibration source. 
Hence, the cable from ground to the transmitting antenna is not needed anymore. Furthermore, the RPA is much less dependent 
on wind, and thus it is easier to perform the measurement compared to the balloon-based calibration. The new calibration is performed with a higher repetition rate and with more positions per measurement.\\
During the measurement, the RPA flies straight up to a height of more than $20\,\rm{m}$ and 
then towards the AUT until it is directly above it. Finally, it flies back and lands again at the starting position. A sketch of the setup is shown at the top of Fig.~\ref{fig:Calib}.\\
The RPA used here was an octocopter obtained from the company MikroKopter \cite{OctoXL}. Such an octocopter also has been used for the fluorescence detector \cite{FDCalib} and CROME \cite{CromeCalib} calibrations.
The horizontal octocopter position is measured by GPS and a barometer provides information about the height above ground.
Both are autonomously recorded nearly each second which enables measurements of the VEL with good resolution in zenith angle $\Theta$.
To obtain further improvements of the octocopter position determination an optical method using two cameras taking pictures of the flight was developed. The cameras are placed on orthogonal axes with a distance of
around $100\,\rm{m}$ to the AUT. Canon Ixus 132 cameras \cite{Camera} with a resolution of 16 MegaPixel are utilized. They are set to an autonomous mode where they take pictures every three seconds. 
From these pictures the full flight path of the octocopter can be reconstructed. The method is explained in detail in \cite{OpticalMeth, OpticalMethProc}.
Beside the octocopter position, information about rotation angles (yaw, pitch, roll as defined in Fig.~\ref{fig:Calib}) are recorded during the flight which are later used to determine the orientation
of the transmission antenna with respect to the AUT.\\
The position of the LPDA station was measured by a differential GPS (DGPS) (Hiper V system \cite{DGPS}) and is therefore known with centimeter accuracy.\\
The reference spectrum generator, model RSG1000 produced by the company TESEQ \cite{RSG1000}, is used as the signal generator. 
It continuously produces a frequency comb spectrum between $5\,\rm{MHz}$ and $1000\,\rm{MHz}$ with a spacing of $5\,\rm{MHz}$.
This signal is further amplified in order to accomplish power well above background for the measurement using the LPDA.
The output signal injected into the transmission antenna has been measured twice in the lab using a FSH4 spectrum analyzer from the company Rohde\&Schwarz \cite{FSH4} and using an Agilent N9030A ESA spectrum analyzer 
\cite{Agilent} both with a readout impedance of $50\,\rm{\Omega}$.\\
In an effort to maintain the strict $2.5\,\rm{kg}$ octocopter payload limit, a small biconical antenna
from Schwarzbeck (model BBOC 9217 \cite{Schwarzbeck}) is mounted $0.7\,\rm{m}$ beneath the octocopter. 
This antenna has been calibrated by the manufacturer in the frequency range from $30\,\rm{MHz}$ to $1000\,\rm{MHz}$ 
with an accuracy of $0.5\,\rm{dB}$. This response pattern and its uncertainty comprise all mismatch effects when connecting a $50\,\rm{\Omega}$ signal source to such a transmitting antenna.
The power received at the LPDA during the calibration procedure is measured using the same FSH4 spectrum analyzer as above.\\
The different VEL components mentioned in Eq.~\eqref{eq:VELSpherical} are determined by performing multiple flights in which the orientation of the transmitting antenna is varied
with respect to the AUT. Sketches of the antenna orientations during the flights
are shown on the left side of Fig.~\ref{fig:ExampleVEL}. The horizontal component $|H_{\phi}|$ of the LPDA is measured in the LPDA main axis perpendicular to the LPDA orientation. Then, both antennas are aligned in parallel
for the whole flight. The meridional component $|H_{\theta}|$ is split into two subcomponents: The other horizontally but perpendicular to $\vec{e}_{\phi}$ oriented component $|H_{\theta,\mathrm{hor}}|$ 
and the vertical component $|H_{\theta,\mathrm{vert}}|$. As the orientation of the transmission antennas is the main difference between both measurements, 
the phase $\alpha_{k}$ with $k=(\theta,\mathrm{hor}), (\theta,\mathrm{vert})$ is the same. Then, these two subcomponents are combined to the meridional component $|H_{\theta}|$:
\begin{linenomath}
\begin{equation}
    |H_{\theta}| = \cos(\Theta)|H_{\theta,\mathrm{hor}}| + \sin(\Theta)|H_{\theta,\mathrm{vert}}|.
    \label{eq:HTheta} 
\end{equation}
\end{linenomath}
Both meridional subcomponents are measured in the axis perpendicular to the LPDA main axis.
Therefore, the transmission antenna needs to be rotated by $90\rm{^{\circ}}$ and the flight path needs to start at the $90\rm{^{\circ}}$ rotated position
in comparison to the measurement of $|H_{\phi}|$. For the case of the $|H_{\theta,\mathrm{vert}}|$ measurement the transmitting antenna is vertically aligned.\\
As the receiving power is measured directly at the output of the LPDA amplifier, all matching effects from connecting a transmission line to the LPDA footpoint and the LPDA LNA are taken into account. 
The VEL is calculated using Eq.~\eqref{eq:VEL}.
\section{Calibration Strategy}
To explain the LPDA calibration strategy a measurement of each of the three VEL components is presented. Several flights at different days with different environmental conditions were performed 
and finally combined to give an average LPDA VEL.
Here, one of the measurements of each VEL component is presented to show the reconstruction procedure as well as the statistical precision of the measurements.
Furthermore, the corrections taking into account cable damping, background 
measurements, misalignments of the transmitting antenna and shift of the 
octocopter position are discussed in detail.
Afterwards, an overview of the measurement uncertainties is given.
\subsection{Example Measurement}
\begin{figure}
    \begin{center}
        \raisebox{0.65cm}{\includegraphics[scale=0.26]{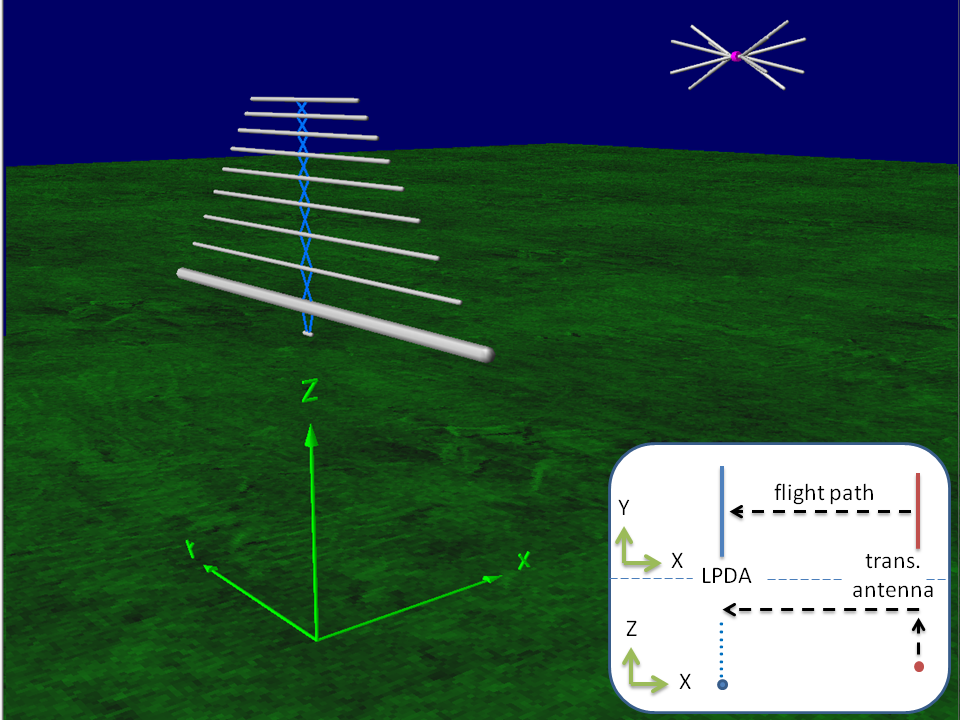}}\includegraphics[scale=0.31]{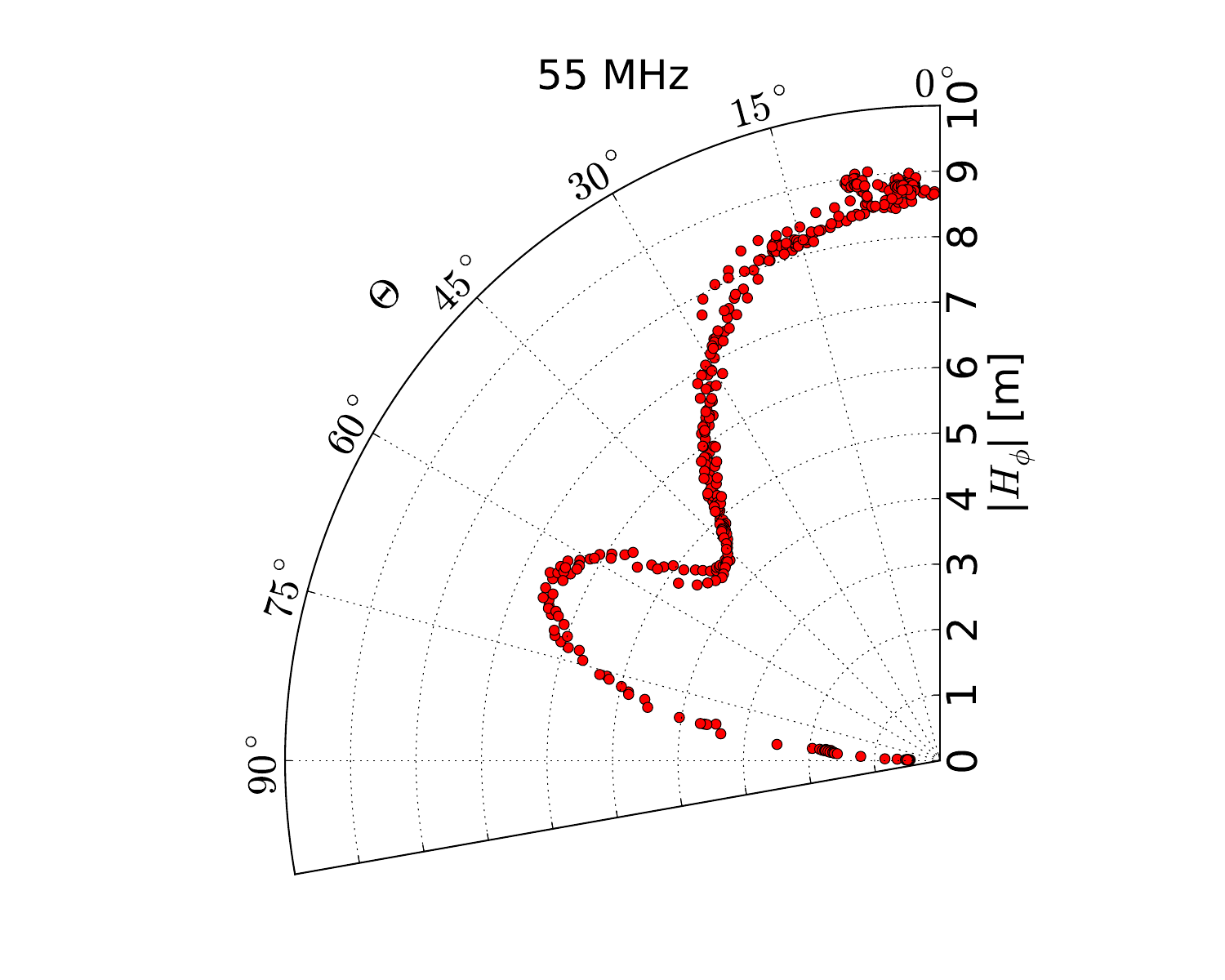}\\
        \raisebox{0.65cm}{\includegraphics[scale=0.26]{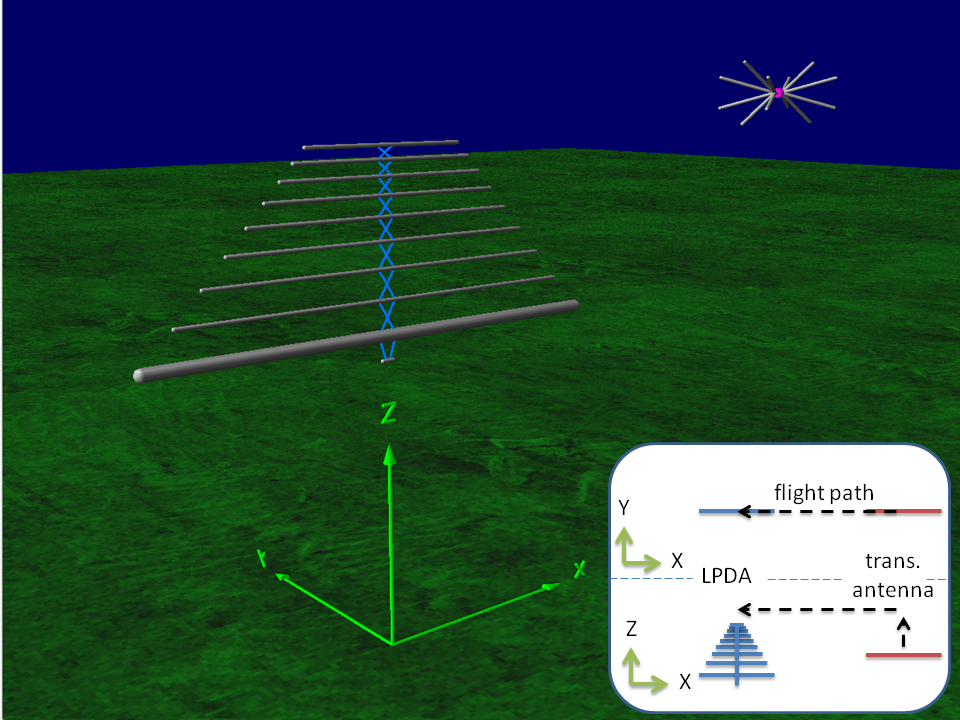}}\includegraphics[scale=0.31]{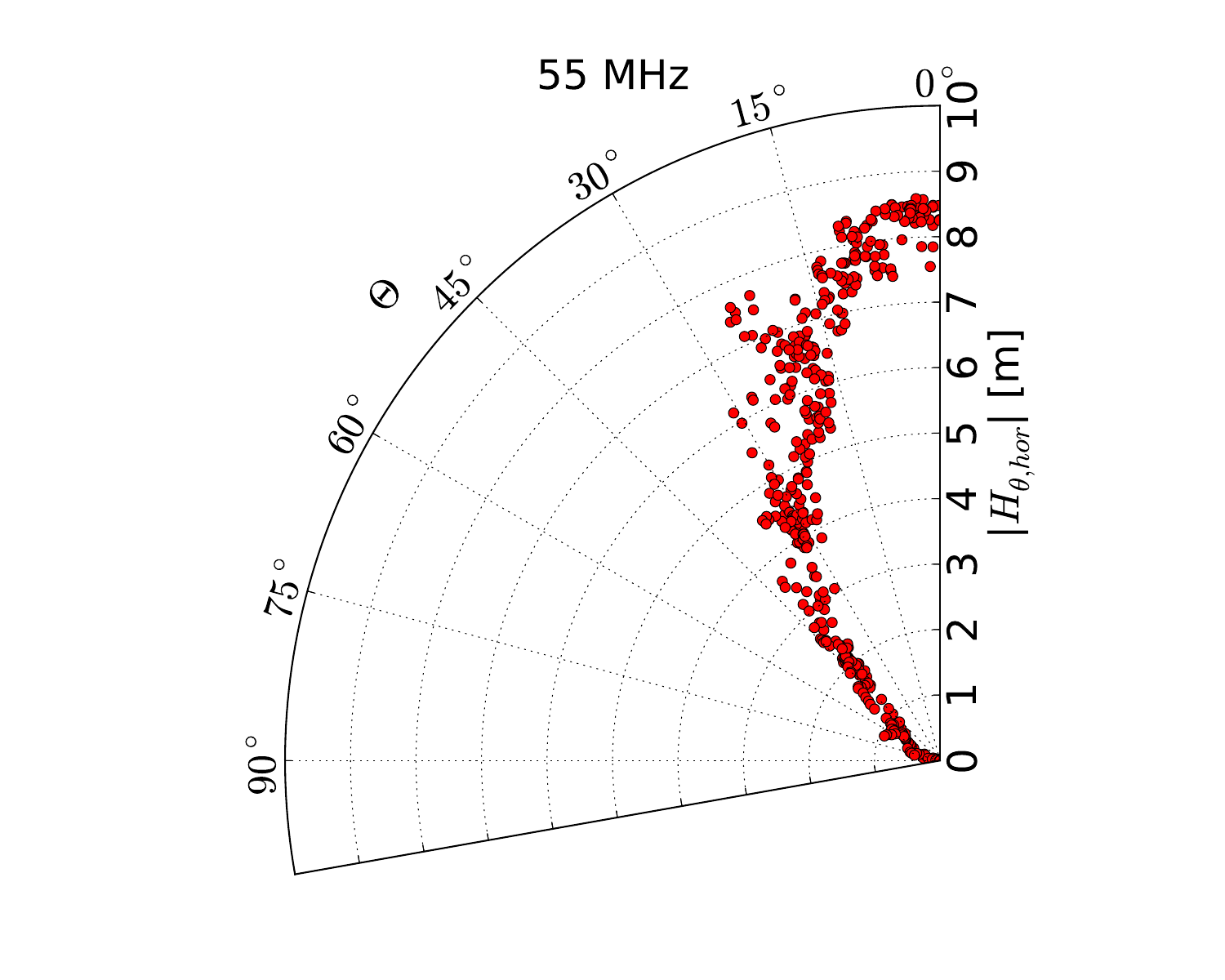}\\\
        \raisebox{0.65cm}{\includegraphics[scale=0.26]{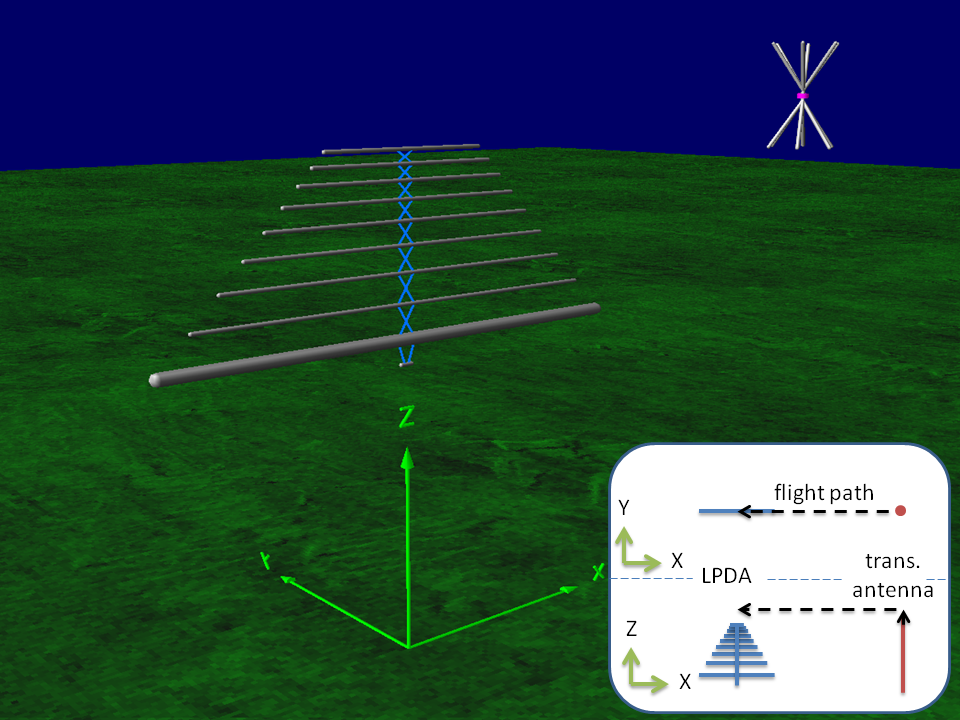}}\includegraphics[scale=0.31]{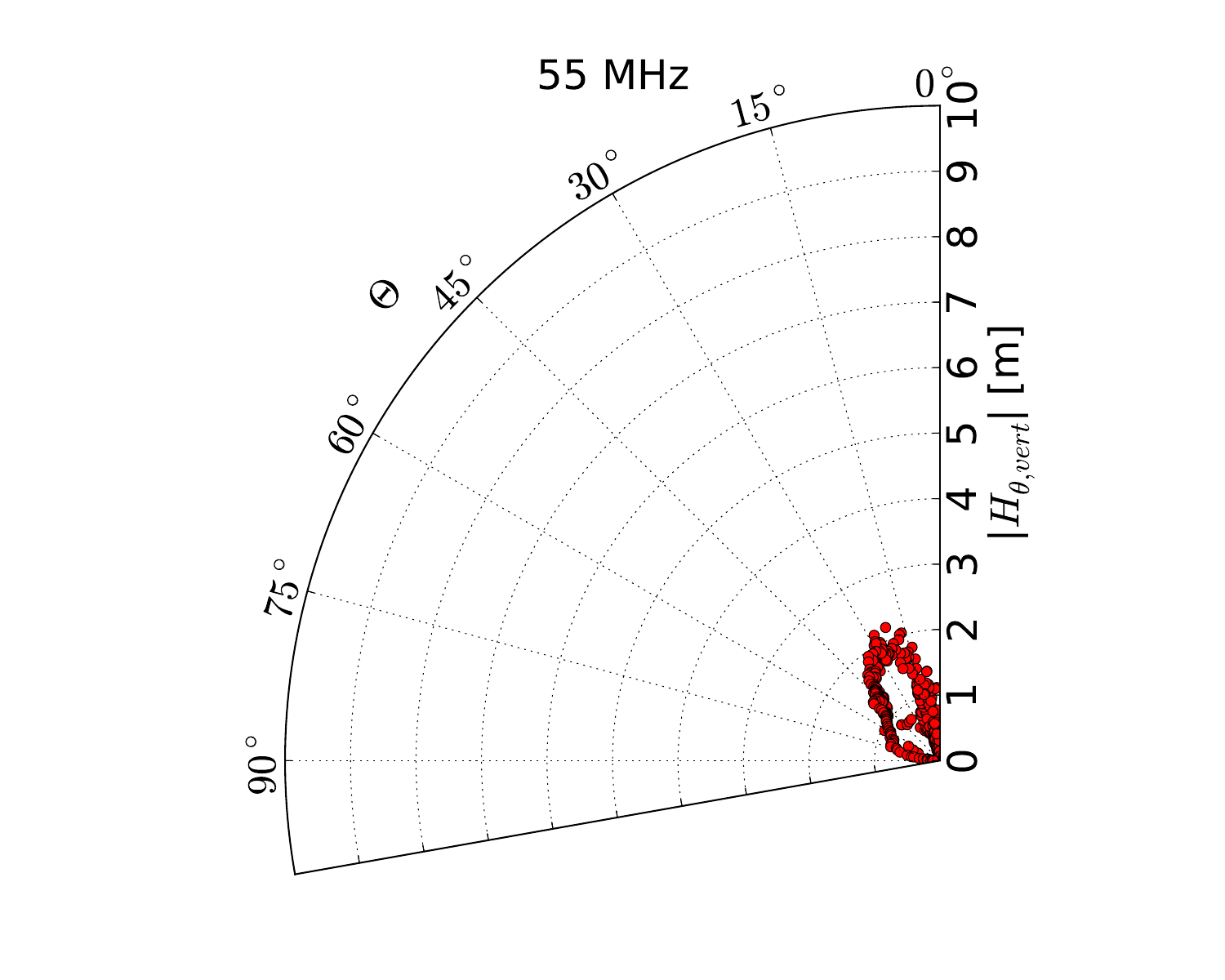}\\
        \caption{\it \textbf{(left)} NEC-2 realization of the setup to simulate the three VEL components \textbf{(from top to bottom)} $|H_{\phi}|$, $|H_{\theta,\mathrm{hor}}|$ and $|H_{\theta,\mathrm{vert}}|$. The meridional component $|H_{\theta}|$ is a combination
        of $|H_{\theta,\mathrm{hor}}|$ and $|H_{\theta,\mathrm{vert}}|$. The distance between transmitting and receiving antenna is reduced and the transmitting antenna
        is scaled by a factor of $3$ to make both antennas visible. For clarity, the LPDA and the transmitting antenna (assumed as a simple dipole) orientations 
        are sketched in the lower right corner of each picture in the XY-plane as well as in the XZ-plane. 
        \textbf{(right)} Measured VEL as function of the zenith angle (red dots) of three example flights for the three VEL components at $55\,\rm{MHz}$. }
    \label{fig:ExampleVEL}
    \end{center}
\end{figure}

In the right diagrams of Fig.~\ref{fig:ExampleVEL} the measured
VEL components $|H_{\phi}|$, $|H_{\theta,\mathrm{hor}}|$ and $|H_{\theta,\mathrm{vert}}|$ at the output of the LPDA LNA as a function of the zenith angle $\Theta$ at $55\,\rm{MHz}$ are shown. 
In the left drawings the respective antenna orientations are visible. 
The antenna response pattern reveals the following features. 
For the VEL component $|H_{\phi}|$, the LPDA is most sensitive in the zenith direction.
The pattern shows a side lobe at around $65\rm{^{\circ}}$. For $|H_{\theta,\mathrm{hor}}|$ the most sensitive direction is the zenith while at larger zenith angles the sensitivity is strongly reduced. At the zenith the components
$|H_{\phi}|$ and $|H_{\theta,\mathrm{hor}}|$ are equal which is expected as the antenna orientations are identical. The fluctuations in $|H_{\theta,\mathrm{hor}}|$ are larger than those in $|H_{\phi}|$ due to the larger dependencies on the octocopter 
rotations. When flying towards the antenna, any acceleration causes a rotation around the pitch angle (Fig.~\ref{fig:Calib}) which does not influence $|H_{\phi}|$. However, for both
meridional subcomponents the pitch angle already changes the transmitting antenna orientation (Fig.~\ref{fig:ExampleVEL}). Therefore, it influences both measurements.
In comparison to the other components $|H_{\theta,\mathrm{vert}}|$ is much smaller. 
Therefore, the LPDA is marginally sensitive to such a signal polarization especially at vertical incoming directions. All these results are frequency dependent. 
\subsection{Corrections}
For the raw VEL determined according to Eq.~\eqref{eq:VEL} corrections for the experimental conditions have to be applied. 
The VEL is averaged in zenith angle intervals of $5\rm{^{\circ}}$. This is motivated by the observed variations in the repeated measurements which were recorded on different days (see e.g. below Fig.~\ref{fig:HorVEL}).
All corrections to the VEL are expressed relative to the measured raw VEL at a zenith
angle of $(42.5\pm2.5)\rm{^{\circ}}$ and a frequency of $55\,\rm{MHz}$ and are listed in Tab.~\ref{tab:Corrections}. The corrections are partly zenith angle and/or frequency dependent.
\begin{table}
  \begin{center}
    \begin{tabular}{lrrr}
      \hline
      \hline
%       \textbf{corrections} & \textbf{$\frac{|H_{\phi}|-|H_{\phi,0}|}{|H_{\phi,0}|}$[\%]} & \textbf{$\frac{|H_{\theta,\mathrm{hor}}|-|H_{\theta,\mathrm{hor},0}|}{|H_{\theta,\mathrm{hor},0}|}$[\%]} & \textbf{$\frac{|H_{\theta,\mathrm{vert}}|-|H_{\theta,\mathrm{vert},0}|}{|H_{\theta,\mathrm{vert},0}|}$[\%]}\\
      \textbf{corrections} & \textbf{$\Delta |H_{\phi}|$ [\%]} & \textbf{$\Delta |H_{\theta,\mathrm{hor}}|$ [\%]} & \textbf{$\Delta |H_{\theta,\mathrm{vert}}|$ [\%]}\\
      \hline
      background noise & $-0.1$ & $-0.5$ & $-0.9$ \\
      cable attenuations & $+44.4$ & $+44.4$ & $+53.2$ \\
      background noise + cable attenuation & $+44.3$ & $+43.7$ & $+51.8$ \\
      octocopter influence & $+0.6$ & $+0.6$ & $-0.2$\\
      octocopter misalignment and misplacement & $+0.3$ & -- & -- \\
      height at take off and landing & $+1.8$ & $+15.8$ & $+5.8$ \\
      height barometric formula & $-5.2$ & $-10.2$ & $-2.5$ \\
      combined height & $-3.6$ & $-5.4$ & $+1.3$ \\
      shift to optical method & $-14.5$ & $-4.8$ & $+0.2$ \\
      combined height + shift to optical method & $-14.6$ & $-5.5$ & $-0.3$ \\
      \hline
      all & $+24.6$ & $+36.4$ & $+51.1$ \\
      \hline
      \hline
    \end{tabular}
    \caption{VEL corrections taking into account different kinds of corrections for the three measured VEL components $|H_{\phi}|$, $|H_{\theta,\mathrm{hor}}|$ and $|H_{\theta,\mathrm{vert}}|$
     of the example flights at a zenith angle of $(42.5\pm2.5)\rm{^{\circ}}$ and a frequency of $55\,\rm{MHz}$ with $\Delta |H_{k}|$ = $\frac{|H_{k}|-|H_{k,0}|}{|H_{k,0}|}$ 
     and $k=\phi, (\theta,\mathrm{hor}), (\theta,\mathrm{vert})$.}
    \label{tab:Corrections}
  \end{center}
\end{table}
The following paragraphs describe the corrections of the raw VEL at the LPDA LNA output from the measurement.
\subsubsection{Background Noise}
During the calibration background noise is also recorded. In a separate measurement the frequency spectrum of the background has been determined and is then subtracted from the calibration signal spectrum.
Typically, the background noise is several orders of magnitude below the signal strength. This is even the case for the component $|H_{\theta,\mathrm{vert}}|$ with lowest LPDA sensitivity. 
For large zenith angles close to $90\rm{^{\circ}}$ and in the case of the component $|H_{\theta,\mathrm{vert}}|$ also for small zenith angles directly above the radio station, however, the background noise and 
the signal can be of the same order of magnitude. In this case, the calibration signal spectrum constitutes an upper limit of the LPDA sensitivity. 
If more than $50\,\rm{\%}$ of the events in a zenith angle bin of $5\rm{^{\circ}}$ are affected, no background is subtracted but half of the 
measured total signal is used for calculating the VEL and a $100\%$ systematic uncertainty on the VEL is assigned.
\subsubsection{Cable Attenuation}
To avoid crosstalk in the LPDA read-out system, the read-out system was placed at a distance of about $25\,\rm{m}$ from the LPDA. The RG58 coaxial cable \cite{RG}, used 
to connect the LPDA to the read-out system, has a frequency-dependent ohmic resistance that attenuates
the receiving power by a frequency-dependent factor $\delta$. To obtain the VEL at the LNA output the cable attenuation is corrected from lab measurements using the FSH4.
\subsubsection{Octocopter Influence}
During the LPDA VEL measurement the transmitting antenna is mounted underneath the octocopter which contains conductive elements and is powered electrically. Therefore, the octocopter itself may change the antenna response
pattern of the transmitting antenna with respect to the zenith angle. To find a compromise between signal reflections at the octocopter and stability during take off, flight and landing, the distance 
between transmitting antenna and octocopter has been chosen to be $0.7\,\rm{m}$. The influence has been investigated by simulating the antenna response pattern of the transmitting antenna with and without
mounting underneath an octocopter. It is found that the average gain of the transmission antenna changes by $0.05\,\rm{dB}$ \cite{PHD}. At a zenith angle of $(42.5\pm2.5)\rm{^{\circ}}$ and a frequency of $55\,\rm{MHz}$
the octocopter influences the transmitting antenna VEL with $0.6\,\rm{\%}$.
\subsubsection{Octocopter Misalignments and Misplacements}
\label{subsec:OctcoMis}
Misalignments and misplacements of the octocopter during the calibration flight have a direct impact on the transmitting antenna position and orientation changing the signal polarization at the position of the AUT.
For this investigation the orientation of the transmission antenna is assumed to 
correspond to a dipole, which holds to a good approximation.
The electric field $\vec{E}_{t}$ emitted from a dipole antenna with orientation $\hat{A}_{t}$ in the direction $\hat{n}$ in the far-field region is 
proportional to $\vec{E}_{t} \sim ( \hat{n} \times \hat{A}_{t} ) \times \hat{n}$, and the amplitude is given by $|\vec{E}_{t}| = \sin(\alpha)$. Here, $\alpha$ describes the smallest angle between the transmitting
antenna alignment $\hat{A}_{t}$ and the direction from the transmitting antenna to the AUT denoted as $\hat{n}$ (see lower sketch of Fig.~\ref{fig:Calib}).
The orientation of the transmitting antenna $\hat{A}_{t}$ is 
calculated by an intrinsic rotation of the initial orientation of the transmitting antenna rotating first by the yaw angle $G$, then by the pitch angle $P$ and finally, by the roll angle $R$. 
The AUT sensitivity $\eta$ to the emitted electric field is then calculated by the absolute value of the scalar product of the electric field and the AUT orientation $\hat{A}_{r}$: 
$\eta = |\vec{E}_{t} \cdot \hat{A}_{r}| = \sin(\alpha) \cos(\beta)$ with $\beta$ describing the smallest angle between $\vec{E}_{t}$ and $\hat{A}_{r}$ (see lower sketch of Fig.~\ref{fig:Calib}).
Finally, the correction factor $\epsilon$ of the power measured at the AUT is determined by the square of the quotient of the nominal and the real value of $\eta$.
In case of the horizontal component $|H_{\phi}|$ the VEL is systematically shifted to larger values for all zenith angles and frequencies due to the octocopter misalignment and misplacement. 
The correction factor $\epsilon$ is used to determine the horizontal VEL $|H_{\phi}|$. 
In both meridional subcomponents the VEL becomes small at large zenith angles and strongly dependent on the antenna alignments. 
Therefore, in the meridional subcomponents $|H_{\theta,\mathrm{hor}}|$ and $|H_{\theta,\mathrm{vert}}|$ the effects of the octocopter misalignment and misplacement are included in the systematic uncertainties.
\subsubsection{Octocopter Flight Height}
\label{par:CopterHeight}
\begin{figure}
    \begin{center}
        {\includegraphics[scale=0.3]{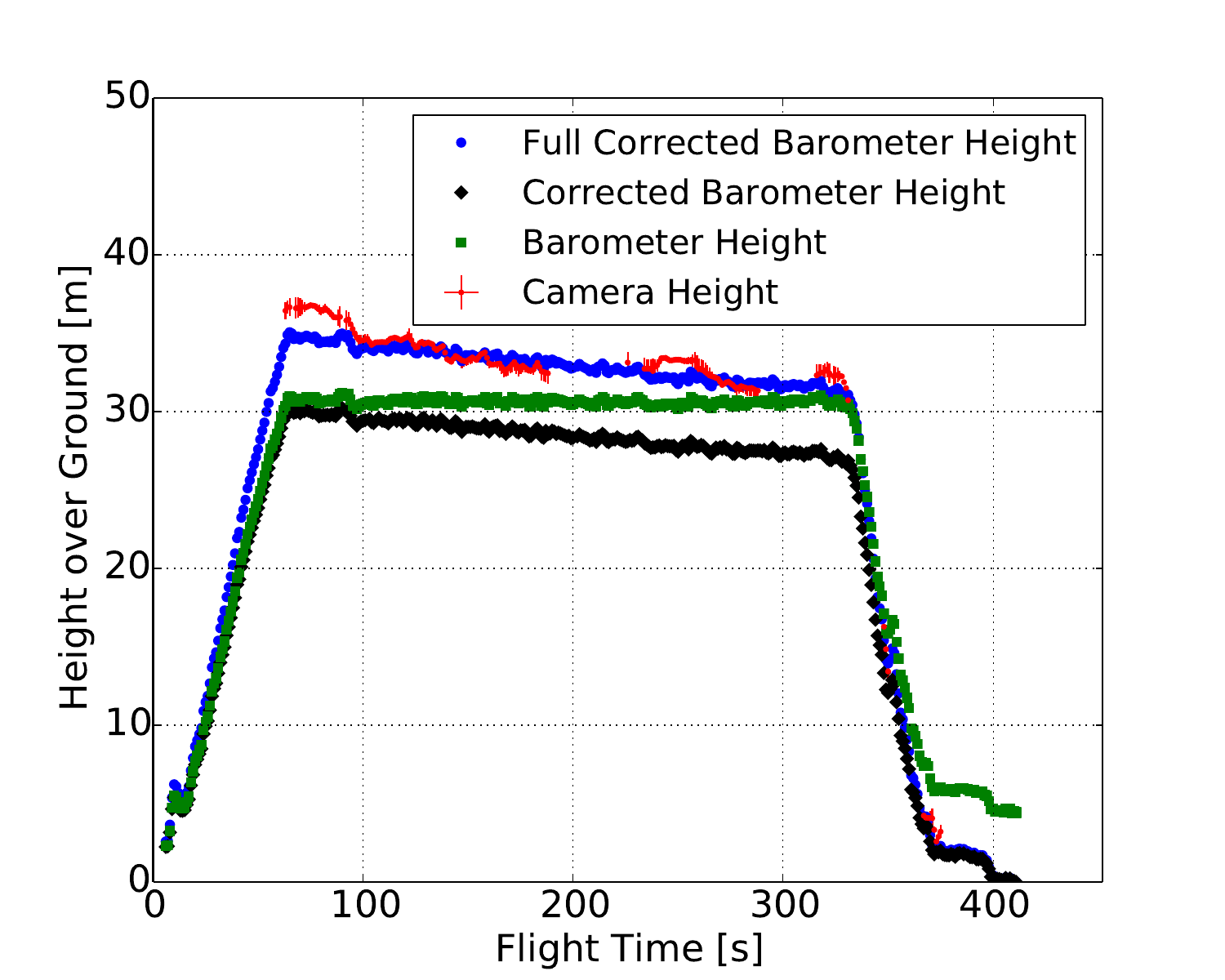}}\includegraphics[scale=0.3]{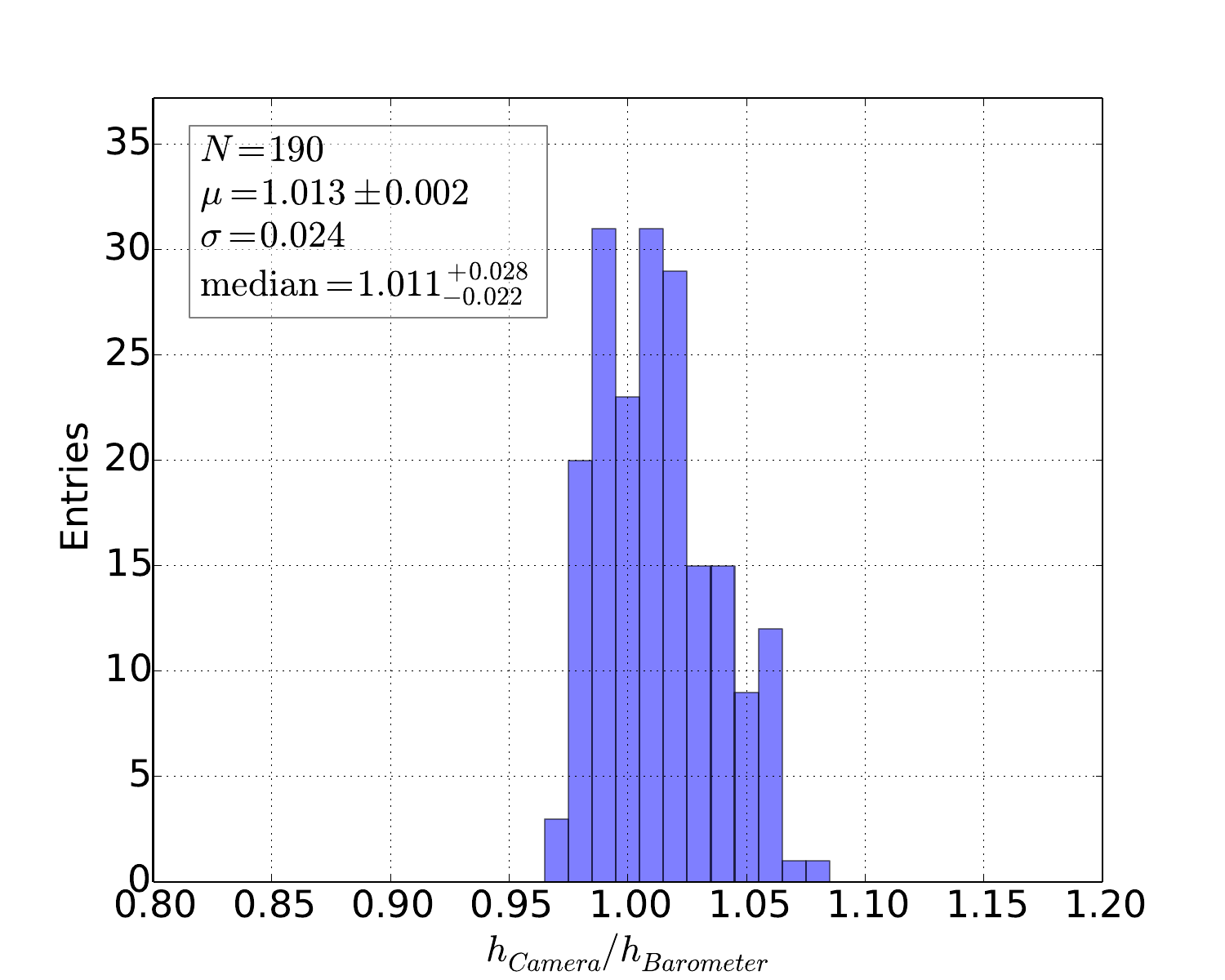}
        \caption{\it  \textbf{(left)} Corrections for the measured octocopter height with the raw
data denoted by the green rectangles. The black diamonds
refer to the height after linear correction for the start and end
positions. The blue circular symbols show the corrections for the linear
barometric formula used in the octocopter electronics. The octocopter height determined by the optical method is denoted by the red dots. All measurements
are shown as a function of the flight time.  \textbf{(right)} Histogram of the quotient of the full corrected barometer height and measured height from the optical method.}
    \label{fig:CopterHeight}
    \end{center}
\end{figure}

The octocopter flight height is determined by a barometer measuring the change of air pressure $\Delta p$ during the flight. The octocopter software assumes a linear dependency of $\Delta p$ and the octocopter 
flight height over ground $h_{raw}$. Two corrections have been applied to the raw flight height. Firstly, it was observed that the flight height differs at take off and landing. Therefore, a linear time dependent
correction is applied which constrains the flight height over ground at take off and landing to zero. Secondly, AERA is located at a height of about $1550\,\rm{m}$ above sea level. Therefore, such a linear relation between 
$\Delta p$ and $h_{raw}$ used by the octocopter software is not precise enough. A more realistic calculation considering an exponential model of the barometric formula \cite{BarometricFormula} as well as the height and
latitude dependent gravitation is used to determine the more precise octocopter height $h_{octo}$. An inverse quadratic relation between gravitation and the height above sea level with a value at sea level of 
$g(0) = 9.797\,\rm{\frac{m}{s^{2}}}$ at the latitude of AERA is taken into account. The raw octocopter height as well as the height after all corrections of the $|H_{\phi}|$ example flight are shown on 
the left side of Fig.~\ref{fig:CopterHeight} in comparison to the octocopter height determined with the optical method. Both methods agree at the level of $1.1\,\rm{\%}$ in the median. 
The quotient of the octocopter height measured by the camera method and by the full corrected barometer method is shown in the histogram on the right side of Fig.~\ref{fig:CopterHeight}.
The optical method is used to correct for the small difference.
\subsubsection{Octocopter Position Shift from Optical Method Position 
Reconstruction}
While the octocopter position measured by the built-in sensors (air pressure, GPS) is recorded nearly each second, the cameras used in the optical method take pictures of the flight every $3\,\rm{s}$. Furthermore, it turned out that the
fluctuations of the built-in sensors are smaller in comparison to the optical method.
Nevertheless, the systematic uncertainties of the octocopter position reconstruction using the optical method are still much smaller. The uncertainties are described in detail in the following
subsection. To combine both advantages of high statistics and small uncertainties, the octocopter position measured by the built-in sensors is taken and then shifted towards the position measured with the optical
method. Therefore, the octocopter position in the XY-plane is shifted by the median distance and the octocopter height measured by the barometer is shifted by the median factor between both methods.
For the $|H_{\phi}|$ example flight the octocopter XY-position measured by GPS is shifted by $0.83\,\rm{m}$ to the west and $3.22\,\rm{m}$ to the south.
The full corrected flight height measured by the barometer is shifted by $1.1\,\rm{\%}$.
\subsection{Uncertainties}
In this subsection the statistical and systematic uncertainties are discussed using the $|H_{\phi}|$ example flight at a middle frequency of $f=55\,\rm{MHz}$ and a zenith angle bin of $(\Theta=42.5\pm2.5)\,\rm^{\circ}$
as mentioned above. 
This zenith angle is chosen as most events at AERA are reconstructed coming from this direction. While some systematic uncertainties are stable between flights,
e.g., measurement of the power injected to the transmitting antenna or the transmitting antenna response pattern, others are flight dependent, e.g., the octocopter position and the measurement of the 
receiving power at the AUT. 
The VEL relative uncertainties are listed in Tab.~\ref{tab:Uncertainties}. These individual uncertainties are described in detail in the following subsections.
The constant systematic uncertainties add quadratically to $6.3\,\rm{\%}$ and the flight dependent systematic 
uncertainty is $6.9\,\rm{\%}$.
\newcommand\tw{0.5cm}
\newcommand\tww{0.2cm}
\begin{table}[tbp]
   \begin{center}
      \begin{tabularx}{\textwidth}{Xrrr}
      \hline \hline 
      \rule{0pt}{3ex}\textbf{source of uncertainty / \%} & systematic  & statistical \\
      \hline
      \rule{0pt}{4ex}\textbf{flight dependent uncertainties} & \textbf{6.9} & \textbf{2.7}\\
      \hspace*{\tww}      transmitting antenna XY-position & $1.5$ & $1.0$ \\
      \hspace*{\tww}      transmitting antenna height & $0.1$ & $0.6$ \\
      \hspace*{\tww}      transmitting antenna tilt & $<0.1$ & $<0.1$ \\
      \hspace*{\tww}      size of antenna under test & $1.4$ & - \\
      \hspace*{\tww}      uniformity of ground & $<0.1$ & - \\
      \hspace*{\tww}      RSG1000 output power & $2.9$ & $2.3$ \\
      \hspace*{\tww}      influence of octocopter & $<0.1$ & - \\
      \hspace*{\tww}      electric-field twist & $0.4$ & $0.2$ \\
      \hspace*{\tww}      LNA temperature drift & $1.0$ & $0.6$ \\
      \hspace*{\tww}      receiving power & $5.8$ & - \\
      \hspace*{\tww}      background & $0.4$ & - \\
  
      \rule{0pt}{4ex}\textbf{global uncertainties} & \textbf{6.3} & \textbf{$<$0.1} \\
      \hspace*{\tww}      injected power & $2.5$ & $<0.1$ \\
      \hspace*{\tww}      transmitting antenna gain & $5.8$ & - \\
      \hspace*{\tww}      cable attenuation & $0.5$ & $<0.1$ \\
  
      %\rule{0pt}{-5ex}\\
      \hline 
      \rule{0pt}{3ex}\textbf{all / \%}  & \textbf{9.3} &  \textbf{4.7} \\ 
      \hline \hline
      \end{tabularx}
   \caption{Uncertainties of the horizontal VEL $|H_{\phi}|$ of the example flight at $55\,\rm{MHz}$ and $(42.5\pm2.5)\,\rm^{\circ}$.
   While the overall systematic uncertainty is the quadratic sum of each single systematic uncertainty,
   the overall statistical uncertainty is described by the observed signal fluctuation during the measurement.
   The statistical uncertainty of each source of uncertainty describes the expected uncertainty,
   e.g., from the manufacturer's information.}
   \label{tab:Uncertainties}
   \end{center}
\end{table}
\subsubsection{Transmitting Antenna Position}
\begin{figure}
    \begin{center}
	\includegraphics[scale=0.29]{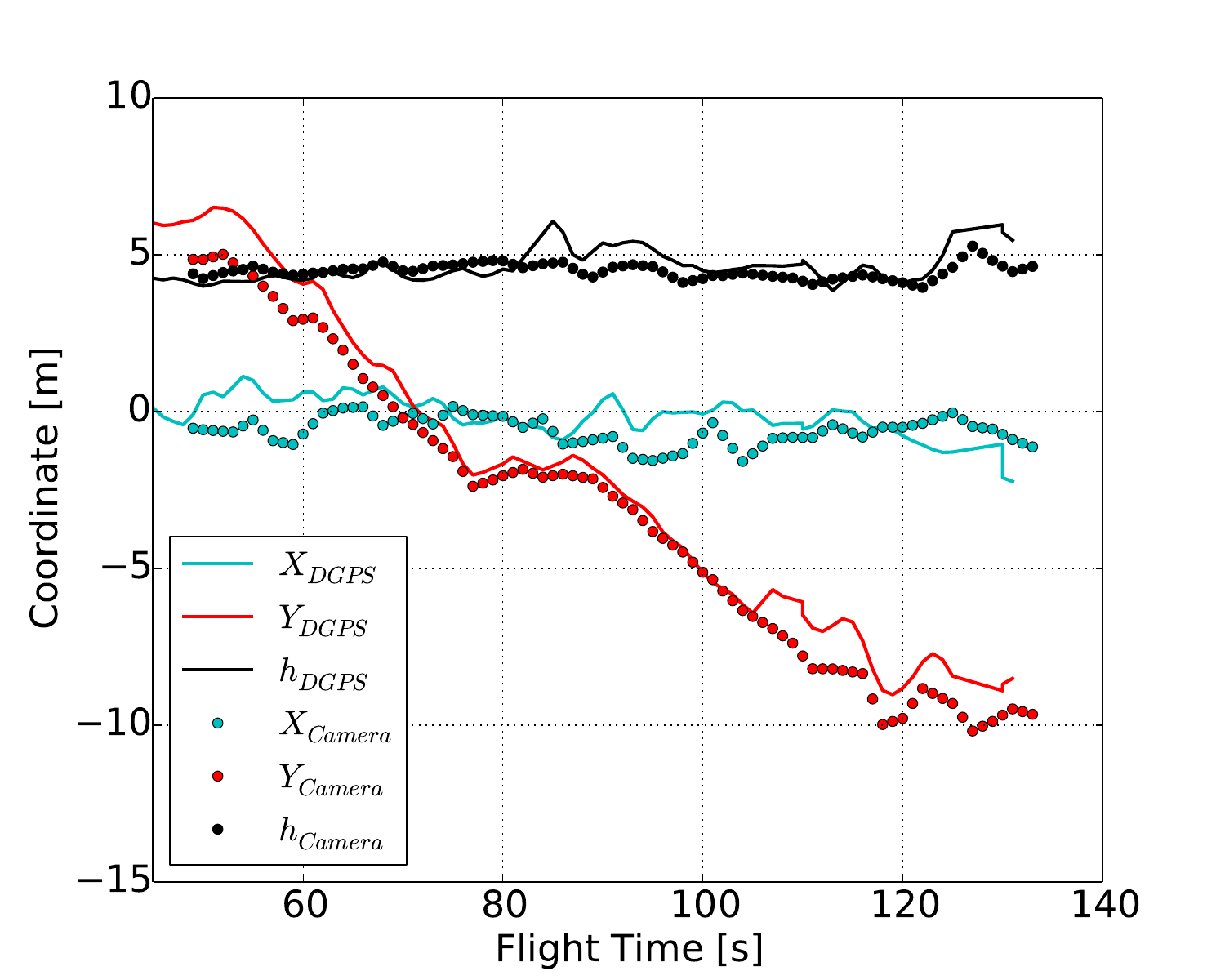}
	\includegraphics[scale=0.29]{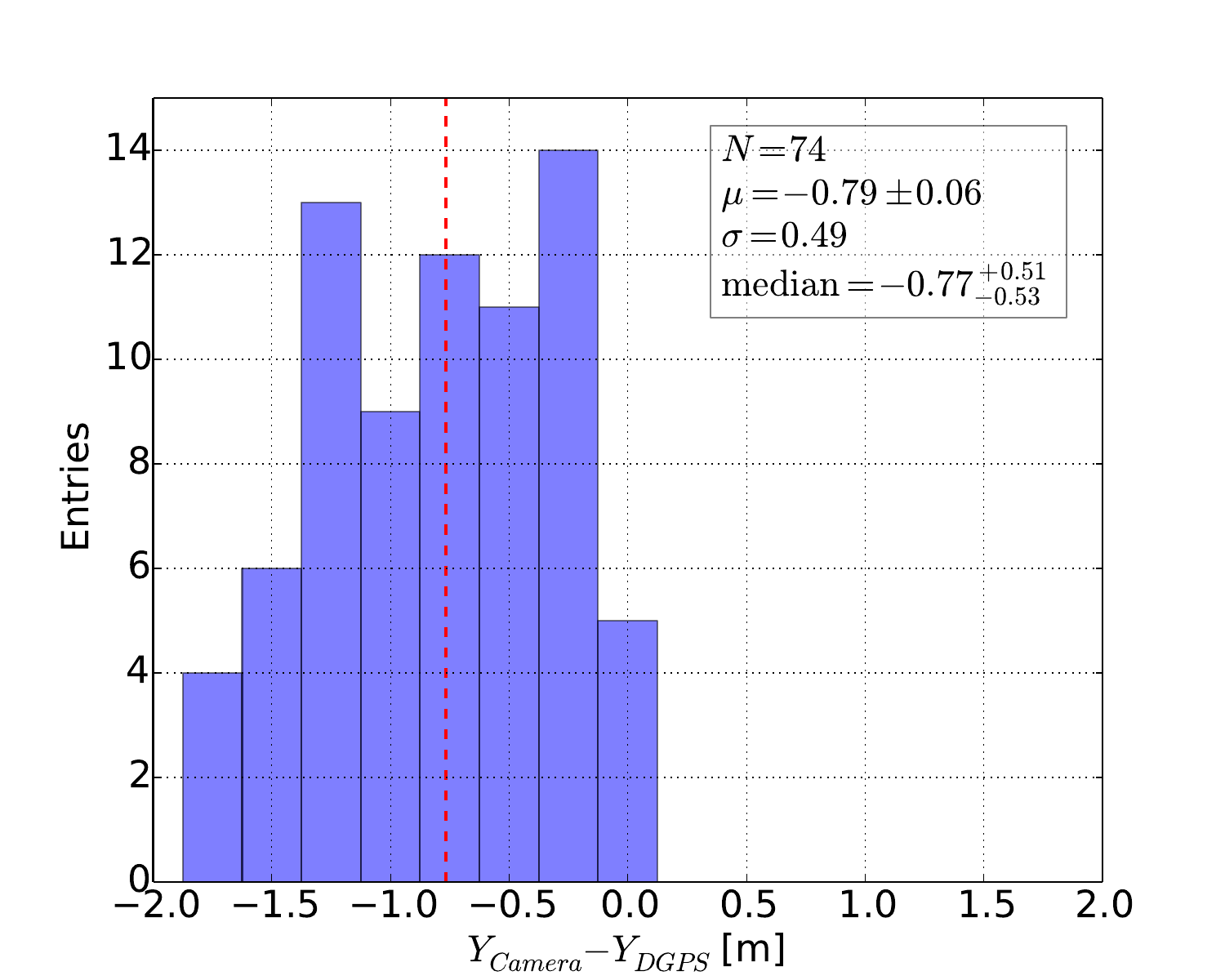}
        \includegraphics[scale=0.29]{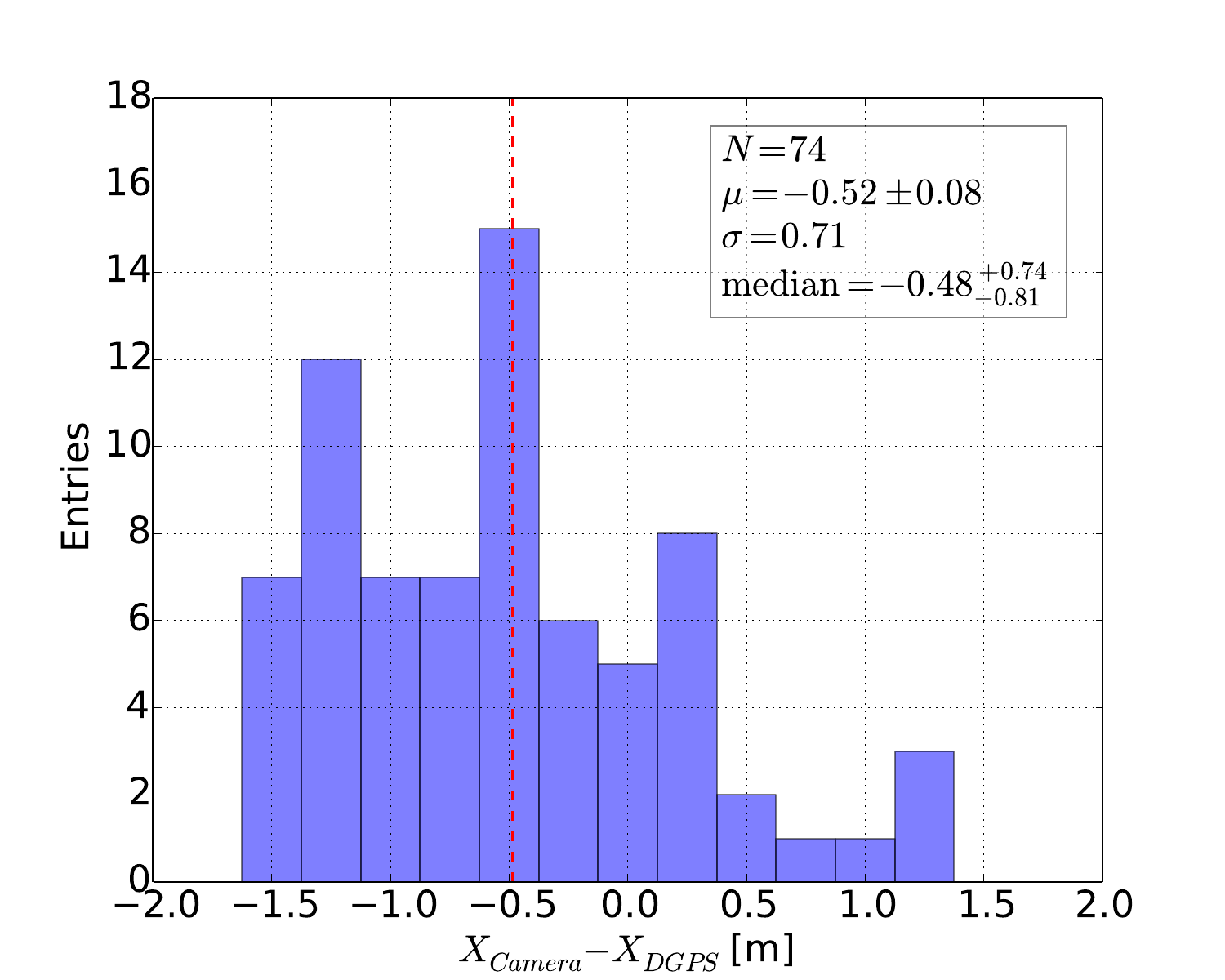}        
        \includegraphics[scale=0.29]{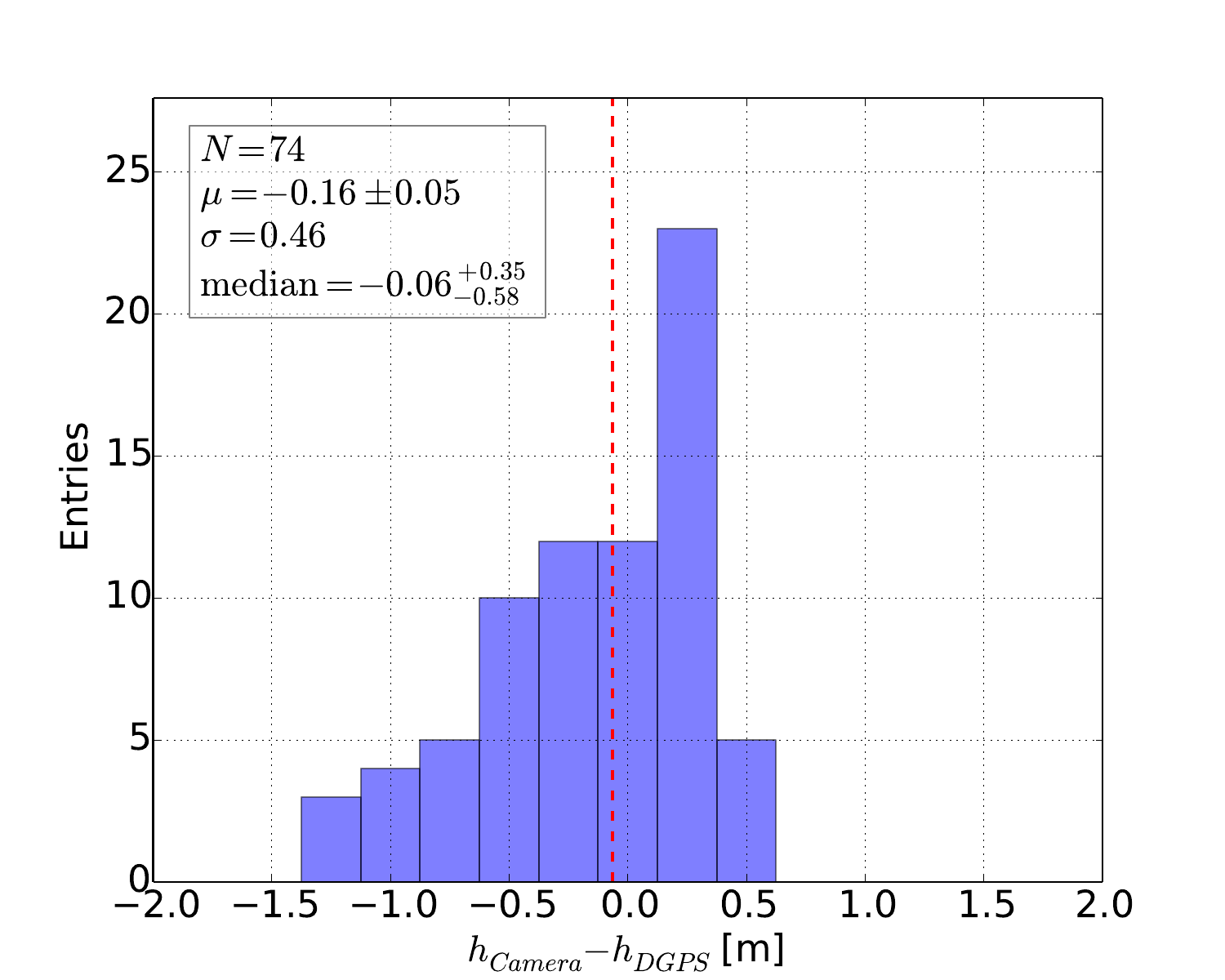}
        \caption{\it Comparison of the octocopter position measured with the optical method and with an additional DGPS mounted at the octocopter during one flight. \textbf{(upper left)} Raw position data measured with 
        DGPS (lines) and the optical method (dots) as function of the flight time. The distance between the reconstructed octocopter position measured by optical method and DGPS
        in X and Y direction are shown in the \textbf{(upper right)} and \textbf{(lower left)} figure. The difference of the octocopter height measured by the barometer and DGPS is shown in the \textbf{(lower right)} figure.
        The systematic uncertainty in the XY-plane of the octocopter position is calculated by the quadratic sum of both median values (red dashed lines) in X and Y direction.
        Similarly, the median of the height difference of both measurement setups is taken as systematic uncertainty of the octocopter height.}
    \label{fig:Camera-DGPS}
    \end{center}
\end{figure}

\begin{figure}
    \begin{center}
        {\includegraphics[scale=0.29]{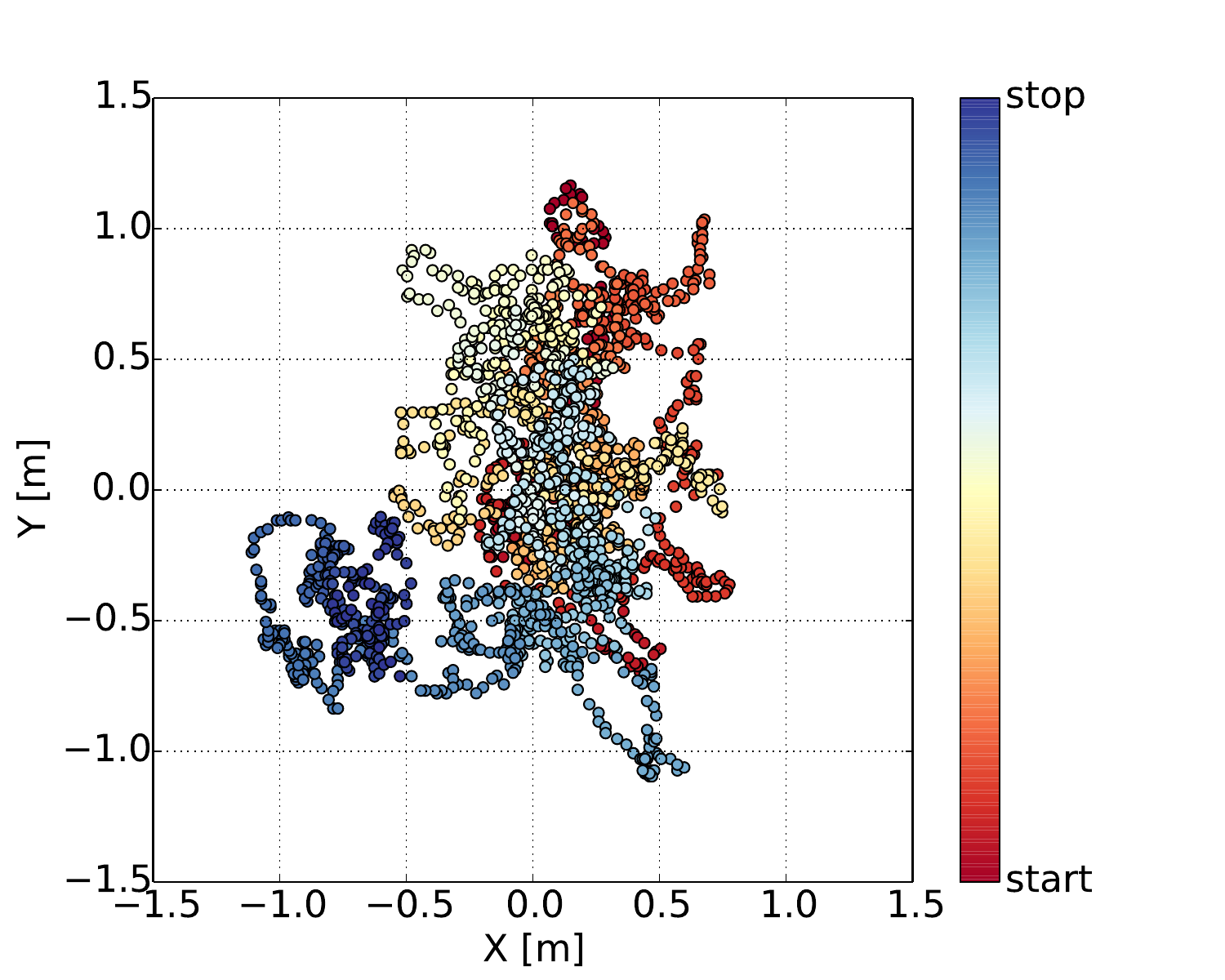}}
        {\includegraphics[scale=0.29]{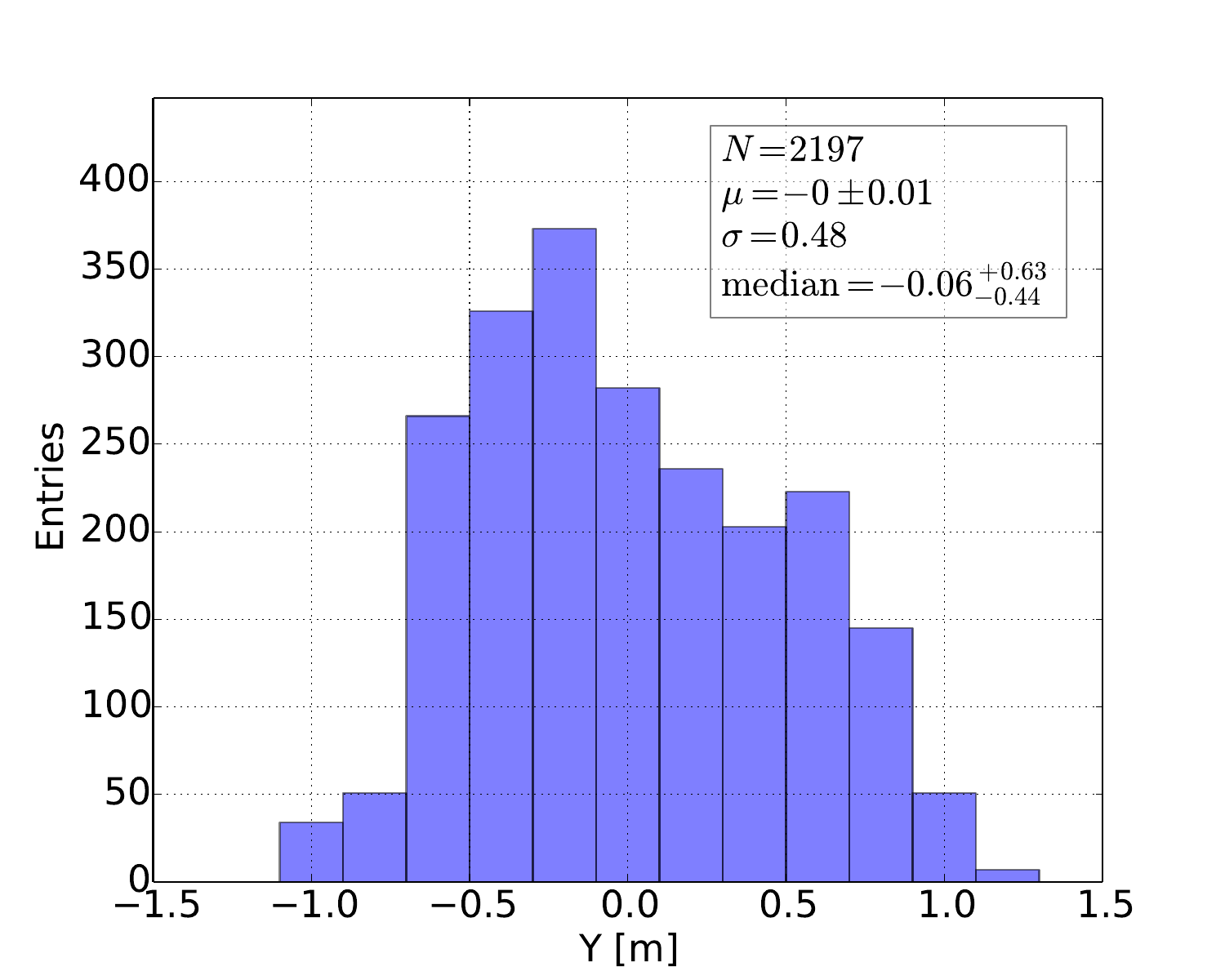}}
        {\includegraphics[scale=0.29]{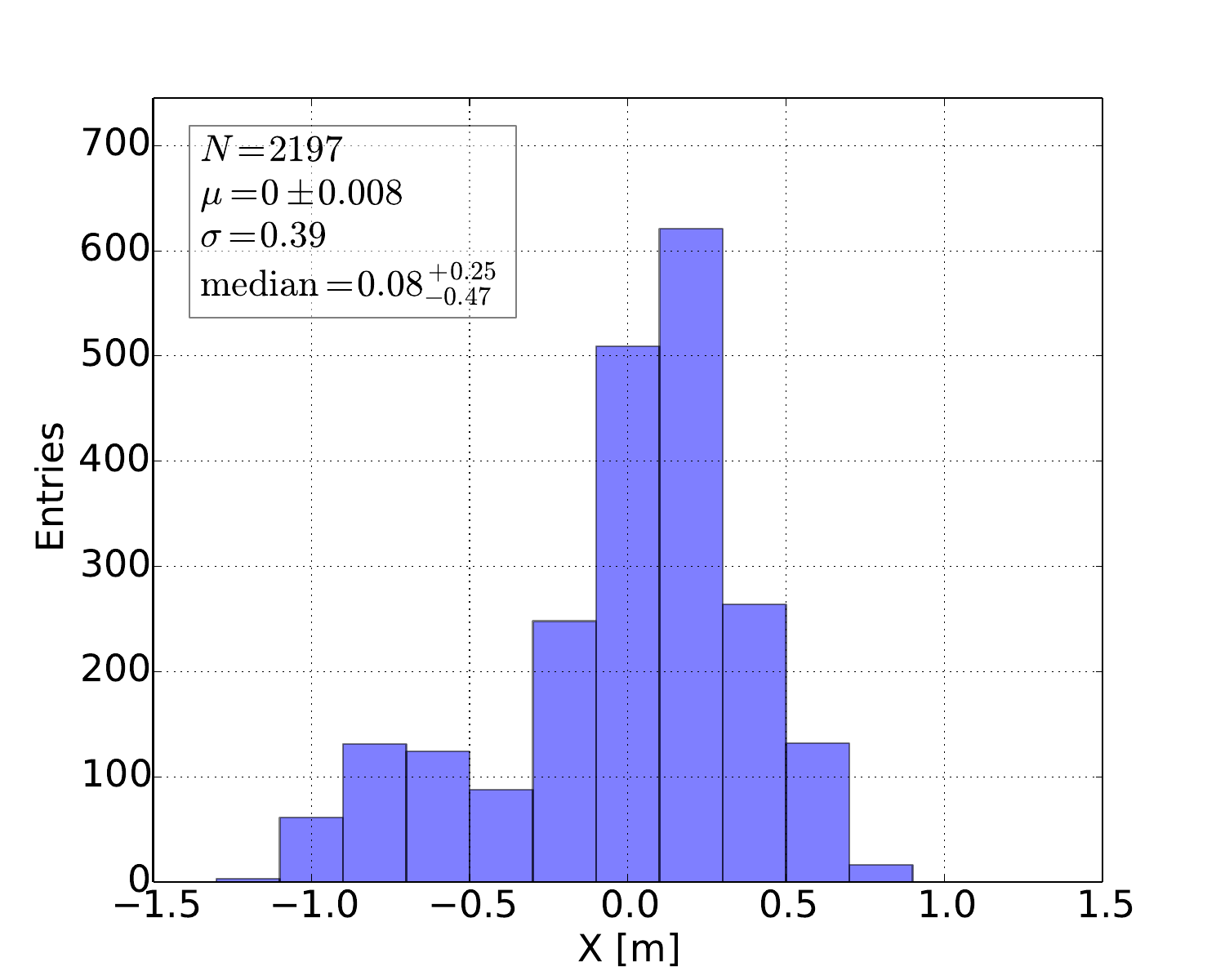}}
        {\includegraphics[scale=0.29]{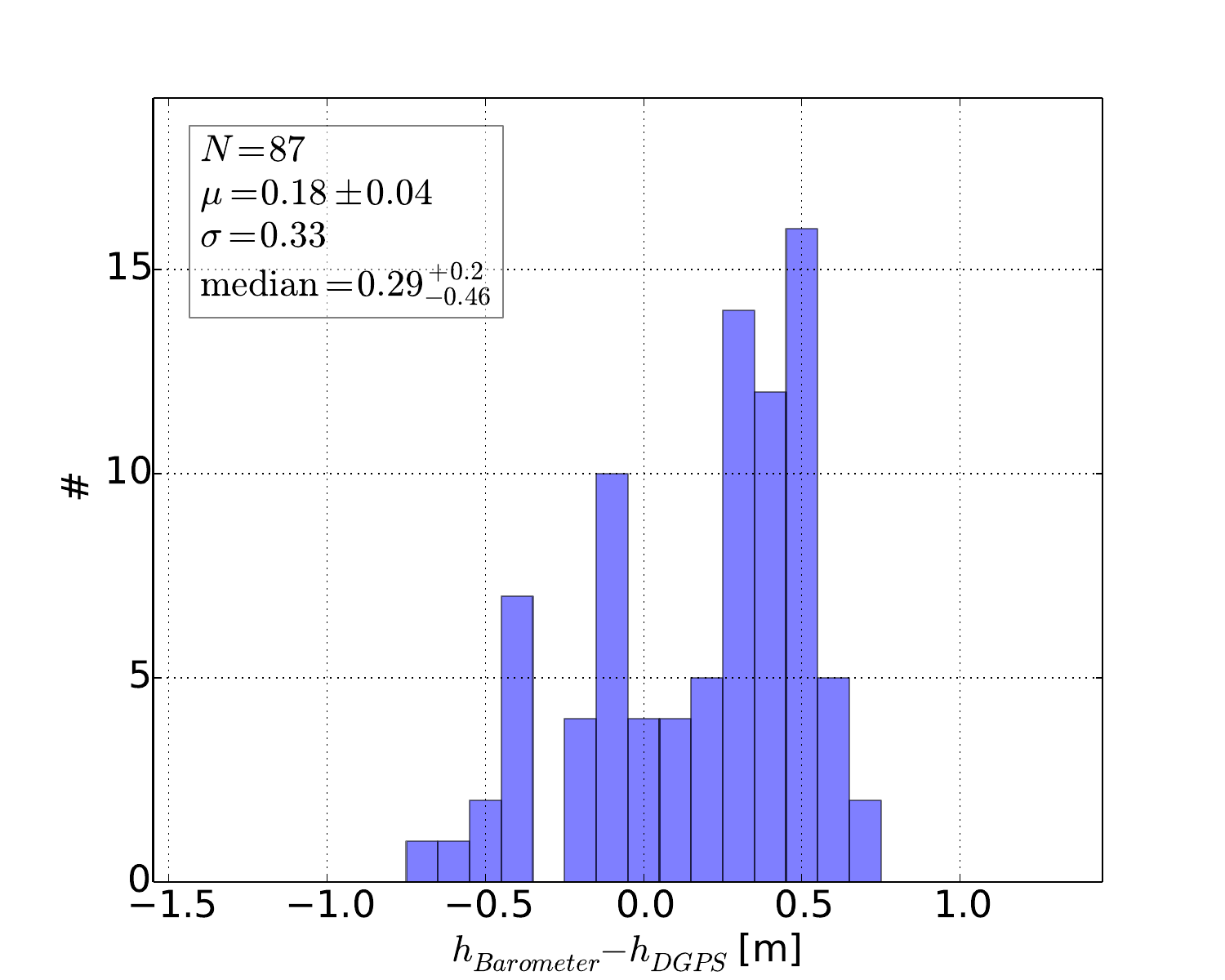}}
    \caption{\it The statistical uncertainties of the octocopter position reconstruction using the built-in sensors. The uncertainty of the horizontal position is determined in a measurement while the 
    octocopter is on ground and does not move. \textbf{(upper left)} Measured octocopter GPS-position with respect to the average position at $(0,0)$. Color coded is the time.
    \textbf{(upper right)} Histogram of the distance between measured and average position in Y direction.
    \textbf{(lower left)} Histogram of the distance between measured and average position in X direction.
    \textbf{(lower right)} The statistical uncertainty of the octocopter height 
    measured with the barometer is determined by comparing the measured flight height with the height measured using a DGPS. Then, uncertainties arising from the height corrections are taken into account. 
    The histogram of the octocopter height difference over ground measured with the barometer compared to the DGPS measurement is shown.}
    \label{fig:CopterPosition}
    \end{center}
\end{figure}

The systematic uncertainty of the position reconstruction of the optical method was determined by comparing the reconstructed octocopter position with the 
position measured by a DGPS which gives an accurate position 
determination.
The combined mass of the transmission antenna and the additional DGPS exceeds the maximal payload capacity of the octocopter. Therefore, a separate flight with DGPS but without transmitting antenna 
and source generator was performed. The octocopter positions measured with the optical method and the DGPS are compared in Fig.~\ref{fig:Camera-DGPS}.
The systematic uncertainty of the octocopter position in the XY-plane is calculated using the quadratic sum of both median values (red dashed lines) in the X and Y direction which is smaller than
$1\,\rm{m}$. Equally, the systematic uncertainty of the octocopter height is $\sigma_{h}=0.06\,\rm{m}$. The influence on the VEL is determined by shifting the reconstructed octocopter position by these
uncertainties and redoing the VEL calculation given in Eq.~\eqref{eq:VEL} of each zenith angle bin separately for the XY-plane and the height. The VEL systematic uncertainty is given by half the difference of the upper
and lower shift of the VEL. The systematic uncertainty on the VEL at a zenith angle of $\Theta = 42.5\rm{^{\circ}} (2.5\rm{^{\circ}}, 72.5\rm{^{\circ}}) \pm 2.5\rm{^{\circ}}$
due to the octocopter's XY-position is $1.5\,\rm{\%}$ ($0.2\,\rm{\%}$, $2.9\,\rm{\%}$) and due to the octocopter's height is $0.1\,\rm{\%}$ ($0.2\,\rm{\%}$, $<0.1\,\rm{\%}$). \\
The statistical uncertainty of the octocopter's built-in sensors is determined in the following way.
The flight height measured by the barometer has to be corrected as described in section \ref{par:CopterHeight}
which causes further uncertainties during the flight. The statistical uncertainty of the octocopter height measured with the barometer is then determined by comparing the measured height
with the height measured by the DGPS (lower right panel of Fig.~\ref{fig:CopterPosition}). The statistical uncertainty are found to be $\sigma = 0.33\,\rm{m}$ which results in a $0.6\,\rm{\%}$ uncertainty in the VEL.
The horizontal position uncertainties are determined in a measurement where the octocopter remains stationary on the ground. The measurement is presented in Fig.~\ref{fig:CopterPosition}.
The diagrams show a statistical uncertainty of $\sigma = \sqrt{0.48^{2}+0.39^{2}}\,\rm{m} = 0.6\,\rm{m}$ in the XY-plane which results in a $1.0\,\rm{\%}$ uncertainty in the VEL.
All these uncertainties are smaller than those
of the optical method described by the widths of the distributions shown in Fig.~\ref{fig:Camera-DGPS} where the octocopter positions measured with DGPS and the camera method are compared.\\
The transmission antenna is mounted at a distance of $s_{Ant}=0.7\,\rm{m}$ beneath the octocopter. Hence, a tilt of the octocopter, described by the pitch and the roll angle,
changes the position in the XY-plane of the transmission antenna as well as its height over ground. In the case of the example flight, the average pitch (roll) angle of the octocopter is $-0.6\rm{^{\circ}}$ ($0.9\rm{^{\circ}}$)
which lead to a systematic uncertainty smaller than $0.1\,\rm{\%}$ at $55\,\rm{MHz}$ and $(42.5\pm2.5)\,\rm^{\circ}$.
\subsubsection{Size of AUT}
The size of the LPDA in the z-direction is $1.7\,\rm{m}$. The interaction point of the signal at each frequency is set to the center of the LPDA. Therefore, there is a systematic uncertainty
in the height interval between transmitting antenna and AUT which is conservatively estimated to be $0.85\,\rm{m}$. For the example flight, this systematic results in a VEL
systematic uncertainty of $1.4\,\rm{\%}$ at $55\,\rm{MHz}$ and $(42.5\pm2.5)\,\rm^{\circ}$.
\subsubsection{Uniformity of Ground Height}
The ground height above sea level at the octocopter starting position and at the LPDA is measured by DGPS. The ground is not completely flat but varies at the level of a few $\rm{cm}$
over a distance of $5\,\rm{m}$ which is incorporated as additional uncertainty on the height. The resulting influence on the VEL is less than $0.1\,\rm{\%}$.
\subsubsection{Emitted Signal towards the Antenna Under Test}
The uncertainty of the emitted signal contains effects from the power output of the RSG1000, the injected
power into the transmission antenna, the transmission response pattern, the influence of the octocopter on the pattern as well as the misalignment and misplacement of the transmitting antenna which changes the 
emitted power transmitted towards the AUT and lead to a twist of the signal polarization at the AUT.\\
The manufacturer of the RSG1000 states a signal stability of $0.2\,\rm{dB}$ measured at a constant temperature of $20\,\rm{^{\circ}}$ which results in a statistical uncertainty of $2.3\,\rm{\%}$ in the VEL.
The calibration measurements were performed at temperatures between $15\,\rm{^{\circ}C}$ and $25\,\rm{^{\circ}C}$. Here, the manufacturer denotes a systematic uncertainty of $0.25\,\rm{dB}$ due to temperature shifts which
results in $2.9\,\rm{\%}$ in the VEL.\\
The injected power from the RSG1000 to the transmission antenna is measured twice in the lab using the FSH4 spectrum analyzer averaged over $100$ samples and a Agilent N9030A ESA spectrum analyzer 
averaged over $1000$ samples. The systematic uncertainty of the FSH4 measurement is $0.5\,\rm{dB}$ and the systematic uncertainty of the Agilent N9030A ESA measurement is $0.24\,\rm{dB}$. 
Both are combined yielding a total systematic uncertainty of $0.22\,\rm{dB}$ in the VEL. As there is a quadratic relation between injected power and the VEL (refer to Eq.~\eqref{eq:VEL}) 
the systematic uncertainty on the VEL is $2.5\,\rm{\%}$. The statistical uncertainties of these measurements are small due to the number of samples
and can be neglected.\\
The antenna manufacturer specifies a systematic uncertainty of the transmitting antenna pattern of $0.5\,\rm{dB}$ which results in a systematic uncertainty on the VEL of $5.8\,\rm{\%}$.
The influence of the octocopter on the transmission antenna pattern investigated with simulations is small \cite{PHD} and, therefore, a systematic uncertainty due to the octocopter influence on the transmission
antenna pattern can be neglected.\\
Misalignment and misplacement of the transmitting antenna lead to a twist of the signal polarization and furthermore, altered the signal strength at the AUT.
The AUT sensitivity to an electric field is given by $\eta = \sin(\alpha) \cos(\beta)$ with the angles $\alpha$ and $\beta$ as described in section \ref{subsec:OctcoMis}.
Both angles, and therefore $\eta$, depend on the octocopter rotation angles as 
well as on the octocopter position.
The angle $\beta$ linearly depends on $\alpha$ and on the AUT orientation which is known with a precision of $1\rm{^{\circ}}$. 
The uncertainty of all three octocopter rotation angles is estimated to be $1\rm{^{\circ}}$. In the case of the horizontal VEL
the uncertainty of $\alpha$ is described by the quadratic sum of two octocopter rotation angles and the angle which arises from the octocopter position uncertainties as well as the size of the AUT. For 
the example flight, the resulting influence on the VEL is $0.4\,\rm{\%}$ at $55\,\rm{MHz}$ and $(42.5\pm2.5)\,\rm^{\circ}$.
In contrast, both meridional subcomponents are not corrected for the octocopter misalignment and misplacement. Here, the octocopter misalignment and misplacement is completely included in the systematic uncertainty.
Therefore, the systematic uncertainty of the VEL due to an octocopter misalignment and misplacement is larger for both meridional subcomponents than in the case of the horizontal component. 
The systematic uncertainty on the VEL is calculated in the same way but using the nominal values of $\alpha$ and $\beta$ in each zenith angle bin of $5\rm{^{\circ}}$ instead. As
$\beta$ linearly depends on $\alpha$, only a further uncertainty on $\alpha$ given by the difference between the measured median values and nominal values of $\alpha$ is needed, quadratically added
and then propagated to the systematic uncertainty on the VEL.
In case of both meridional subcomponents, both angles $\alpha$ and $\beta$ depend on the zenith angle. Hence, this systematic uncertainty is strongly zenith angle dependent for both meridional subcomponents.\\
The uncertainties of the injected power to the transmitting antenna and the transmitting antenna pattern limit the overall calibration accuracy.
In comparison to other calibration campaigns at LOFAR or Tunka-Rex, a RSG1000 were used as signal source as well but a different transmitting antenna. Both RSG1000 signal sources  
differ on a percent level only. However, the manufacturer of the transmitting antenna used at LOFAR and Tunka-Rex states a systematic uncertainty of the transmitting antenna pattern of $1.25\,\rm{dB}$ \cite{HillerPHD}. Hence, the 
AERA calibration has a significantly smaller systematic uncertainty due to the more precise calibration of the transmitting antenna.\\
\subsubsection{Received Signal at the Antenna Under Test}
Within the uncertainty of the received signal all uncertainty effects of the received power at the AUT including the full signal chain from the LPDA to the spectrum analyzer as well as the LNA and cables are considered. 
In the following a drift of the LPDA LNA gain due to
temperature fluctuations, the uncertainty of the received power using the FSH4 and the influence of background noise as well as the uncertainty of the cable attenuation measurements are discussed.\\
The LPDA LNA gain depends on the temperature. The gain temperature drift was measured in the laboratory and was determined to $0.01\,\rm{dB/K}$ using the FSH4 in the vector network analyzer mode \cite{PHD}.
The calibration measurements were performed at temperatures between $15\,\rm{^{\circ}C}$ and $25\,\rm{^{\circ}C}$ which results in a systematic uncertainty of $1\,\rm{\%}$ in the VEL due to temperature drifts of the LNA.
The measurements of the LPDA LNA gain due to temperature fluctuations using the FSH4 show fluctuations of the LNA gain at the level of $0.1\,\rm{dB}$ which results in an expected statistical uncertainty of $0.6\,\rm{\%}$ in the VEL.\\
The event power is measured using the FSH4 spectrum analyzer. The manufacturer states a systematic uncertainty of $0.5\,\rm{dB}$. The systematic uncertainty in the VEL
is then $5.8\,\rm{\%}$. Also the background noise is measured using the FSH4 in spectrum analyzer mode.
The systematic uncertainty of the VEL considering event power (P) and background noise (B) is $\sqrt{\frac{P^{2}+B^{2}}{P^{2}-B^{2}}} \frac{0.5}{2}\,\rm{dB}$. If the background noise is of
the same order of magnitude as the measured event power for more than $50\,\rm{\%}$ of events in a $5\rm{^{\circ}}$ zenith angle bin, the systematic uncertainty for 
this zenith angle bin is set to $100\,\rm{\%}$. For the example flight, the systematic due to background noise results in an additional VEL
systematic uncertainty of $0.4\,\rm{\%}$ at $55\,\rm{MHz}$ and $(42.5\pm2.5)\,\rm^{\circ}$. A further background influence on the measured signal at the LPDA
due to the communication between the remote control and the octocopter is not expected, as they communicate at $2.4\,\rm{GHz}$ and 
the LPDA is sensitive in the frequency range from $30\,\rm{MHz}$ to $80\,\rm{MHz}$.\\
The attenuation of the cable is measured with the FSH4 in network analyzer mode transmitting a signal with a power of $0\,\rm{dBm}$ and averaged over $100$ samples. Therefore, the statistical uncertainty
can be neglected.
The manufacturer states a systematic uncertainty of $0.04\,\rm{dB}$ for transmission measurements with a transmission higher than $-20\,\rm{dB}$ which applies in case of the cables.
This results in a systematic uncertainty of $0.5\,\rm{\%}$ in the VEL.
\subsection{Simulation of the Experimental Setup}
\begin{figure}
    \begin{center}
        {\includegraphics[scale=0.29]{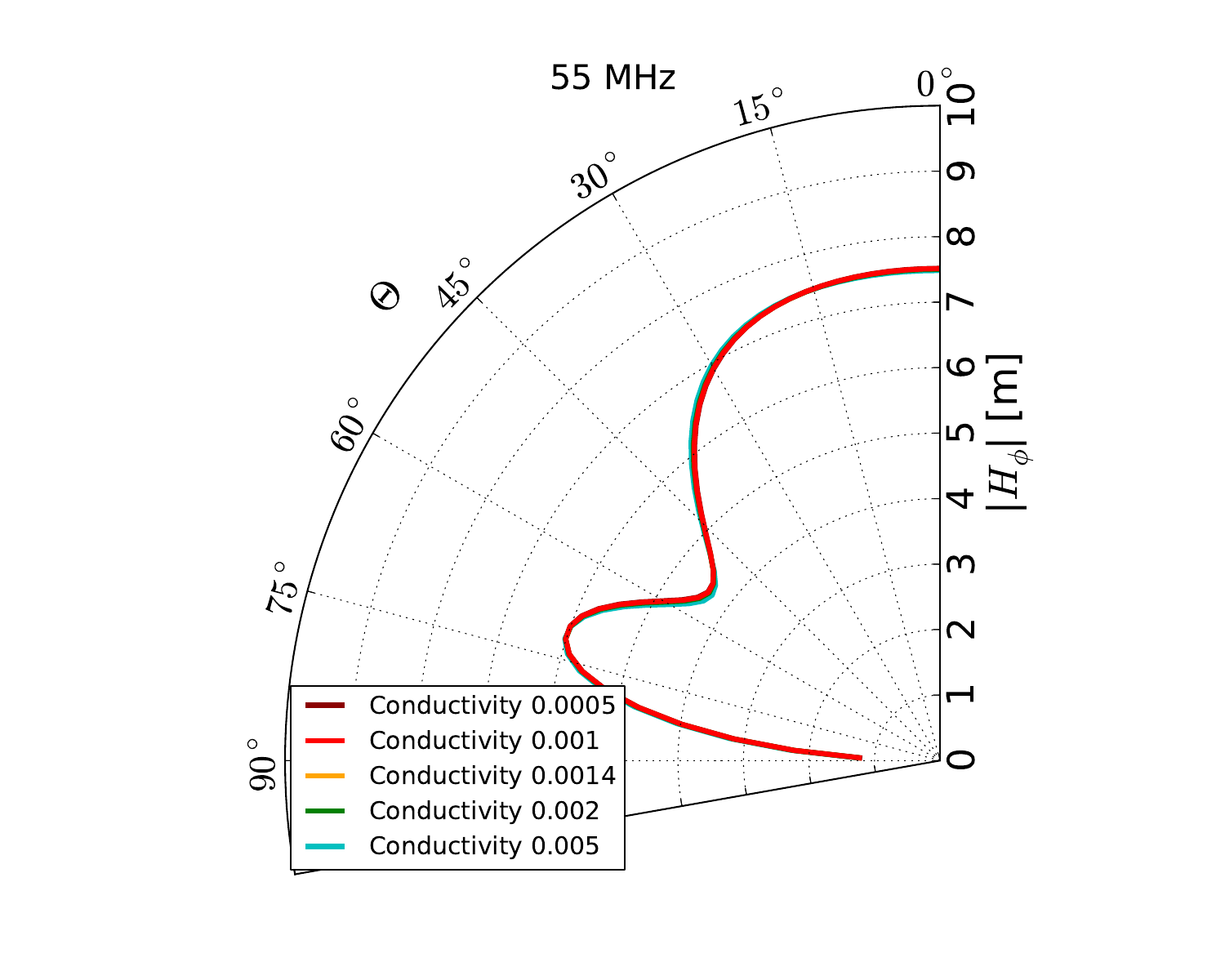}}{\includegraphics[scale=0.29]{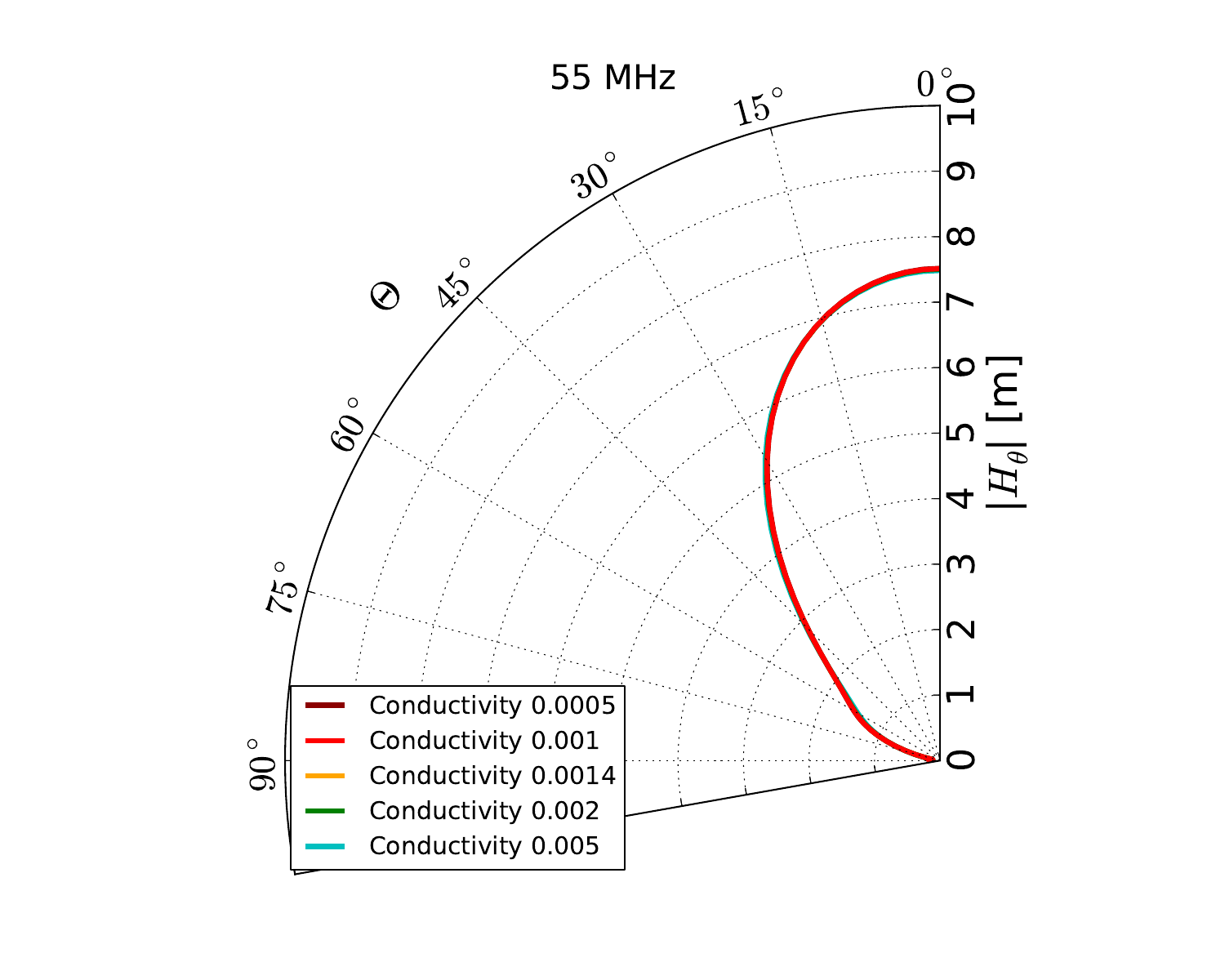}}
        \vspace{0.1cm}
        {\includegraphics[scale=0.29]{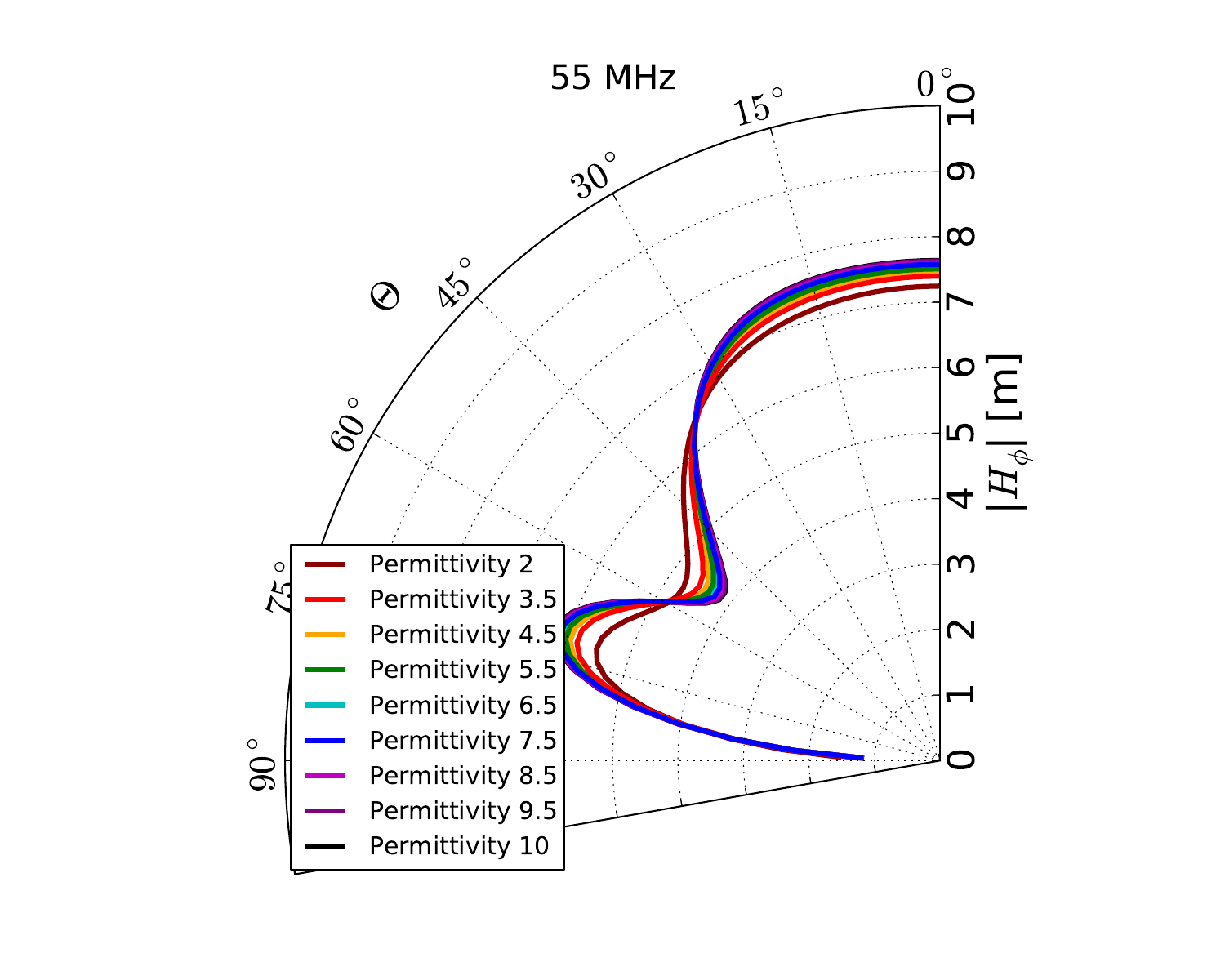}}{\includegraphics[scale=0.29]{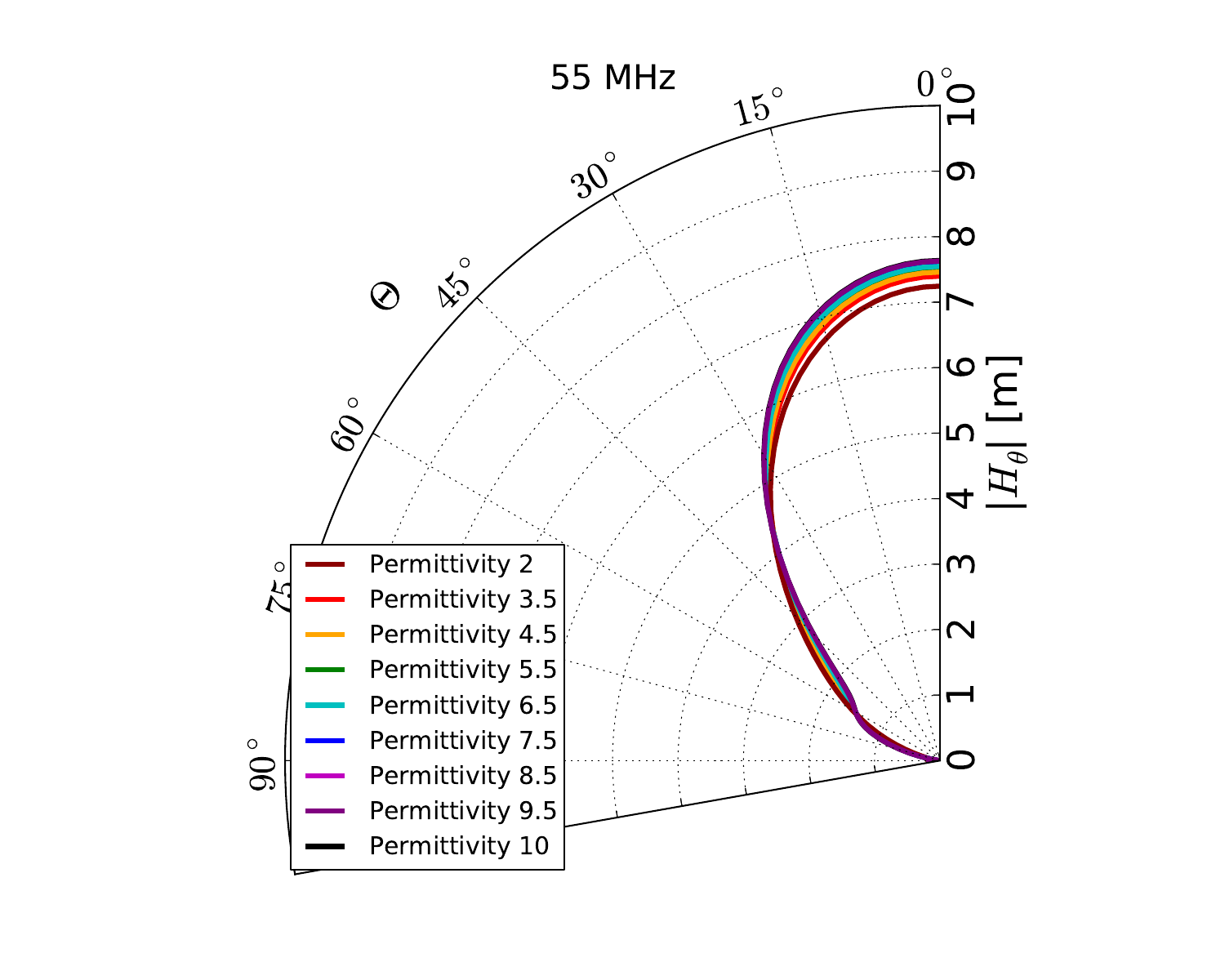}}
        \caption{\it Simulations of the VEL for different ground conditions. A variation in conductivity is shown in the upper diagrams whereas a variation in permittivity is shown in the lower diagrams. In the \textbf{(left)}
        diagrams the horizontal VEL $|H_{\phi}|$ and in the \textbf{(right)} diagrams the meridional VEL $|H_{\theta}|$ as function of the zenith angle $\Theta$ at $55\,\rm{MHz}$ are shown.}
    \label{fig:NEC2}
    \end{center}
\end{figure}

The calibration measurement is simulated using the NEC-2 simulation code. Here, the AUT, the transmission antenna and realistic ground properties are taken into account. At standard ground conditions the 
ground conductivity is set to be $0.0014\,\rm{S/m}$ which was measured at the AERA site. Values of the conductivity of dry sand, which is the typical ground consistency at AERA, 
are reported here \cite{Geo-Auger, Conductivity}. Measurements of the ground permittivity at the AERA site yield values between 
$2$ and $10$ depending on the soil wetness \cite{PHD}. The standard ground permittivity is set to be $5.5$ in the simulation. The distance between both antennas is set to be $30.3\,\rm{m}$. 
The VEL is calculated using Eq.~\eqref{eq:VEL} modified with Eq.~\eqref{eq:SimVEL} considering the manufacturer information for the response pattern of the transmitting antenna as well as the transfer function from the AUT output to
the system consisting of the transmission line from the LPDA footpoint to the LNA and the LNA itself. To investigate the simulation stability several simulations with varying antenna separations
and changing ground conditions were performed \cite{PHD}.
Antenna separations ranging from $25\,\rm{m}$ to $50\,\rm{m}$ were simulated and did not change the resulting VEL of the LPDA.
Hence, the simulation confirms that the measurement is being done in the far-field region. Furthermore, the influence of different ground conditions is investigated. Conductivity and permittivity
impact the signal reflections on ground. The LPDA VEL is simulated
using ground conductivities ranging from $0.0005\,\rm{\frac{S}{m}}$ to $0.005\,\rm{\frac{S}{m}}$ and using ground permittivities ranging from $2$ to $10$.
Within the given ranges the conductivity and permittivity independently influence the signal reflection properties of the ground.
In Fig.~\ref{fig:NEC2} the simulations of the horizontal and meridional VEL for these different ground conditions as function of the zenith angle at $55\,\rm{MHz}$ are shown.
Different ground conductivities do not change the LPDA response pattern. In contrast the influence of the ground permittivity on the antenna response is slightly higher.
In the case of an applied ground permittivity of $2$ and of $10$, the influence 
on the horizontal VEL is at the level of $1\,\rm{\%}$ averaged over all 
frequencies and zenith angles with a scatter of less than $6\,\rm{\%}$.
The influence of the ground permittivity on the electric-field reconstruction is discussed in section \ref{subsec:UncerCR}.\\
Simulations of an electronic box beneath the LPDA show influences on the horizontal antenna VEL smaller than $0.3\,\rm{\%}$ which is negligible compared to the influence of the ground permittivity \cite{PHD}. 
\section{Measurement of the LPDA Vector Effective Length}
In this section, the reproducibility and the combination of all measurements performed on different days and under different environmental conditions are discussed.
Furthermore, the combined results of the LPDA VEL are compared to the values obtained from the NEC-2 simulation.
\subsection{Horizontal Vector Effective Length}
\begin{figure}
    \begin{center}
        {\includegraphics[scale=0.29]{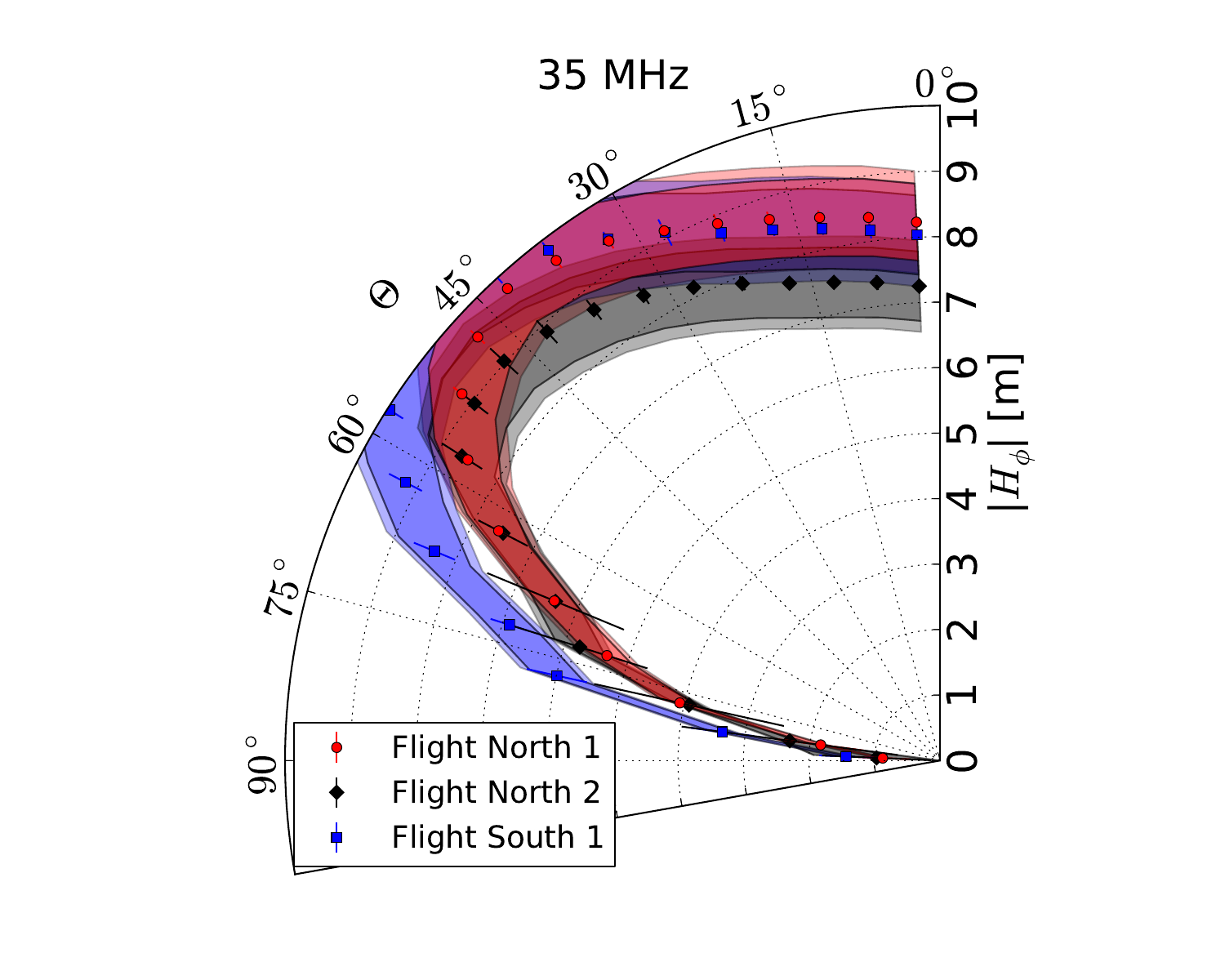}}{\includegraphics[scale=0.29]{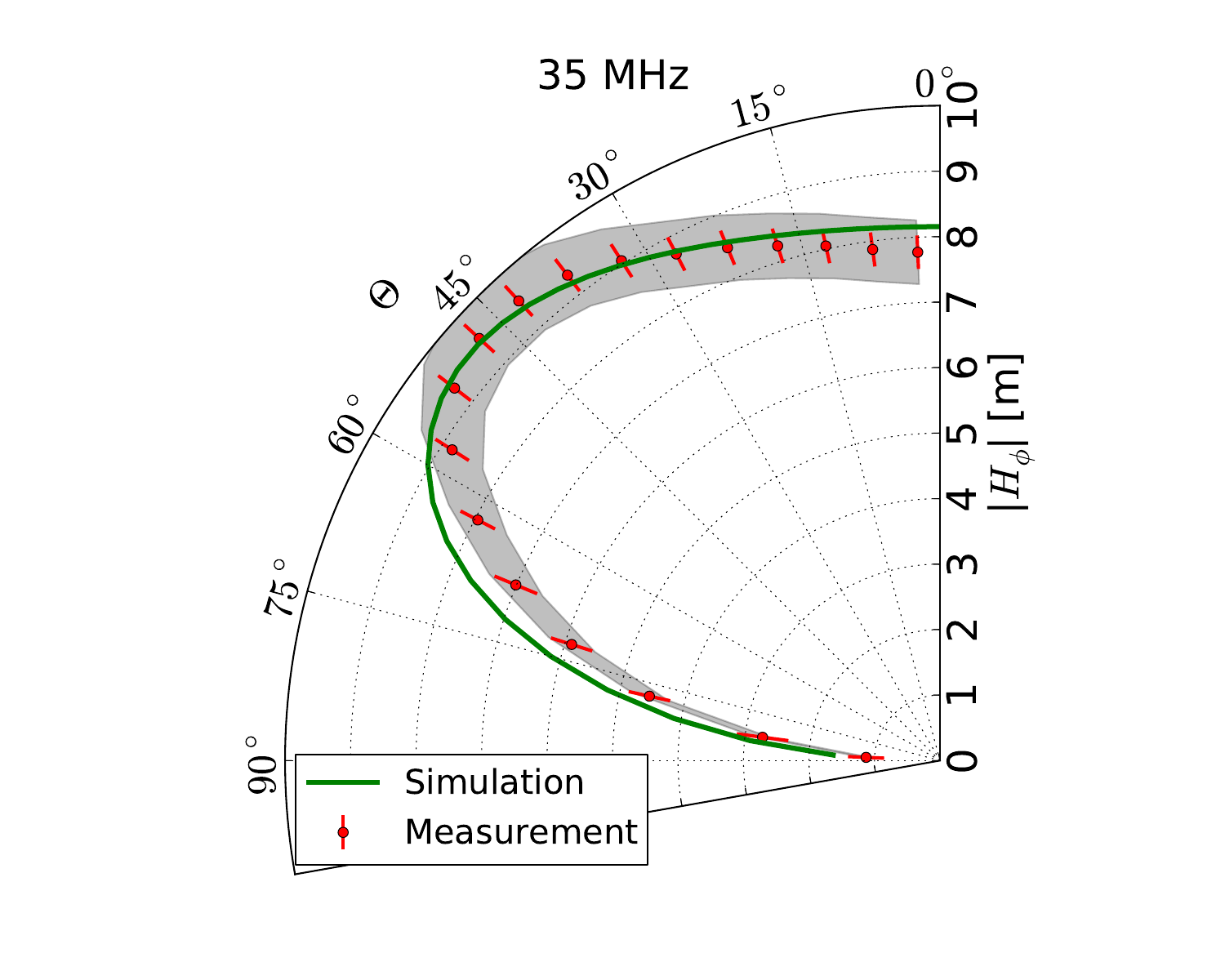}}\\
        \vspace{0.1cm}
        {\includegraphics[scale=0.29]{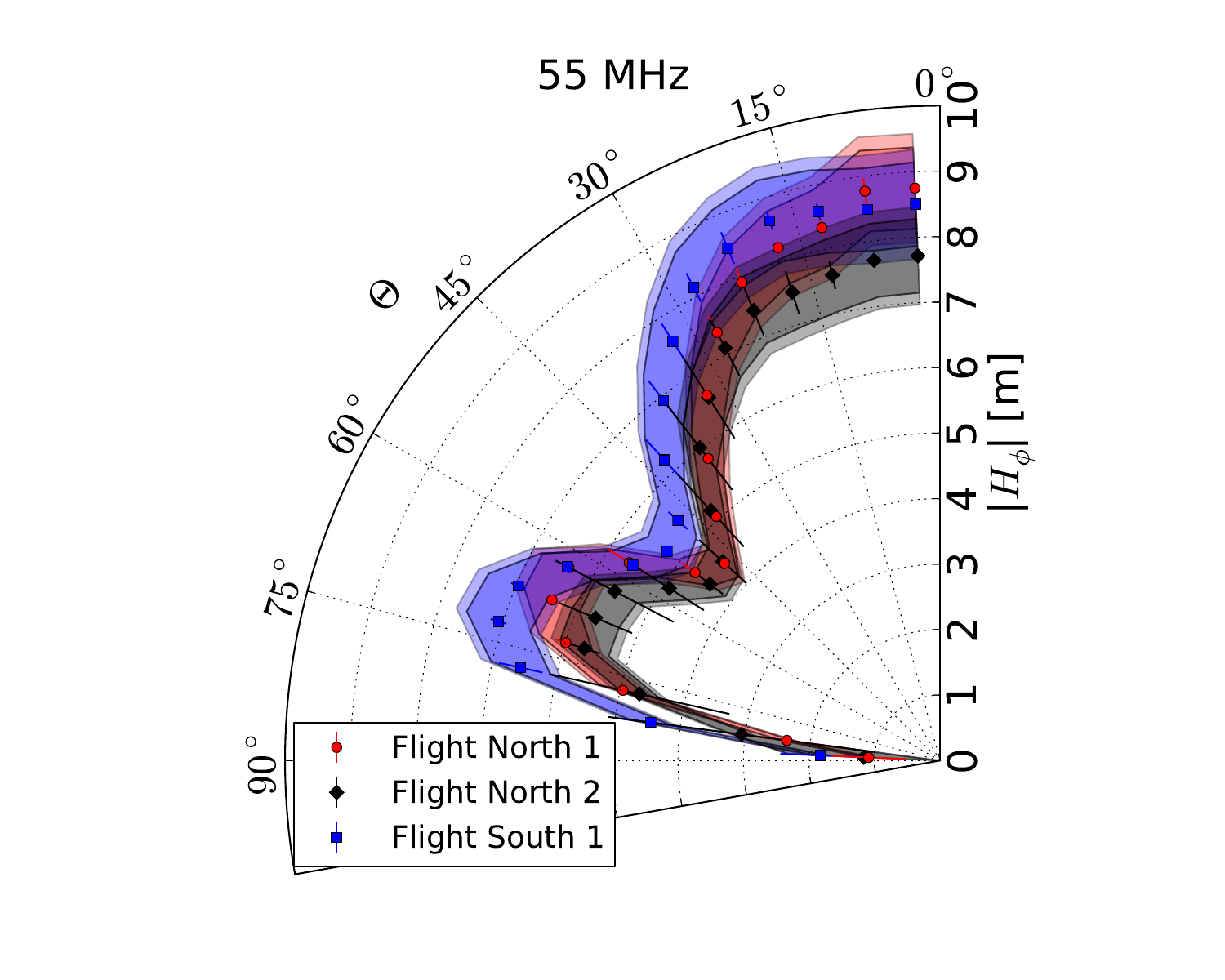}}{\includegraphics[scale=0.29]{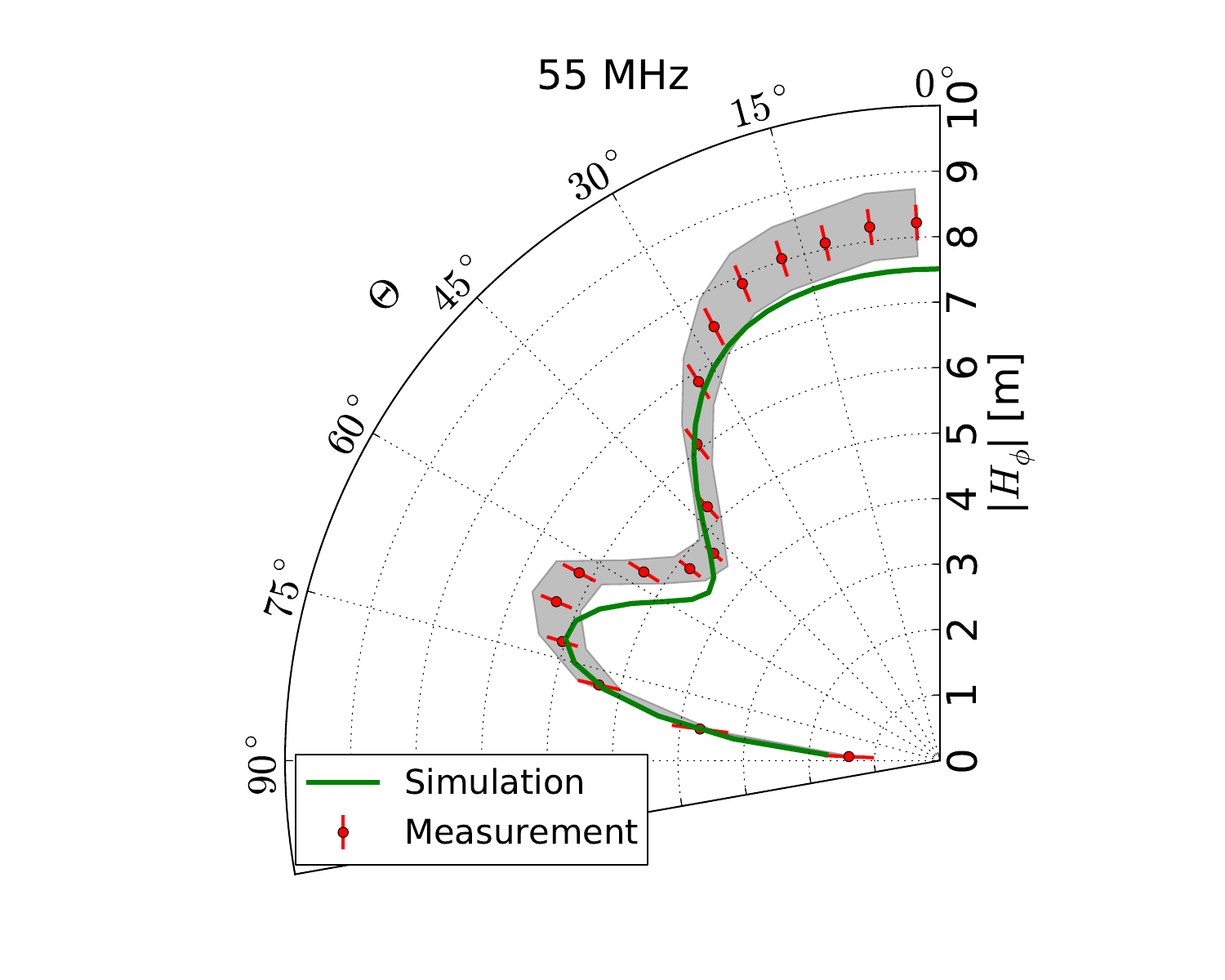}}\\
        \vspace{0.1cm}
        {\includegraphics[scale=0.29]{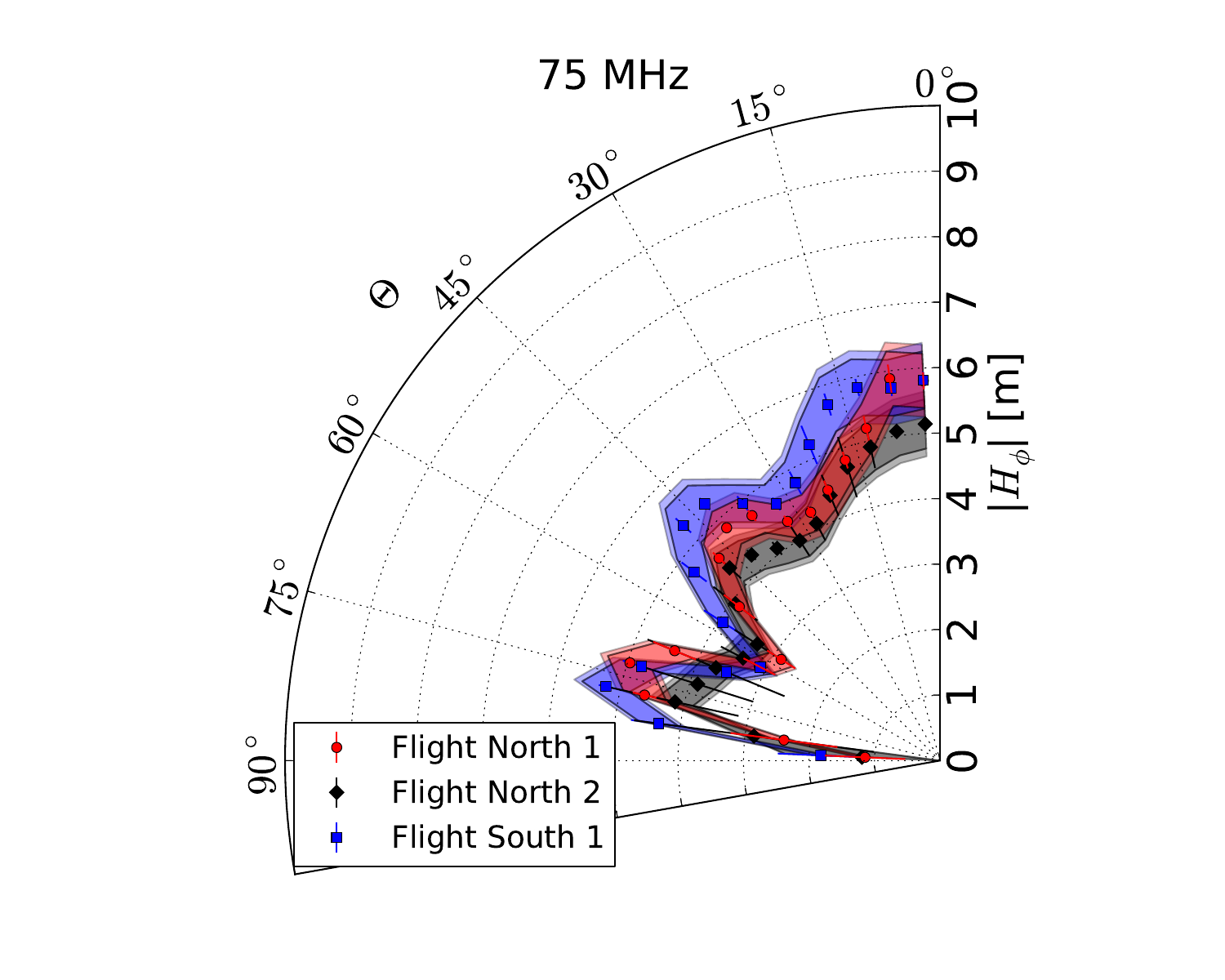}}{\includegraphics[scale=0.29]{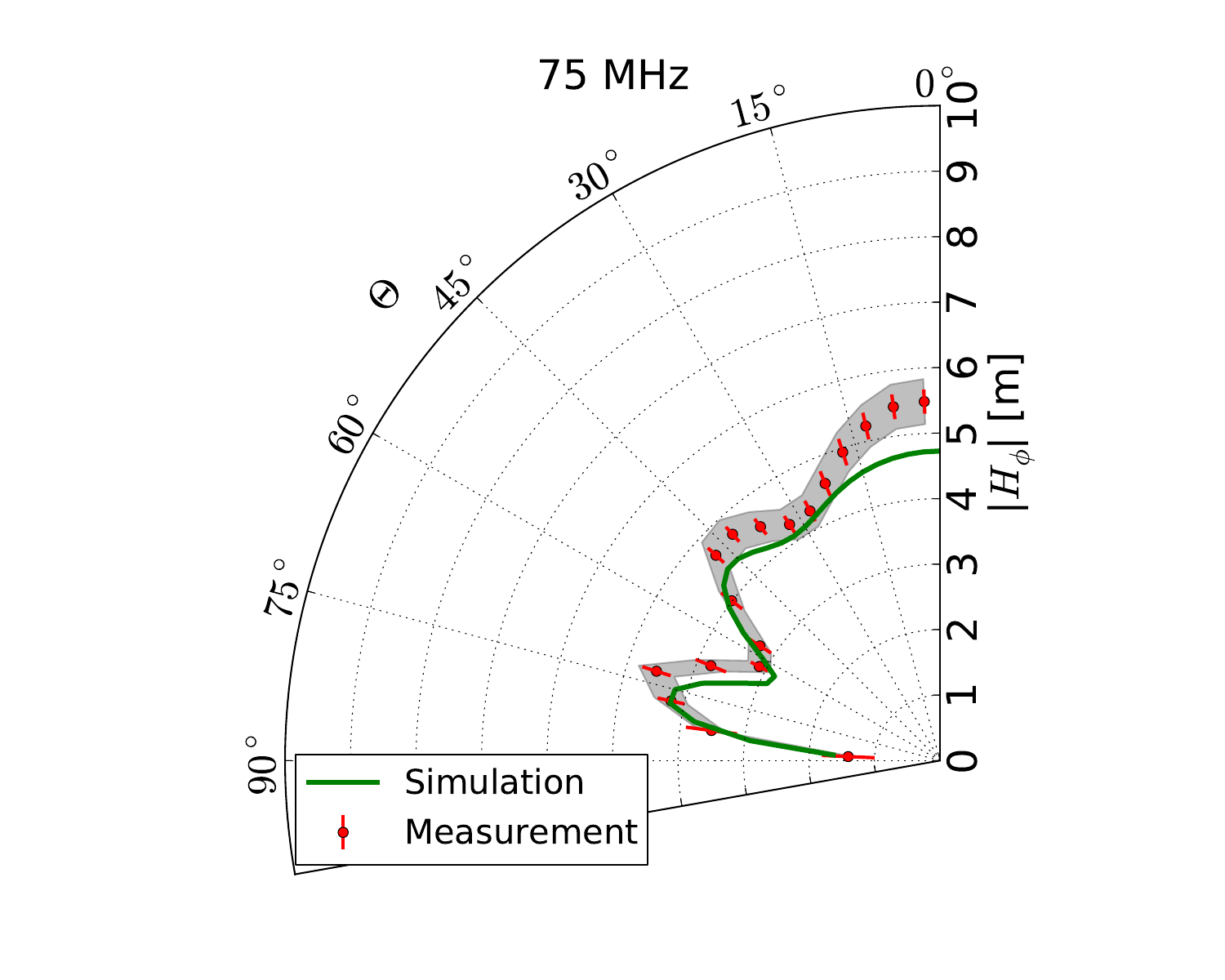}}
        \caption{\it \textbf{(left)} Mean horizontal VEL $|H_{\phi}|$ (dots) and standard deviation (error bars) of three different measurements and \textbf{(right)} the overall combinations in comparison 
        to the simulation (green curve) as a function of the zenith angle in $5\rm{^{\circ}}$ bins at \textbf{(from top to bottom)} $35\,\rm{MHz}$, $55\,\rm{MHz}$ and $75\,\rm{MHz}$.
        The colored bands in the left diagrams describe the constant (light color) and flight-dependent (dark color) systematic uncertainties of each flight.
        The gray band in the right diagrams describes the constant systematic uncertainties whereas the statistical and flight-dependent systematic uncertainties
        are combined within the error bars.}
    \label{fig:HorVEL}
    \end{center}
\end{figure}

\begin{figure}
    \begin{center}
        {\includegraphics[scale=0.3]{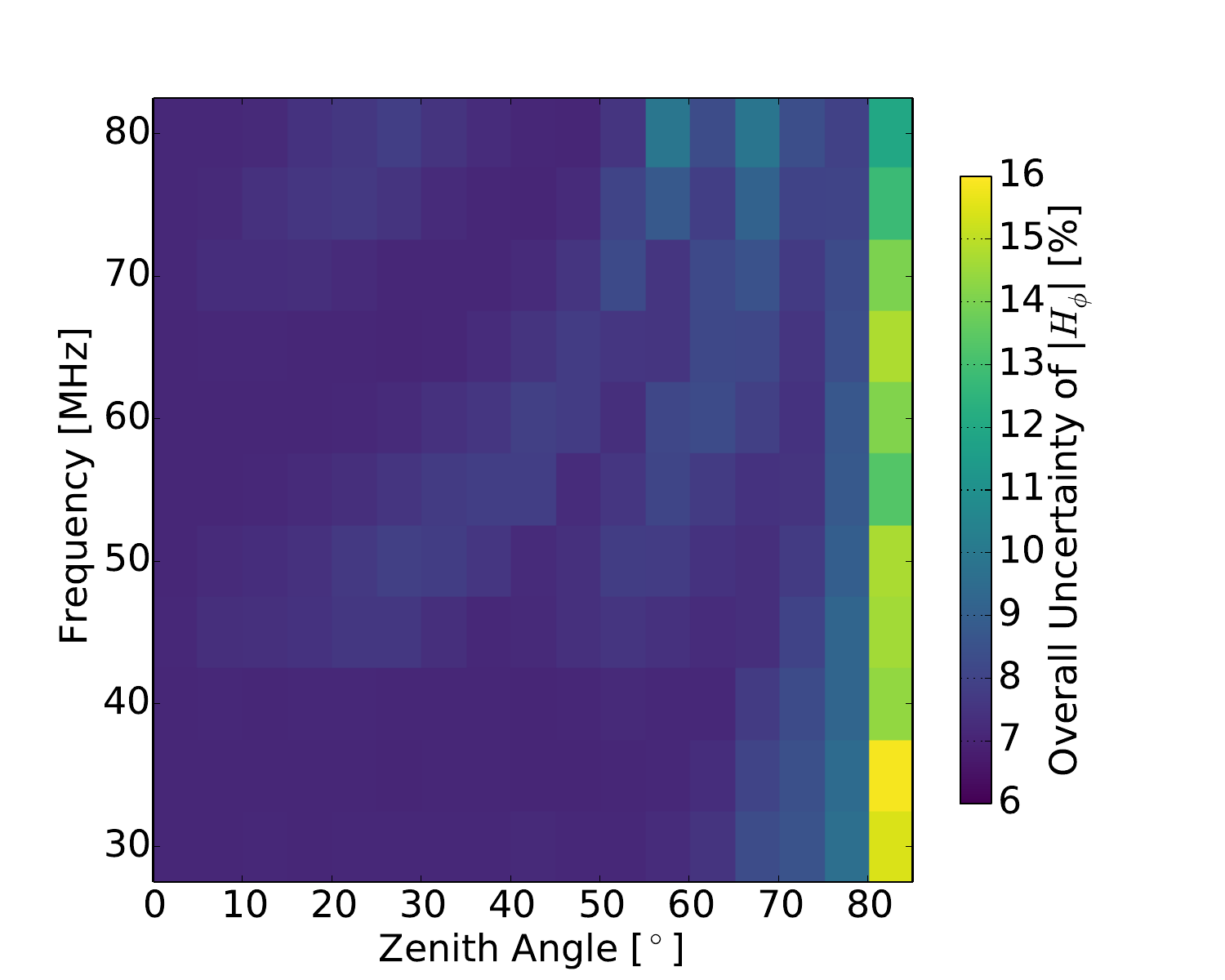}}{\includegraphics[scale=0.3]{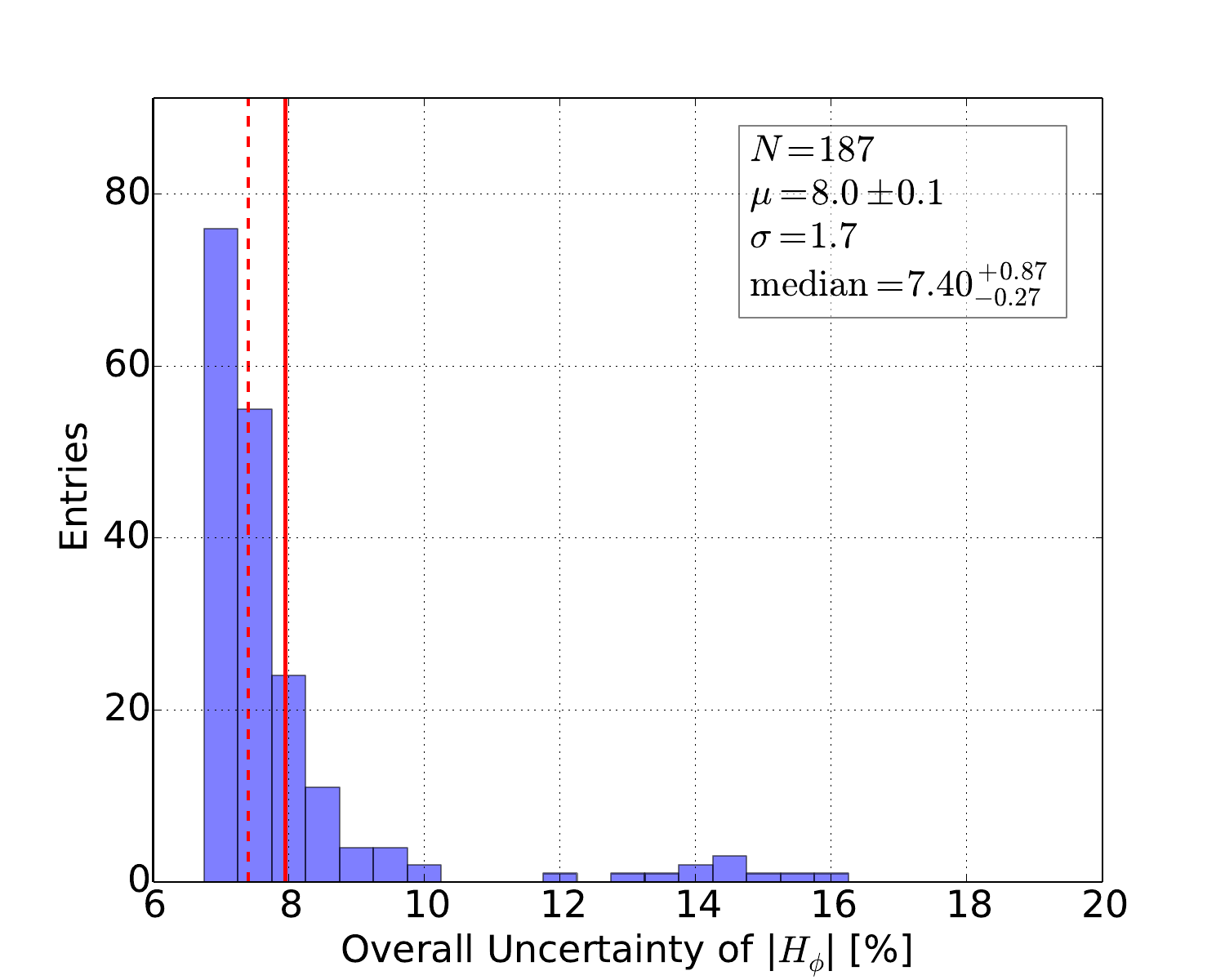}}\\
        \caption{\it \textbf{(left)} Overall uncertainty of the horizontal VEL $|H_{\phi}|$ including statistical and systematic uncertainties for all frequencies as a function of the zenith angle $\Theta$ up to 
        $85\rm{^{\circ}}$ in $5\rm{^{\circ}}$ bins. \textbf{(right)} Histogram of all overall uncertainties for all frequencies and all zenith angle bins previously mentioned. The median (average value $\mu$) is marked
        as red dashed line (red line).}
    \label{fig:HorVELError}
    \end{center}
\end{figure}

\begin{figure}
    \begin{center}
        {\includegraphics[scale=0.29]{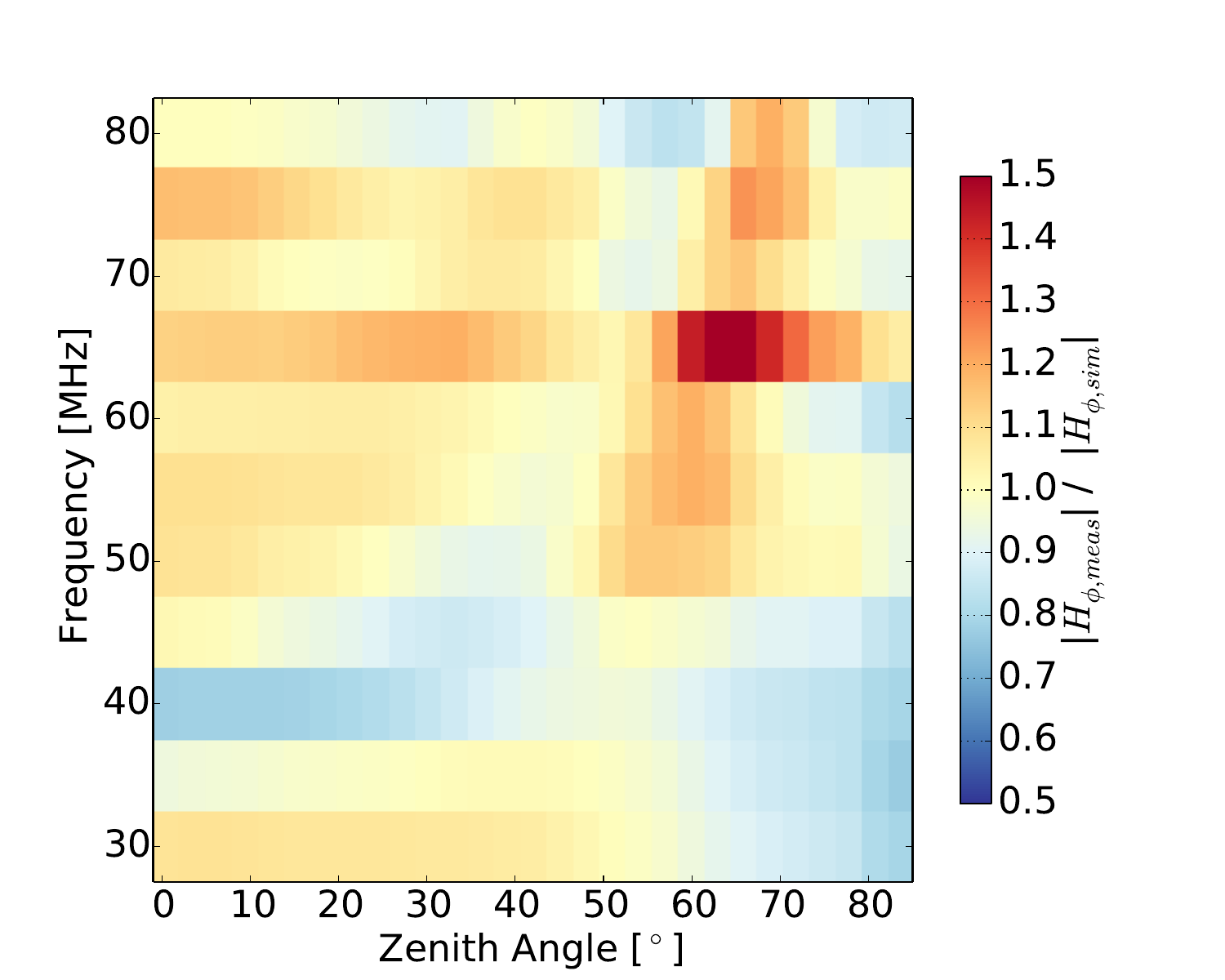}}
        \vspace{0.1cm}
        {\includegraphics[scale=0.29]{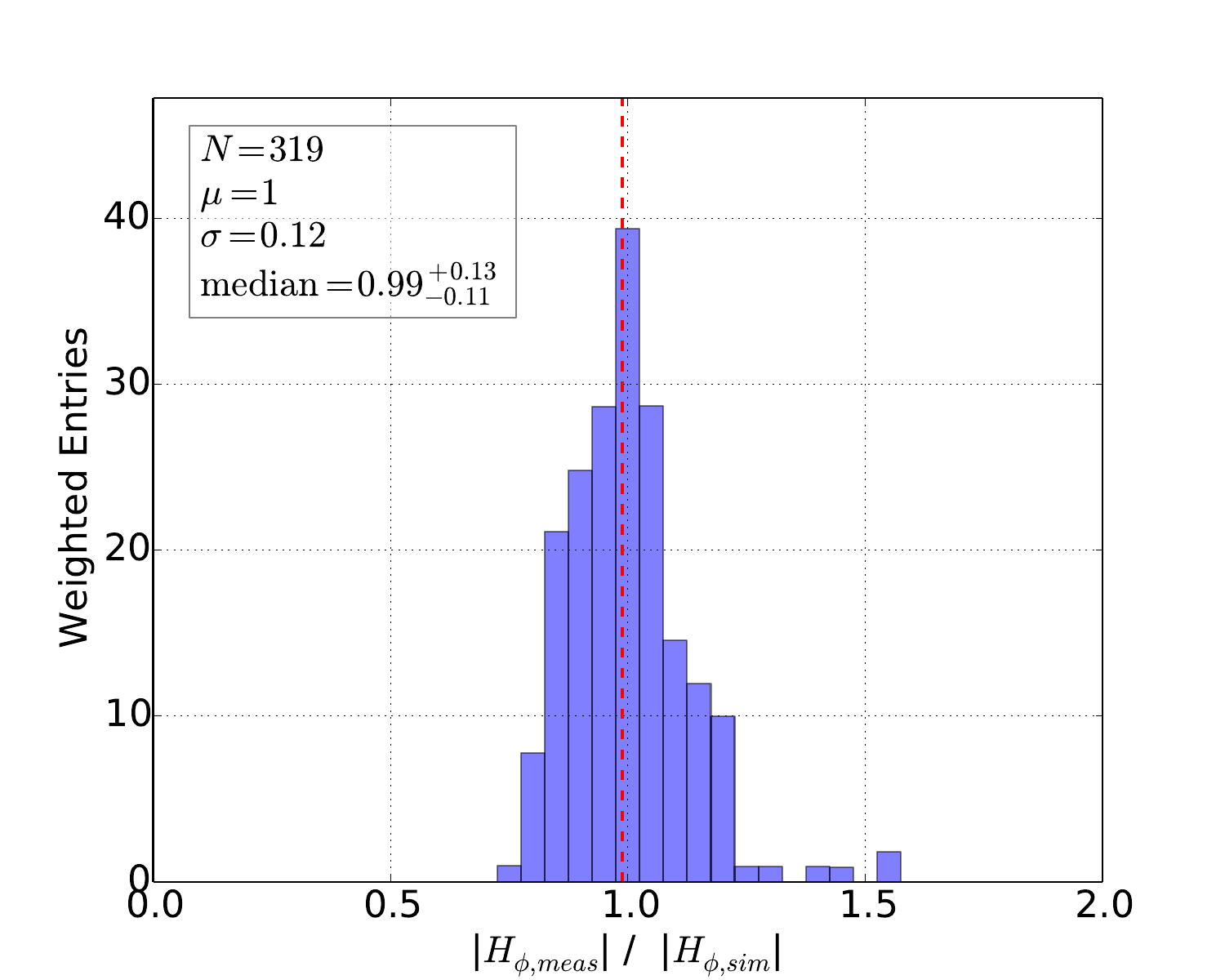}}
        \vspace{0.1cm}
        {\includegraphics[scale=0.29]{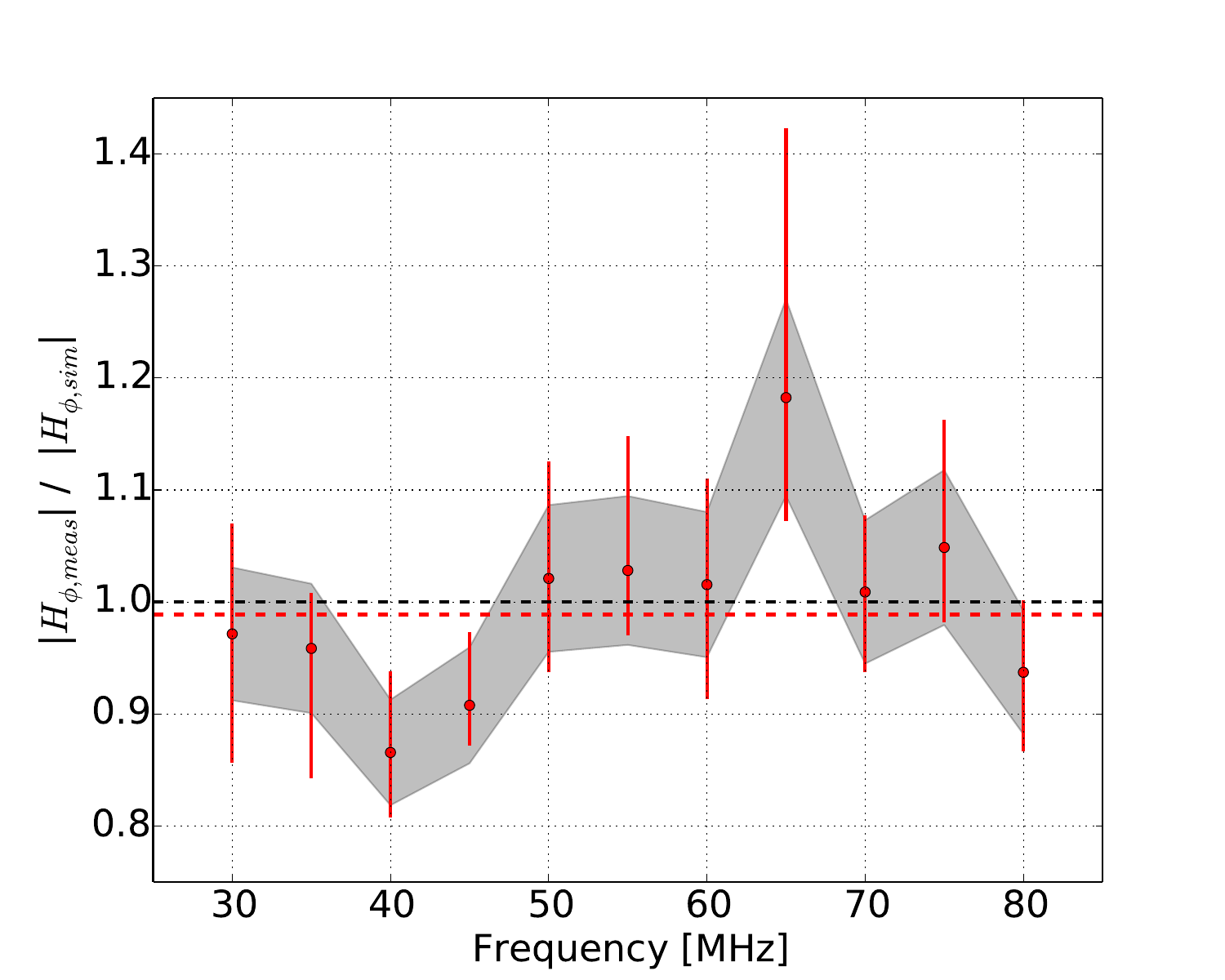}}
        \vspace{0.1cm}
        {\includegraphics[scale=0.29]{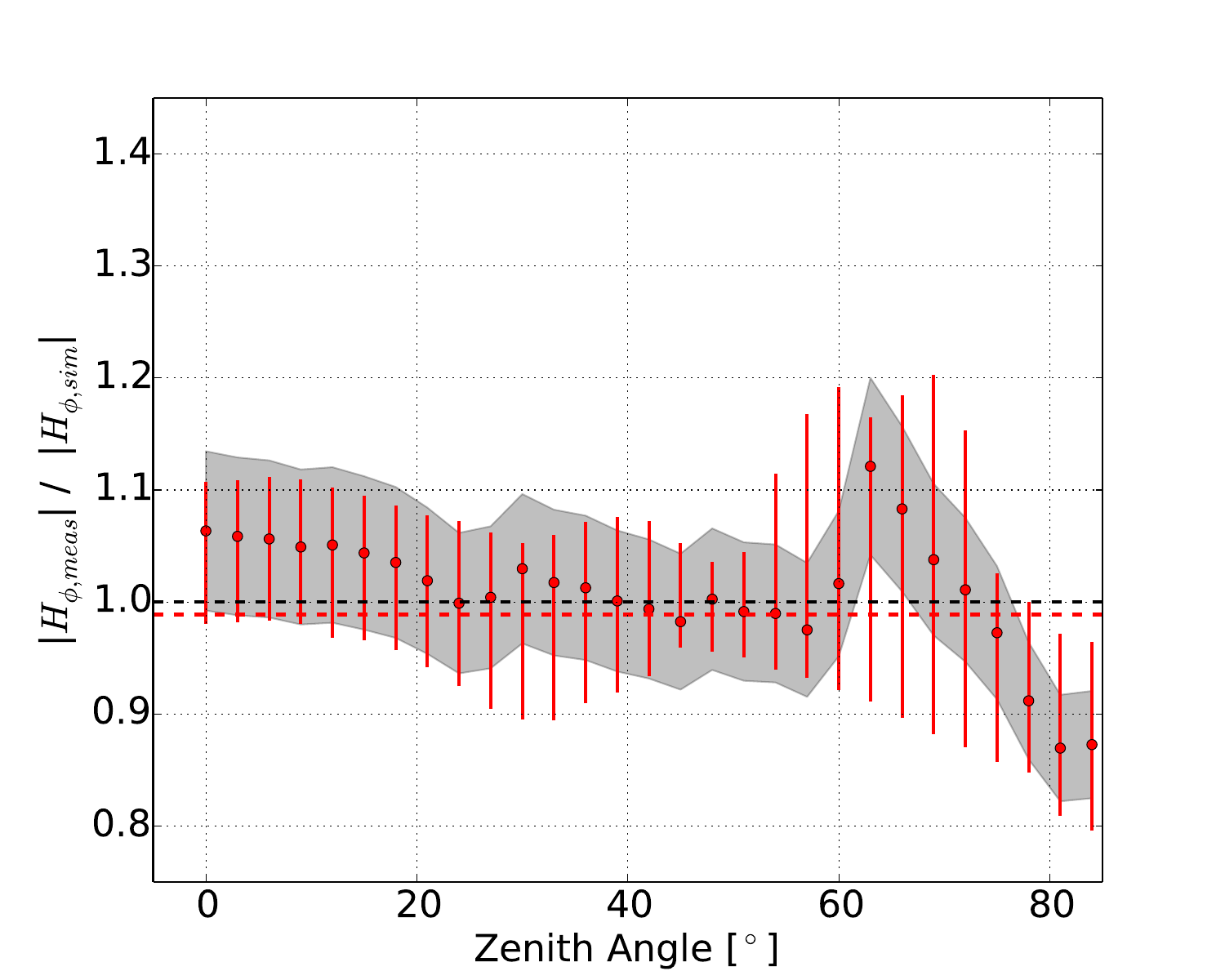}}
        \caption{\it Comparison of the combined horizontal VEL $|H_{\phi}|$ with the simulation. \textbf{(top left)} Ratio of the combination of all measurements and simulation for all frequencies as a function of 
        the zenith angle $\Theta$ up to $84\rm{^{\circ}}$ in $3\rm{^{\circ}}$ bins.
        \textbf{(top right)} Histogram of all ratios of the combination of all measurements and simulation for all frequencies and all zenith angle bins previously mentioned weighted with $\sin(\Theta)$.
        The median value is marked as the red dashed line.
        \textbf{(bottom left)} Median (red dots) and the $68\,\rm{\%}$ quantile (red error bars) of the zenith angle weighted ratio distribution 
        as a function of the frequency. \textbf{(bottom right)} Median (red dots) and the $68\,\rm{\%}$ quantile (red error bars) of the ratio distribution 
        as a function of $\Theta$. The gray band indicates the constant systematic uncertainty of the measurement and the red dashed lines mark the overall zenith angle weighted average in both 
        lower diagrams.}
    \label{fig:HorCorrection}
    \end{center}
\end{figure}

Here, the results of the measurements of the horizontal VEL $|H_{\phi}|$ are presented. In total, five independent measurements were performed to determine $|H_{\phi}|$ as a function of the zenith angle $\Theta$.
The horizontal VEL $|H_{\phi}|$ in zenith angle intervals of $5\rm{^{\circ}}$ for three different measurements at $35\,\rm{MHz}$, $55\,\rm{MHz}$ and $75\,\rm{MHz}$ is shown on the left side of Fig.~\ref{fig:HorVEL}. 
The constant systematic uncertainties of each flight are denoted by the light colored band and the flight dependent systematic uncertainties are indicated by the dark colored band.
Compared to the average $\overline{\mathrm{VEL}}$ from $5$ measurements the median value of the ratio $\sigma/\overline{\mathrm{VEL}}$ is $6\,\rm{\%}$ which is well compatible with the estimated uncertainties presented
in Tab.~\ref{tab:Uncertainties}. At the right side of Fig.~\ref{fig:HorVEL} all performed measurements to determine $|H_{\phi}|$ are combined in zenith angle intervals 
of $5\rm{^{\circ}}$, weighted by the quadratic sum of the systematic and the statistical uncertainties of each flight. The gray band describes the constant systematic uncertainties whereas the
statistical and flight-dependent systematic uncertainties are combined within the error bars. 
The constant systematic uncertainty of the combined horizontal VEL is $6.3\,\rm{\%}$ and the uncertainties considering flight dependent systematic and statistical uncertainties for the combined horizontal VEL 
result in $4.7\,\rm{\%}$ at a zenith angle of $(42.5\pm2.5)\rm{^{\circ}}$ and a frequency of $55\,\rm{MHz}$. The overall uncertainty of the determined LPDA VEL in the horizontal polarization adds quadratically 
to $7.9\,\rm{\%}$. The overall uncertainty of all other arrival directions and frequencies are shown on the left side of Fig.~\ref{fig:HorVELError}.
On the right side of Fig.~\ref{fig:HorVELError} a histogram of all overall uncertainties for all frequencies and all zenith angles up to $85\rm{^{\circ}}$ is shown.
For larger zenith angles the LPDA loses sensitivity and the systematic uncertainty exceeds $20\,\rm{\%}$. Therefore, angles beyond $85\rm{^{\circ}}$ are not considered in the following discussion.
Taking all intervals of the frequencies and zenith angles with equal weight the
median overall uncertainty including statistical and systematic uncertainties is $7.4^{+0.9}_{-0.3}\,\rm{\%}$.
The green curve in Fig.~\ref{fig:HorVEL} marks the simulation of $|H_{\phi}|$. The agreement between the combined measurements and the simulation of $|H_{\phi}|$ is illustrated in the diagram of their ratio versus
zenith angle $\Theta$ and frequency $f$ in the upper left panel of Fig.~\ref{fig:HorCorrection}. 
In the upper right panel of Fig.~\ref{fig:HorCorrection} all ratios are filled into a histogram with entries weighted by $\sin(\Theta)$
in consideration of the decrease in field of view at small zenith angles. The combined measurement and the simulation agree to within $1\,\rm{\%}$ in the median.
The fluctuation described by the $68\,\rm{\%}$ quantile  
is at the level of $12\,\rm{\%}$. The two lower panels of Fig.~\ref{fig:HorCorrection} show the median ratio as a function of the frequency (left) and as a function
of the zenith angle (right). In both cases, the red error bars mark the $68\,\rm{\%}$ quantile of the distributions.
\subsection{Meridional Vector Effective Length}
\begin{figure}
    \begin{center}
        {\includegraphics[scale=0.29]{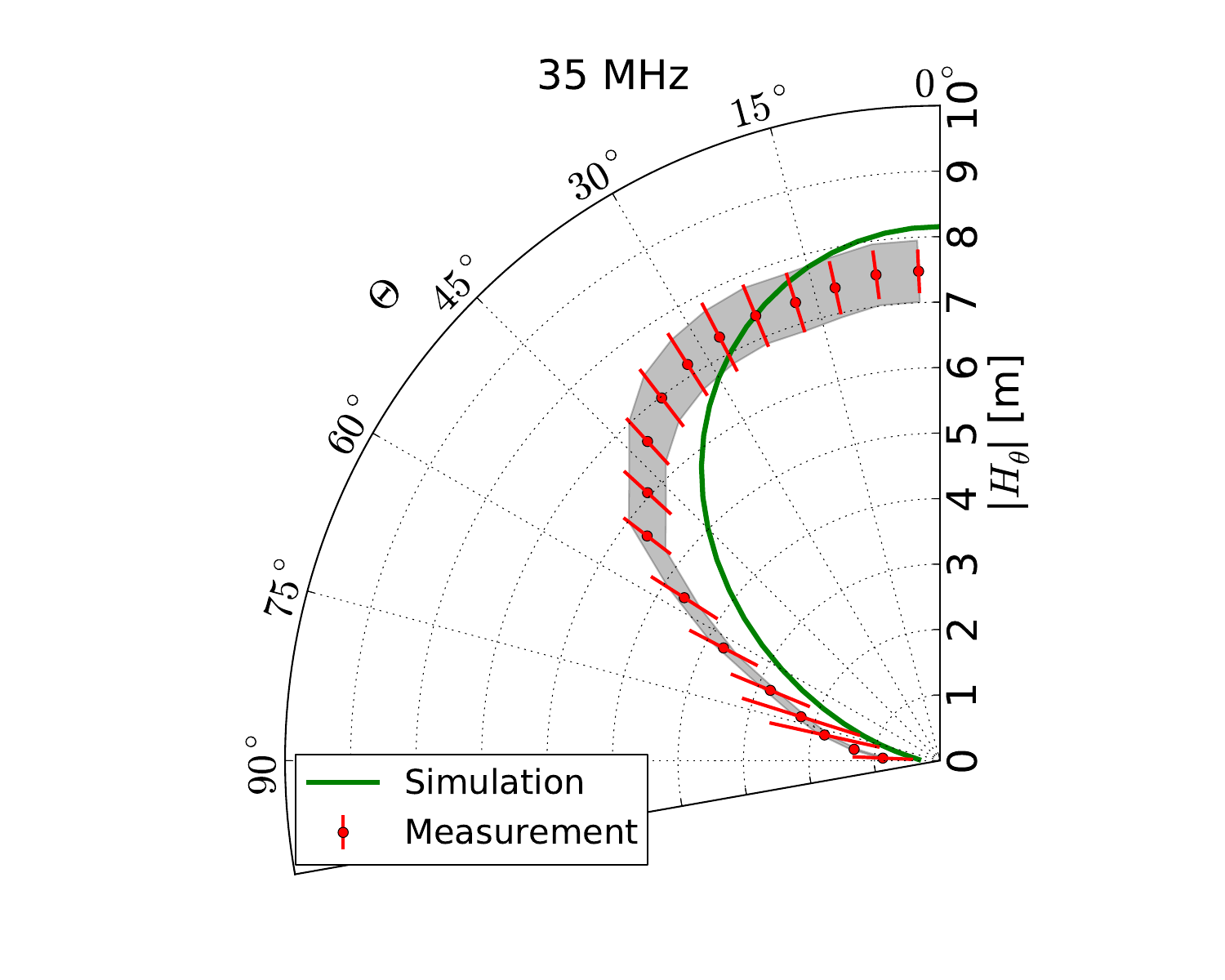}}\\
        {\includegraphics[scale=0.29]{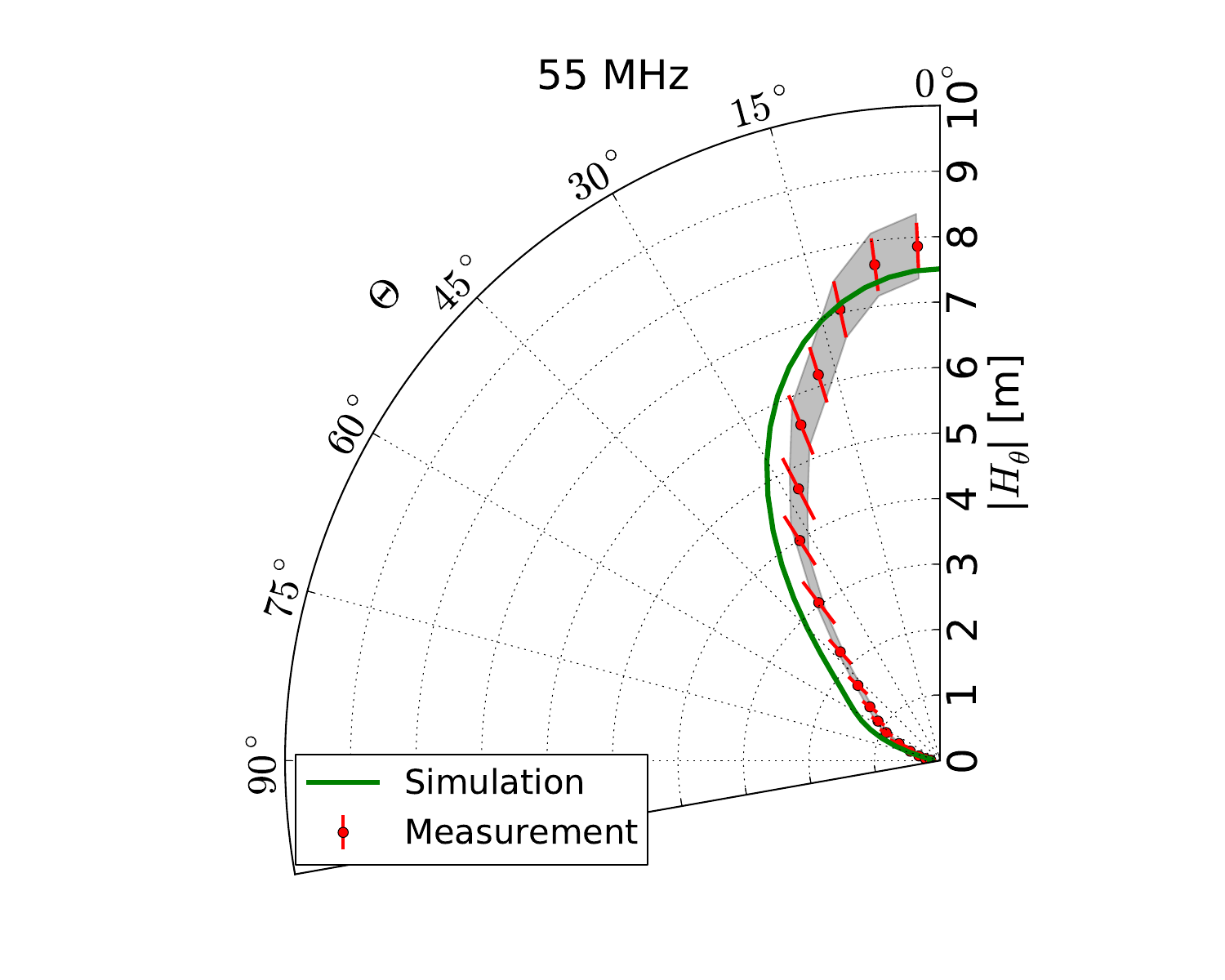}}\\
        {\includegraphics[scale=0.29]{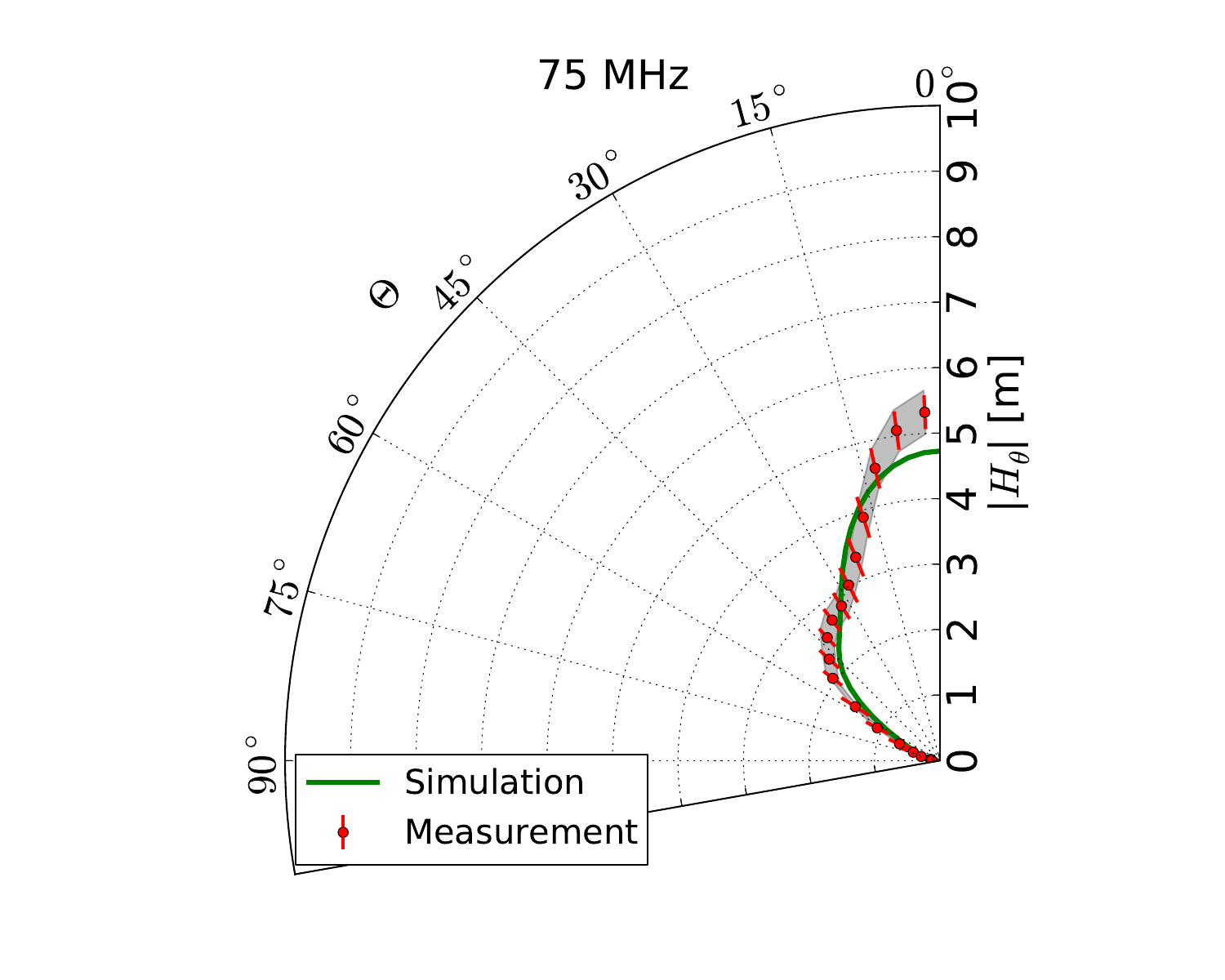}}
        \caption{\it Combination of all measurements of the meridional VEL $|H_{\theta}|$ (red dots) as a function of the zenith angle $\Theta$ in comparison to the simulation (green curve) 
        for three different frequencies \textbf{(from top to bottom)} $35\,\rm{MHz}$, $55\,\rm{MHz}$ and $75\,\rm{MHz}$.
        The gray band describes the constant systematic uncertainties whereas the statistical and flight-dependent systematic uncertainties are combined within the error bars.}
    \label{fig:VerVEL}
    \end{center}
\end{figure}

\begin{figure}
    \begin{center}
        {\includegraphics[scale=0.3]{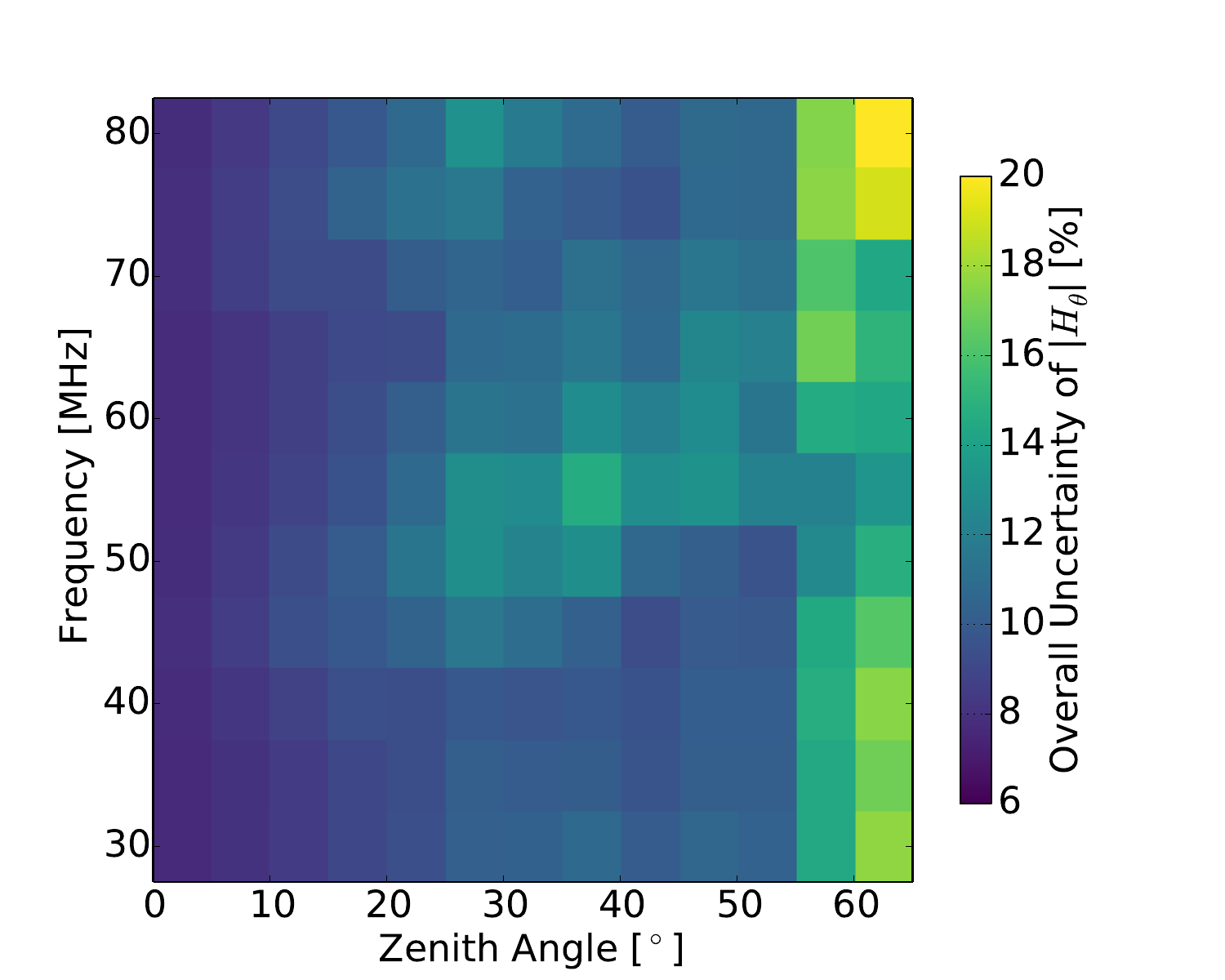}}{\includegraphics[scale=0.3]{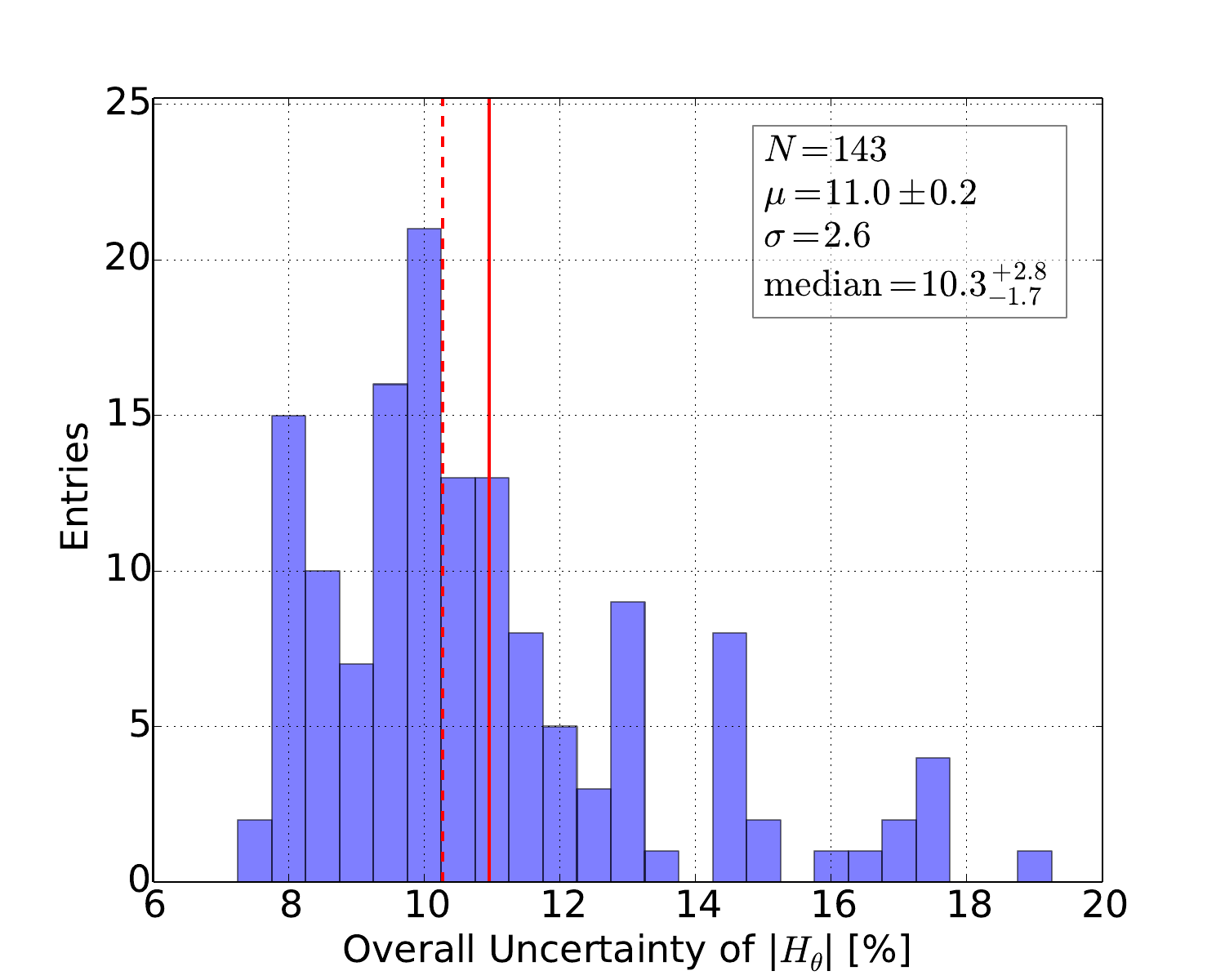}}\\
        \caption{\it \textbf{(left)} Overall uncertainty of the meridional VEL 
$|H_{\theta}|$ including statistical and systematic uncertainties for all 
frequencies as a function of the zenith angle $\Theta$ up to $65\rm{^{\circ}}$ 
        in $5\rm{^{\circ}}$ bins. \textbf{(right)} Histogram of all overall 
uncertainties for all frequencies and all $\Theta$ up to $65\rm{^{\circ}}$. The 
median (average value $\mu$) is marked
        as red dashed line (red line).}
    \label{fig:VerVELError}
    \end{center}
\end{figure}

\begin{figure}
    \begin{center}
        {\includegraphics[scale=0.29]{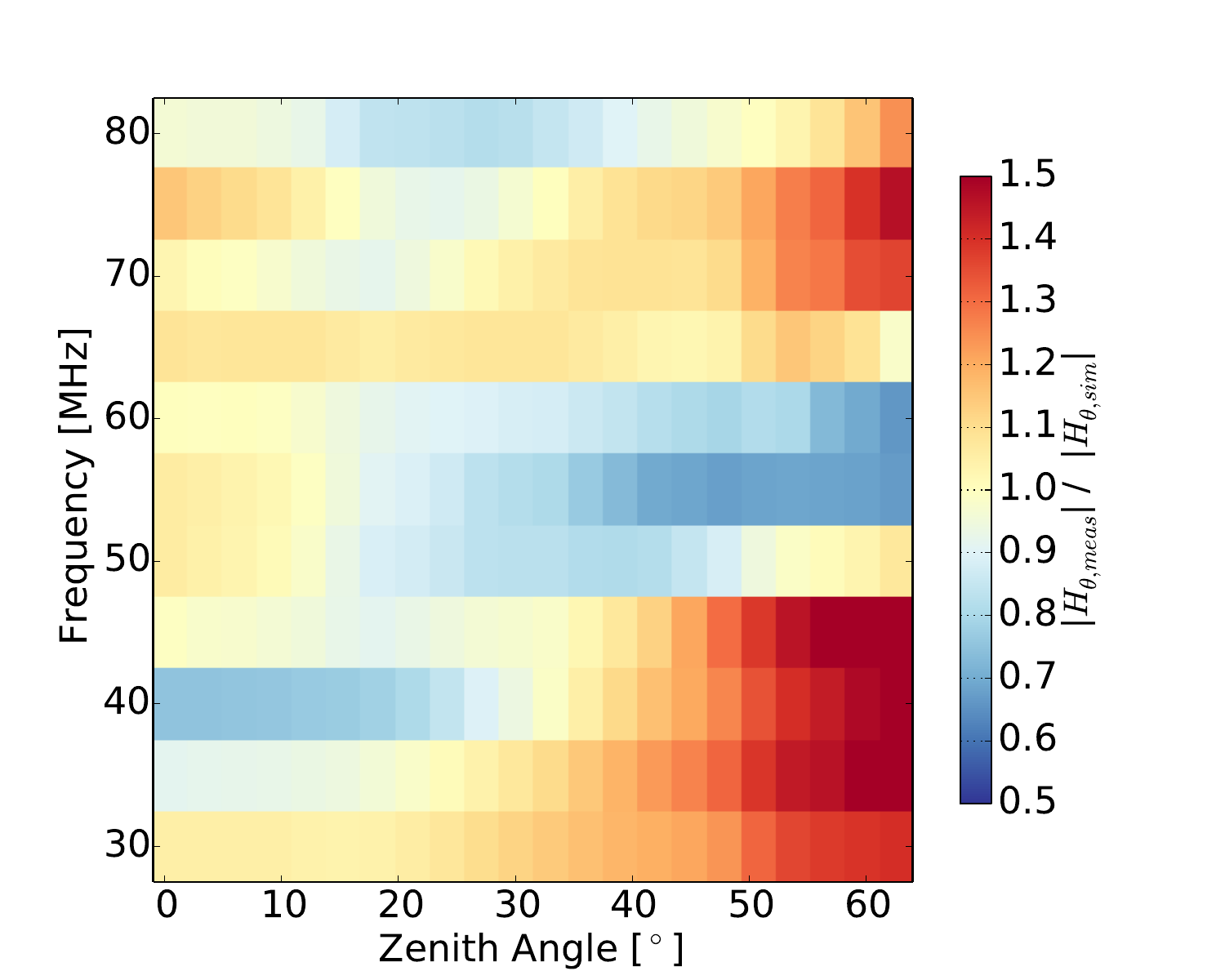}}
        \vspace{0.1cm}
        {\includegraphics[scale=0.29]{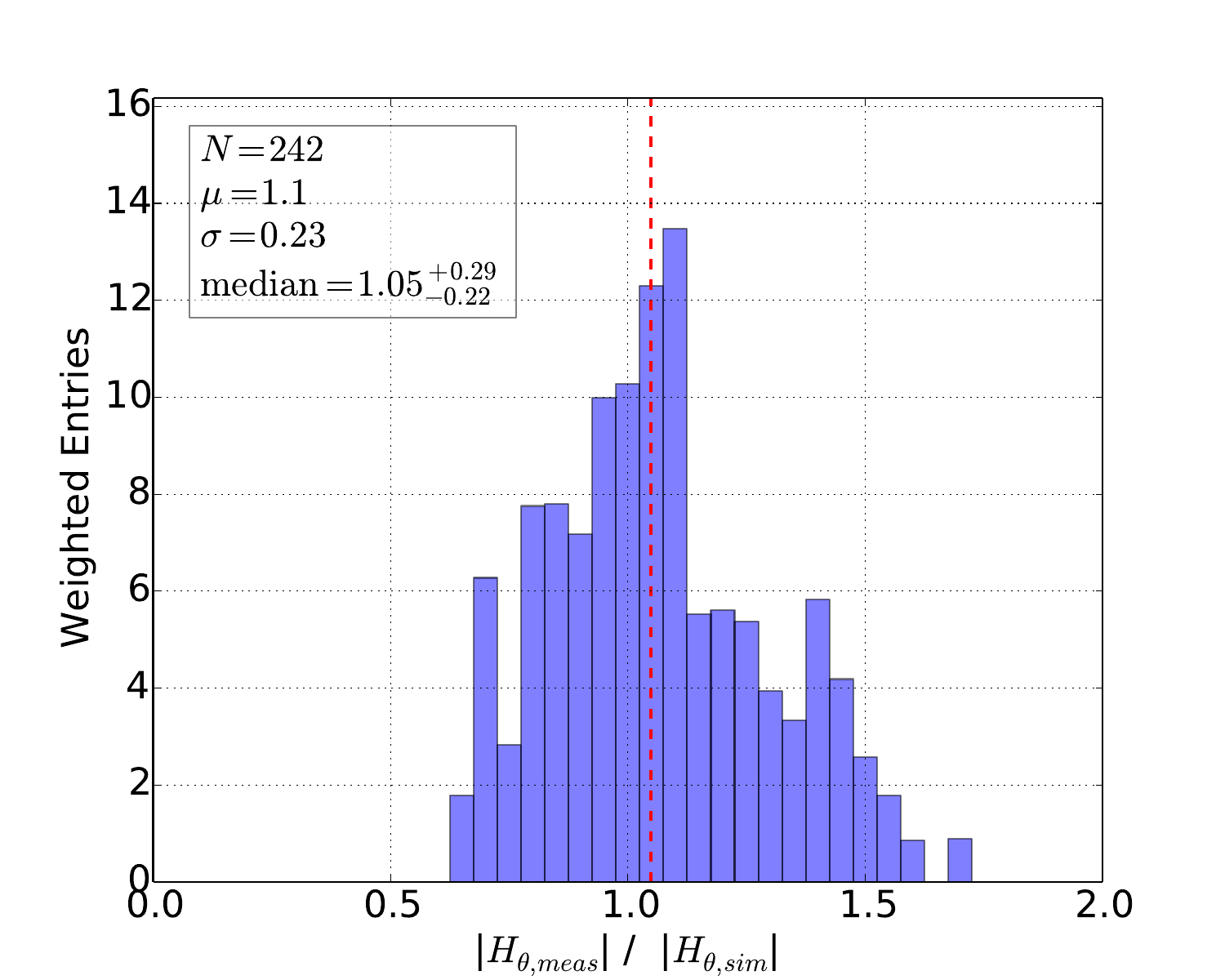}}
        \vspace{0.1cm}
        {\includegraphics[scale=0.29]{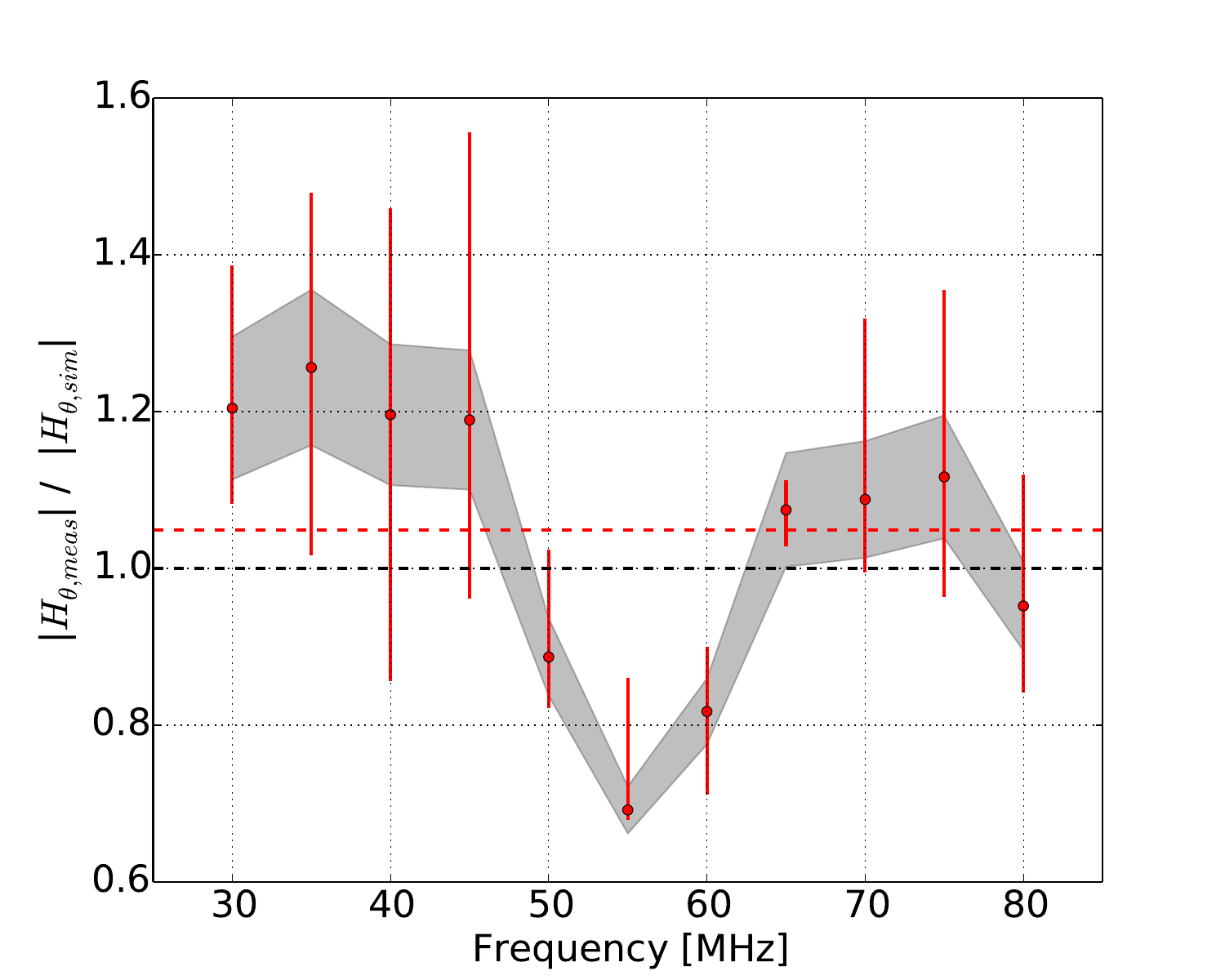}}
        \vspace{0.1cm}
        {\includegraphics[scale=0.29]{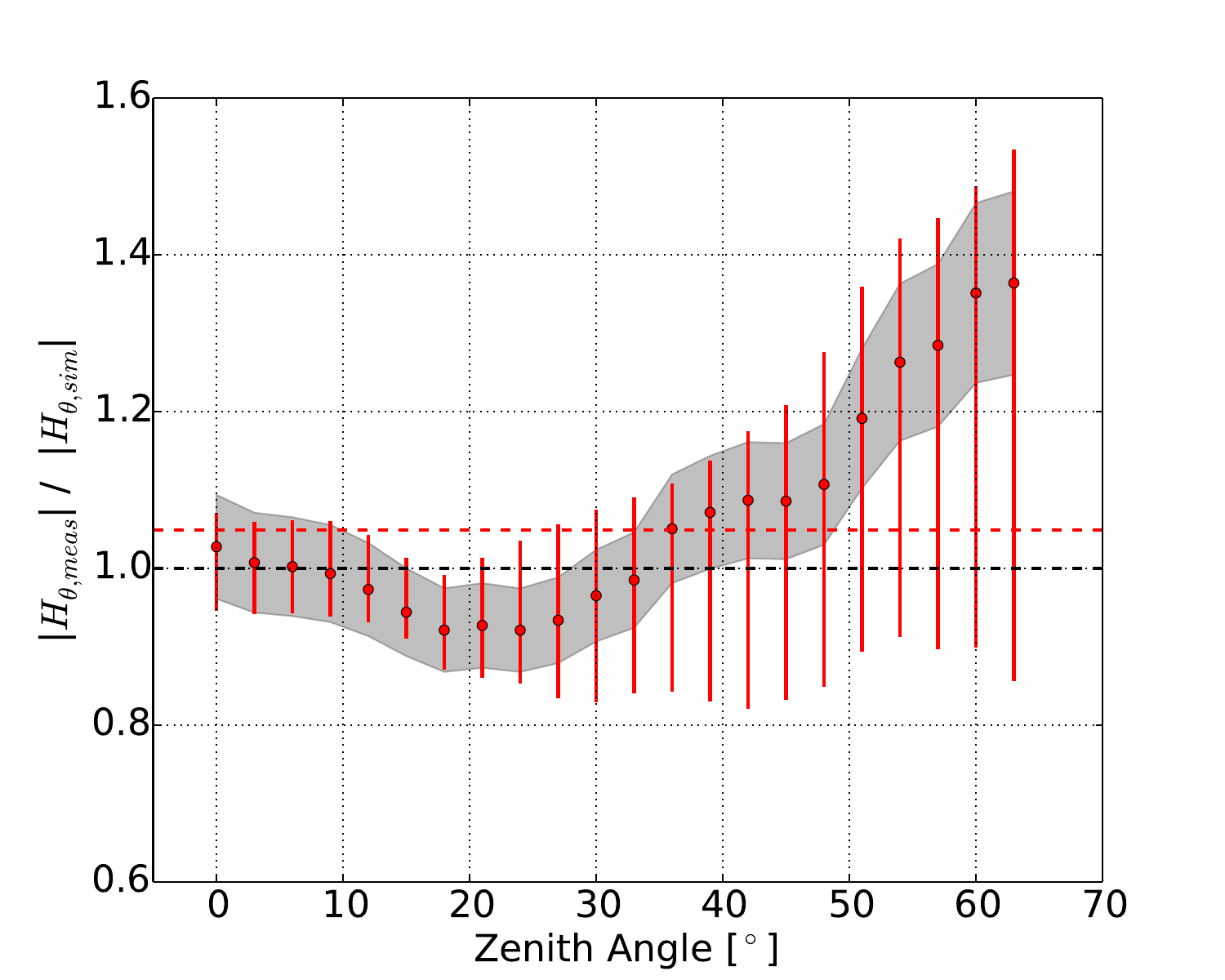}}
        \caption{\it Comparison of the combined meridional VEL $|H_{\theta}|$ with the simulation. \textbf{(top left)} Ratio of combination of all measurements and simulation 
        for all frequencies as a function of the zenith angle $\Theta$ up to $63\rm{^{\circ}}$ in $3\rm{^{\circ}}$ bins.
        \textbf{(top right)} Histogram of all ratios of the combination of all measurements and simulation for all frequencies and all zenith angle bins previously mentioned weighted with $\sin(\Theta)$.
        The median value is marked as the red dashed line.
        \textbf{(bottom left)} Median (red dots) and the $68\,\rm{\%}$ quantile (red error bars) of the zenith angle weighted correction factor distribution 
        as a function of the frequency. \textbf{(bottom right)} Median (red dots) and the $68\,\rm{\%}$ quantile (red error bars) of the ratio distribution 
        as a function of $\Theta$. The gray band indicates the constant systematic uncertainty of the measurement and the red dashed lines mark the overall zenith angle weighted average in both lower diagrams.}
    \label{fig:VerCorrection}
    \end{center}
\end{figure}

In this subsection, the results of the meridional VEL $|H_{\theta}|$ are discussed. For both subcomponents $|H_{\theta,\mathrm{hor}}|$ and $|H_{\theta,\mathrm{vert}}|$ three independent measurements were taken and averaged.
The averaged components are combined to determine $|H_{\theta}|$ as a function 
of the zenith angle $\Theta$ using Eq.\eqref{eq:HTheta}. In Fig.~\ref{fig:VerVEL} all performed measurements of $|H_{\theta}|$ are combined in zenith angle intervals of $5\rm{^{\circ}}$, weighted by the quadratic sum of the
systematic and the statistical uncertainties of each flight. The gray band describes the constant systematic uncertainties whereas
the statistical and flight-dependent systematic uncertainties are combined within the red error bars.
The constant systematic uncertainty of the combined VEL is $6.3\,\rm{\%}$. The uncertainties considering flight
dependent systematic and statistical uncertainties of the combined VEL result in $11.2\,\rm{\%}$  at a zenith angle of $(42.5\pm2.5)\rm{^{\circ}}$ and a frequency of $55\,\rm{MHz}$.
The overall uncertainty of the determined LPDA VEL in the meridional polarization adds quadratically to $12.9\,\rm{\%}$.
The overall uncertainty of all other arrival directions and frequencies are shown on the left side of Fig.~\ref{fig:VerVELError}. On the right side of Fig.~\ref{fig:VerVELError}, a histogram of
all overall uncertainties for all frequencies and all zenith angles up to $65\rm{^{\circ}}$ is shown.
For larger zenith angles the LPDA loses sensitivity and the systematic uncertainty exceeds $20\,\rm{\%}$. Therefore, these angles are not considered in the following discussion.
Taking all intervals of the frequencies and zenith angles with equal weight the
median overall uncertainty including statistical and systematic uncertainties is $10.3^{+2.8}_{-1.7}\,\rm{\%}$.
This is larger than the uncertainty of the horizontal component $|H_{\phi}|$. The reasons are that firstly, the meridional
component $|H_{\theta}|$ is a combination of two measurements of $|H_{\theta,\mathrm{hor}}|$ and $|H_{\theta,\mathrm{vert}}|$ whereas $|H_{\phi}|$ is directly measured. Secondly, the number of measurements is smaller than in
the case of $|H_{\phi}|$ and thirdly, the horizontal component is corrected for the octocopter misplacement and misalignment in comparison to the meridional subcomponents where this effect is included in the systematic
uncertainties.
The green curve in Fig.~\ref{fig:VerVEL} indicates the simulation of $|H_{\theta}|$. The agreement between the combination of all measurements and the simulations of $|H_{\theta}|$ is illustrated
by the diagram of their ratio versus zenith angle $\Theta$ and frequency $f$ shown in the upper left panel of Fig.~\ref{fig:VerCorrection}. 
In the upper right panel all ratios for all zenith angles and frequencies are filled into a histogram with entries weighted by $\sin(\Theta)$ in consideration of the 
decrease in field of view at small zenith angles. The combined measurement and the simulation agree to within $5\,\rm{\%}$ in the median. The fluctuation described by the $68\,\rm{\%}$ quantile 
is at the level of $26\,\rm{\%}$. The two lower panels of Fig.~\ref{fig:VerCorrection} show the median ratio as a function of the 
frequency (left) and as a function of the zenith angle (right). In both cases, the red error bars mark the $68\,\rm{\%}$ quantile of the distributions.
\subsection{Interpolation to all Arrival Directions and Frequencies}
The horizontal and meridional VEL magnitudes are measured in the LPDA axis with 
highest sensitivity to the respective VEL component (refer to section 
\ref{sec:Calib}) and in frequency bins of $5\,\rm{MHz}$.
To achieve an overall LPDA calibration for all incoming directions and frequencies the measurement is combined with simulations.
The LPDA VEL pattern is simulated using the NEC-2 simulation code. In contrast 
to the previous simulations presented in this work, only the LPDA with the 
following amplifier stage but without the transmitting antenna is taken into 
account. This original simulation of the LPDA pattern is then combined with the 
results from the calibration campaign. The calibrated LPDA VEL pattern is 
obtained by multiplying the full pattern of the simulated VEL with the ratio of 
measured to simulated VEL magnitudes shown in 
Figs.~\ref{fig:HorCorrection}~and~\ref{fig:VerCorrection}.
The ratios are linearly interpolated between the measurements at each zenith angle and each frequency bin. 
The respective frequency and zenith angle dependent ratios are applied at all azimuth angles. With this method, the magnitude of the VEL
is normalized to the calibration measurements, while the VEL phase comes entirely from the original simulation.
\section{Influence on Cosmic-Ray Signal Reconstruction}
In this section the influence of the modified LPDA pattern on the cosmic-ray signal reconstruction is discussed. 
In the first part of this section the influence of the differences between simulated and measured VEL on the electric field as well as on the radiation energy
for one event with a specific arrival direction are presented. In the second part the influence of the uncertainty of both components of the VEL on the electric-field is discussed.
\subsection{Influence of Modified Pattern on one Example Event}
\begin{figure}
    \begin{center}
        {\includegraphics[scale=0.55]{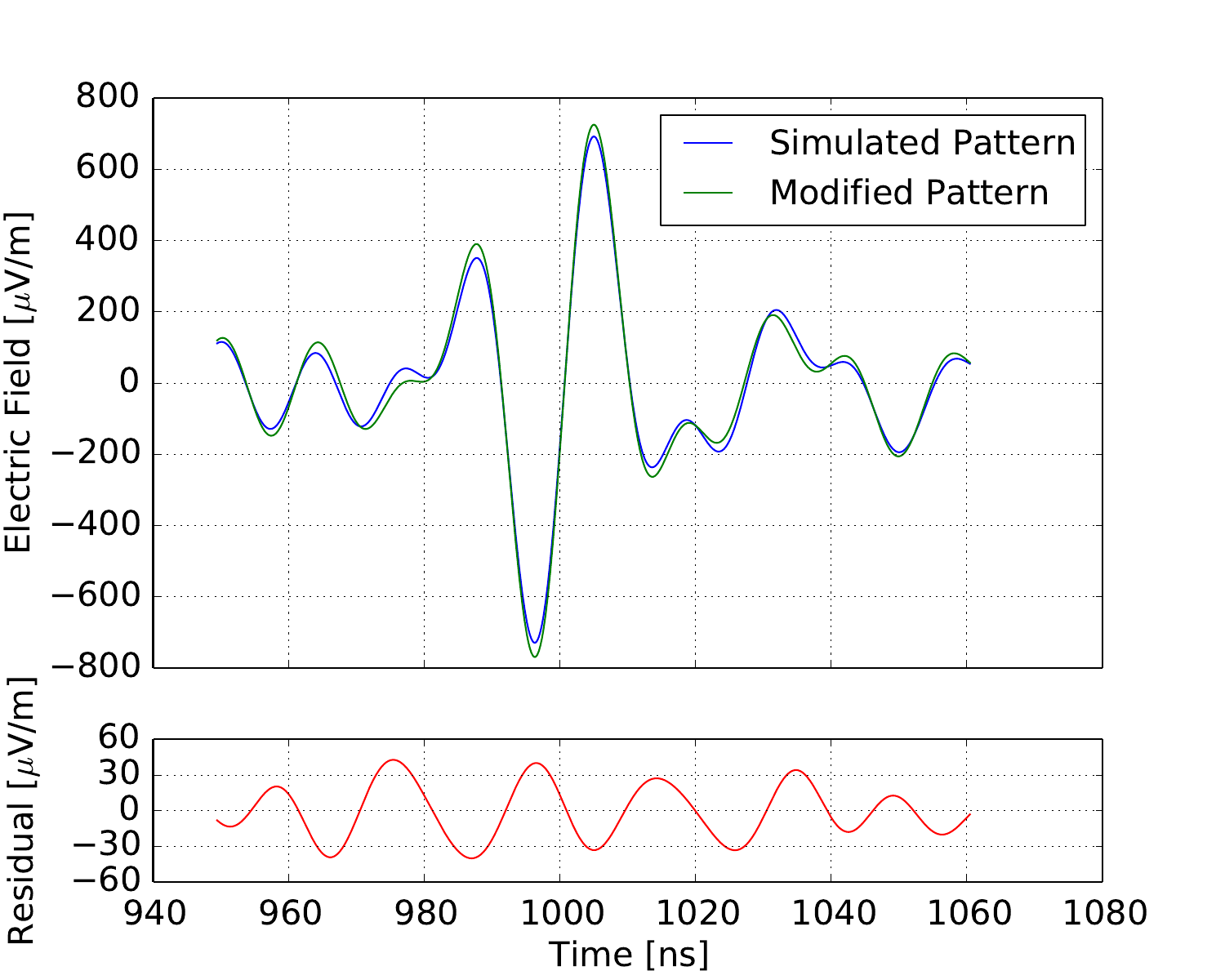}}
        \caption{\it \textbf{(top)} Reconstructed electric-field trace at the station with highest SNR in the east-west polarization of a signal measured at AERA with a zenith angle of $30\rm{^{\circ}}$ and 
        an azimuth angle of $14\rm{^{\circ}}$
        south of east using the simulated LPDA pattern (blue line) and using the modified pattern considering the correction factors between measurement and simulation (green line). The residual between both 
        reconstructed traces as function of the time is shown in the \textbf{(lower)} diagram. The measured energy fluence in the east-west polarization changes from $100\frac{\rm{eV}}{\rm{m^{2}}}$ to 
        $112\frac{\rm{eV}}{\rm{m^{2}}}$. }
    \label{fig:EFieldTrace}
    \end{center}
\end{figure}

To reconstruct the electric field of a measured air shower induced radio signal the Auger software framework 
{\mbox{$\overline{\textrm{Off}}$\hspace{.05em}\protect\raisebox{.4ex}{$\protect\underline{\textrm{line}}$}}} \cite{Offline} is used. To show the influence of the improved VEL,
an air shower measured in $9$ stations at AERA with a zenith angle of $30\rm{^{\circ}}$ and an azimuth angle of $14\rm{^{\circ}}$ south of east is presented as an example. The energy of the primary cosmic ray
is reconstructed to $1.1\times10^{18}\,\rm{eV}$ using information from the surface detector.
In Fig.~\ref{fig:EFieldTrace} the electric field reconstructed at the station with highest signal-to-noise ratio (SNR) is shown once using the simulated antenna response with and once without the corrections owing to the measurements of the VEL magnitude
in both components. For clarity only one polarization component of the electric field is shown.
The general shape of the electric-field trace is the same for both reconstructions. The trace of the modified LPDA pattern exhibits an up to $7\,\rm{\%}$ larger amplitude. 
The measured energy fluence that scales with the amplitude squared in the east-west polarization at this station with highest SNR changes from $100\frac{\rm{eV}}{\rm{m^{2}}}$ to $112\frac{\rm{eV}}{\rm{m^{2}}}$. 
The total energy fluence of all polarizations changes from $141\frac{\rm{eV}}{\rm{m^{2}}}$ using the simulated antenna response pattern to $156\frac{\rm{eV}}{\rm{m^{2}}}$ using the modified antenna 
response pattern which is an effect at the level of $9\,\rm{\%}$. The reconstructed radiation energy of the full event changes from $7.96\,\rm{MeV}$ to $8.54\,\rm{MeV}$. The ratio of these radiation energies is $0.93$. 
% The energy of the primary cosmic ray is reconstructed to $7.8\times10^{17}\,\rm{eV}$ using the simulated antenna response pattern and is reconstructed to $8.1\times10^{17}\,\rm{eV}$ using the modified antenna response 
% pattern, using the calibration function presented in \cite{AERA-Energy-PRL,AERA-Energy-PRD}. They agree at the level of $4\,\rm{\%}$.
\subsection{Uncertainty of the Cosmic-Ray Signal Reconstruction}
\label{subsec:UncerCR}
\begin{figure}
    \begin{center}
        {\includegraphics[scale=0.3]{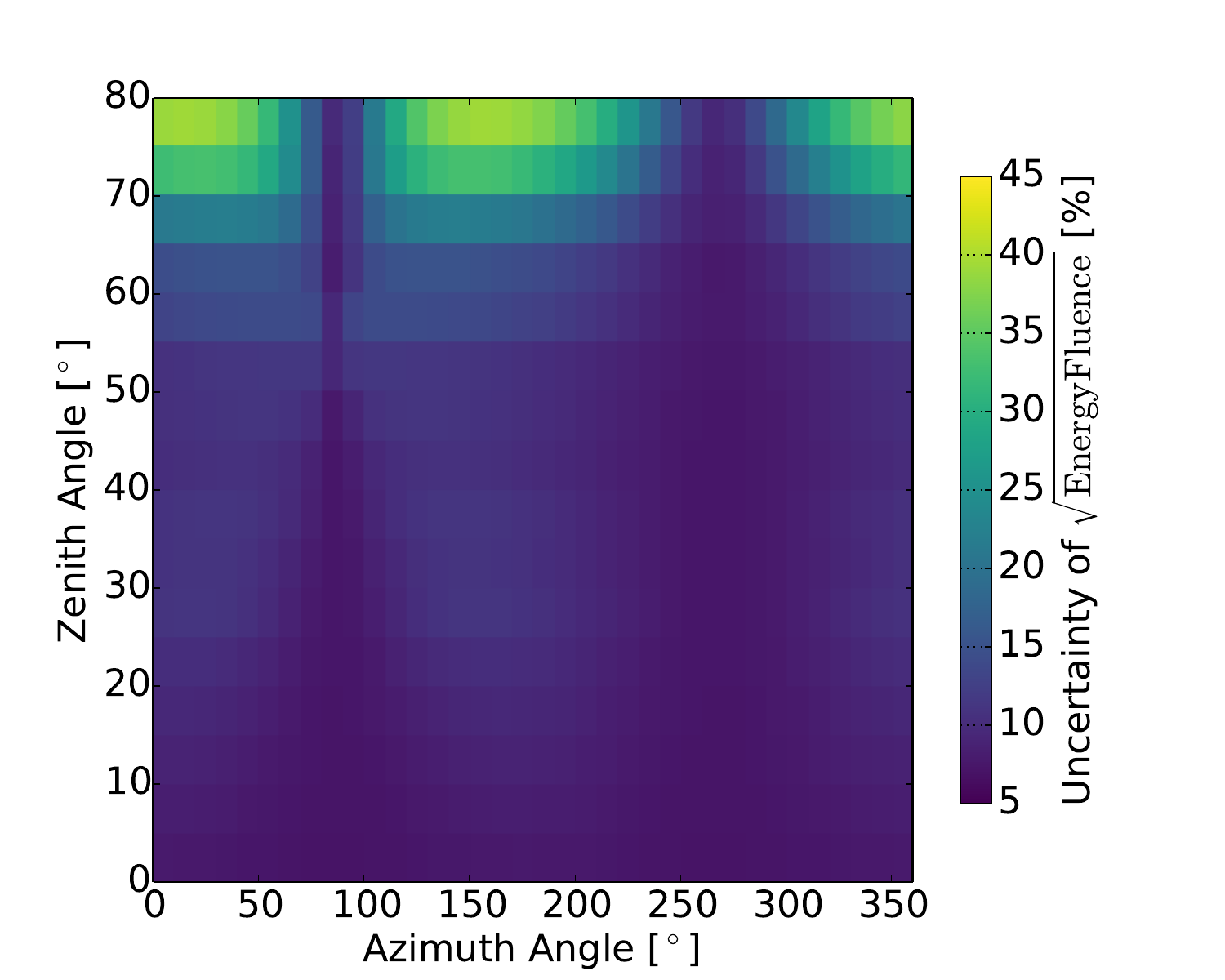}}{\includegraphics[scale=0.3]{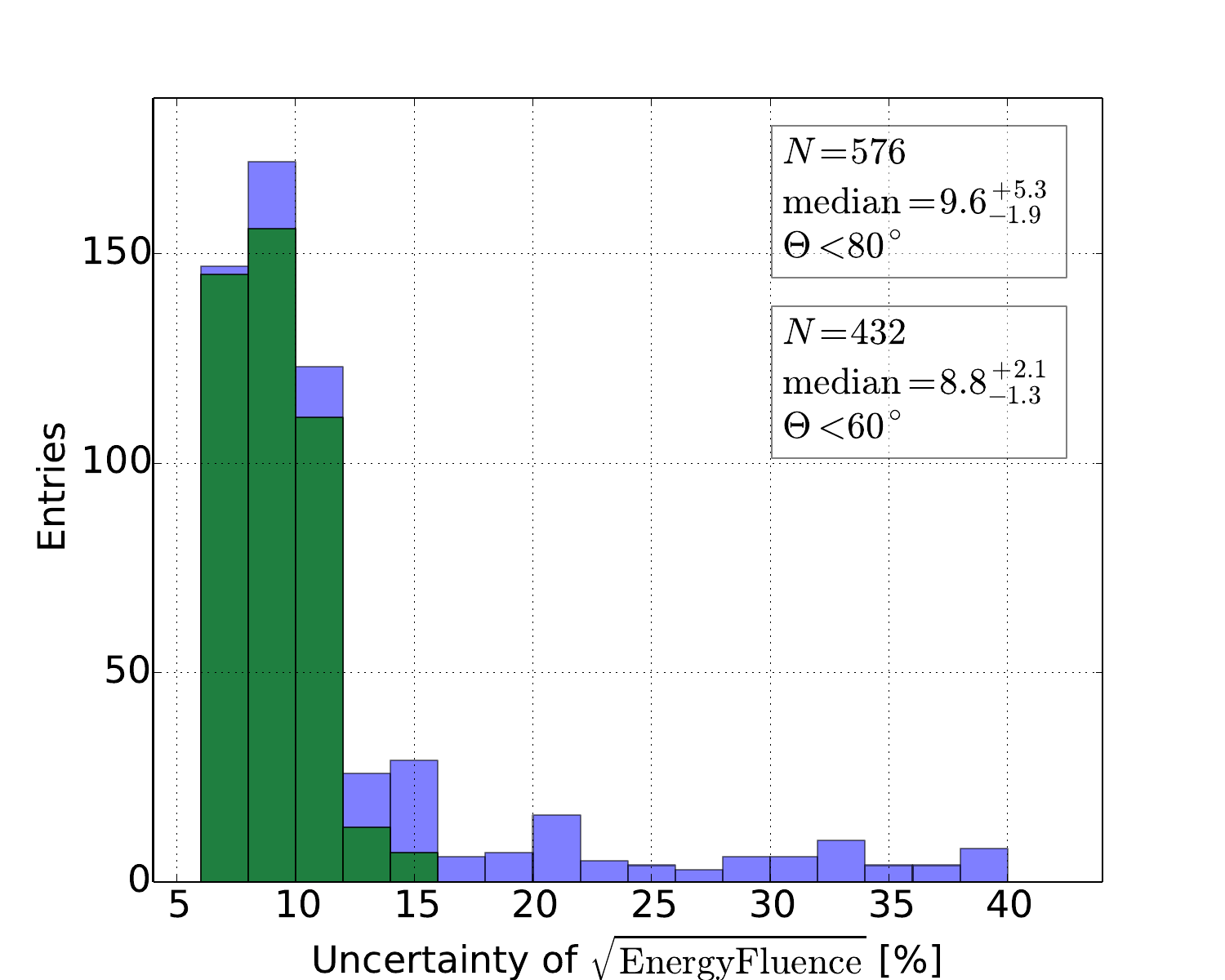}}
        \caption{\it \textbf{(left)} Systematic uncertainty of the square root of the energy fluence for all arrival directions taking into account a signal polarization due to the dominant geomagnetic emission process.
        The square root of the energy fluence is shown because the energy fluence scales quadratically with the electric-field amplitude and the cosmic-ray energy. Hence, the uncertainties of the square root 
        of the energy fluence is the relevant uncertainty in most analyses. \textbf{(right)} Histogram of the systematic uncertainty of the square root of the energy fluence of signals with zenith angles smaller 
        than $80\rm{^{\circ}}$ (blue) and of signals with zenith angles smaller than
        $60\rm{^{\circ}}$ (green).}
    \label{fig:EFieldUncertainty}
    \end{center}
\end{figure}

In this subsection the systematic uncertainty of the cosmic-ray signal reconstruction that results from the overall uncertainty of the antenna VEL magnitude and from the uncertainty due to different ground permittivities is determined. 
In the first case, the VEL magnitude is shifted up and down by one standard deviation of the overall uncertainty. The VEL phase remains unchanged.
In the case of the uncertainty due to different ground permittivities the antenna pattern with a ground permittivity of $2$ and of $10$ are used (see Fig.~\ref{fig:NEC2}).
The respective VEL is denoted as $H^\mathrm{down}$ and $H^\mathrm{up}$. 
The antenna response is applied to a simulated electric-field pulse using once $H^\mathrm{up}$ and once $H^\mathrm{down}$, to obtain the corresponding 
voltage traces $\mathcal{U}^\mathrm{up}$ and $\mathcal{U}^\mathrm{down}$ according to Eq.~\eqref{eq:AntResponse}. Then, the original VEL is used to reconstruct the electric-field pulse once from 
$\mathcal{U}^\mathrm{up}$ and once from $\mathcal{U}^\mathrm{down}$. From the difference of the two resulting electric-field pulses, the systematic uncertainty of the amplitude or the energy fluence can be determined. 
Both uncertainties resulting from the antenna VEL magnitude uncertainty and resulting from different ground permittivities, are then combined quadratically. \\
An additional uncertainty on the electric-field trace can arise due to an 
uncertainty on the VEL phase.
An uncertainty in the VEL phase leads to a signal distortion of the radio pulse
resulting in an increased signal pulse width and a smaller electric-field amplitude or vice versa. However, the energy fluence of the RD pulse which is given by 
the integral over the electric-field trace remains constant.
Hence, a VEL phase uncertainty propagates to an additional uncertainty in the electric-field amplitude whereas the energy fluence does not change due to a VEL phase uncertainty.
Therefore, the systematic uncertainty of the energy fluence due to the VEL uncertainty is discussed in the following.\\
The radio pulse is approximated with a bandpass-limited Dirac pulse and the polarization is adjusted according to the dominant geomagnetic emission process. 
As the uncertainty of the VEL and the polarization of the electric-field pulse depend on the incoming signal direction, different directions in bins of $10\rm{^{\circ}}$ 
in the azimuth angle and in bins of $5\rm{^{\circ}}$ in the zenith angle are simulated. Due to the changing polarization also the relative influences of the $|H_{\phi}|$ and $|H_{\theta}|$ components change with direction.
The resulting systematic uncertainty of the energy fluence is presented in Fig.~\ref{fig:EFieldUncertainty}. The square root of the energy fluence is shown because the energy fluence scales quadratically with the 
electric-field amplitude and the cosmic-ray energy. Hence, the systematic uncertainty of the square root of the energy fluence is the relevant uncertainty in most analyses.
For most regions the systematic uncertainty is at the level of $10\,\rm{\%}$. The uncertainty increases only at large zenith angles ($\theta > 60\rm{^{\circ}}$) due to the increased uncertainty of $|H_{\theta}|$.
An azimuthal pattern appears at $90\rm{^{\circ}}$ and $270\rm{^{\circ}}$. At these azimuth angles the uncertainty is smaller because the electric-field pulse 
is polarized in the $\vec{e}_{\phi}$ component and only the more precise $|H_{\phi}|$ component contributes.
For incoming signal directions with zenith angles smaller than $60\rm{^{\circ}}$ the systematic uncertainty of the square root of the energy fluence owing to the antenna calibration and 
different ground permittivities is at most $14.2\,\rm{\%}$ with a median of $8.8^{+2.1}_{-1.3}\,\rm{\%}$.
\section{Conclusion}
In this work, the results of an absolute antenna calibration are presented
performed on a radio station equipped with a logarithmic periodic dipole
antenna (LPDA). The station belongs to the AERA field of radio stations
at the site of the Pierre Auger Observatory. The calibrated LPDA is representative
of all the LPDAs which are mechanically and
electrically identical at the percent level.\\
The radio stations are used to reconstruct the electric field
emitted by cosmic particle induced air showers which gives, e.g., a
precise measure of the energy contained in the electromagnetic shower.
The accuracy of the reconstructed shower energy is limited by the
uncertainty in the absolute antenna calibration such that reduction of
the uncertainties was most desirable.\\
The frequency and directional dependent sensitivity of the LPDA
has been probed by an octocopter carrying a calibrated radio source with
dedicated polarization of the emitted radio signals. The measured LPDA
response has been quantified using the formalism of the vector effective
length and decomposed in terms of a horizontal and a meridional component.\\
All experimental components involved in the calibration campaign were
quantified with respect to their uncertainties. Special emphasis was put
on the precision in the position reconstruction of the source which was
supported by a newly developed optical system with two cameras used in
conjunction with on-board measurements of inclination, GPS, and
barometric height. To ensure reproducible results, all calibration
measurements were repeated by several flights on different days under
different environmental conditions.\\
The combination of all measurements gives an overall accuracy for the
horizontal component of the vector effective length of $7.4^{+0.9}_{-0.3}\,\rm{\%}$, and
for the meridional component of $10.3^{+2.8}_{-1.7}\,\rm{\%}$. Note that
for air showers with zenith angles below $60\rm{^{\circ}}$
the horizontal component gives the dominant contribution.
The obtained accuracy is to be compared with a previous balloon based
measurement probing a smaller phase space of the horizontal component
with a systematic uncertainty of $12.5\,\rm{\%}$.\\
The measurements of the new calibration campaign enable
thorough comparisons with simulations of the calibration setup including 
ground condition dependencies using the NEC-2 program. The
measurements were used to correct the simulated pattern at multiple
points in the phase space described by arrival direction, frequency and
polarization of the waves. While the median of all correction factors
are close to unity at standard ground conditions, corrections of the
simulated vector effective length
vary with an rms of $12\,\rm{\%}$ for the horizontal component, and with
rms of $26\,\rm{\%}$ for the meridional component.
The simulations have been further used to confirm that the measurements
have been done in the far-field region. Additionally, the LPDA
sensitivity to different ground conditions has been investigated showing
that the LPDA is insensitive to different ground conductivities and the
sensitivity to different permittivity is only at the level of $1\,\rm{\%}$.\\
The effect of the correction factors on the simulated vector
effective length has been demonstrated in the reconstruction of
one example radio event measured with AERA.\\
Finally, the uncertainty of the two VEL components are propagated onto the square root of the energy fluence that is
obtained by unfolding the antenna response from the measured voltage traces. 
For incoming directions up to $60\rm{^{\circ}}$, the expected systematic 
uncertainty in the square root of the energy fluence due to the LPDA 
calibration is $8.8^{+2.1}_{-1.3}\,\rm{\%}$ in the median.

% \appendix
% \section{Some title}
% Please always give a title also for appendices.

\acknowledgments
\begin{sloppypar}
The successful installation, commissioning, and operation of the Pierre Auger Observatory would not have been possible without the strong commitment and effort from the technical and administrative staff in Malarg\"ue. We are very grateful to the following agencies and organizations for financial support:
\end{sloppypar}

\begin{sloppypar}
Argentina -- Comisi\'on Nacional de Energ\'\i{}a At\'omica; Agencia Nacional de Promoci\'on Cient\'\i{}fica y Tecnol\'ogica (ANPCyT); Consejo Nacional de Investigaciones Cient\'\i{}ficas y T\'ecnicas (CONICET); Gobierno de la Provincia de Mendoza; Municipalidad de Malarg\"ue; NDM Holdings and Valle Las Le\~nas; in gratitude for their continuing cooperation over land access; Australia -- the Australian Research Council; Brazil -- Conselho Nacional de Desenvolvimento Cient\'\i{}fico e Tecnol\'ogico (CNPq); Financiadora de Estudos e Projetos (FINEP); Funda\c{c}\~ao de Amparo \`a Pesquisa do Estado de Rio de Janeiro (FAPERJ); S\~ao Paulo Research Foundation (FAPESP) Grants No.\ 2010/07359-6 and No.\ 1999/05404-3; Minist\'erio de Ci\^encia e Tecnologia (MCT); Czech Republic -- Grant No.\ MSMT CR LG15014, LO1305 and LM2015038 and the Czech Science Foundation Grant No.\ 14-17501S; France -- Centre de Calcul IN2P3/CNRS; Centre National de la Recherche Scientifique (CNRS); Conseil R\'egional Ile-de-France; D\'epartement Physique Nucl\'eaire et Corpusculaire (PNC-IN2P3/CNRS); D\'epartement Sciences de l'Univers (SDU-INSU/CNRS); Institut Lagrange de Paris (ILP) Grant No.\ LABEX ANR-10-LABX-63 within the Investissements d'Avenir Programme Grant No.\ ANR-11-IDEX-0004-02; Germany -- Bundesministerium f\"ur Bildung und Forschung (BMBF); Deutsche Forschungsgemeinschaft (DFG); Finanzministerium Baden-W\"urttemberg; Helmholtz Alliance for Astroparticle Physics (HAP); Helmholtz-Gemeinschaft Deutscher Forschungszentren (HGF); Ministerium f\"ur Innovation, Wissenschaft und Forschung des Landes Nordrhein-Westfalen; Ministerium f\"ur Wissenschaft, Forschung und Kunst des Landes Baden-W\"urttemberg; Italy -- Istituto Nazionale di Fisica Nucleare (INFN); Istituto Nazionale di Astrofisica (INAF); Ministero dell'Istruzione, dell'Universit\'a e della Ricerca (MIUR); CETEMPS Center of Excellence; Ministero degli Affari Esteri (MAE); Mexico -- Consejo Nacional de Ciencia y Tecnolog\'\i{}a (CONACYT) No.\ 167733; Universidad Nacional Aut\'onoma de M\'exico (UNAM); PAPIIT DGAPA-UNAM; The Netherlands -- Ministerie van Onderwijs, Cultuur en Wetenschap; Nederlandse Organisatie voor Wetenschappelijk Onderzoek (NWO); Stichting voor Fundamenteel Onderzoek der Materie (FOM); Poland -- National Centre for Research and Development, Grants No.\ ERA-NET-ASPERA/01/11 and No.\ ERA-NET-ASPERA/02/11; National Science Centre, Grants No.\ 2013/08/M/ST9/00322, No.\ 2013/08/M/ST9/00728 and No.\ HARMONIA 5 -- 2013/10/M/ST9/00062; Portugal -- Portuguese national funds and FEDER funds within Programa Operacional Factores de Competitividade through Funda\c{c}\~ao para a Ci\^encia e a Tecnologia (COMPETE); Romania -- Romanian Authority for Scientific Research ANCS; CNDI-UEFISCDI partnership projects Grants No.\ 20/2012 and No.194/2012 and PN 16 42 01 02; Slovenia -- Slovenian Research Agency; Spain -- Comunidad de Madrid; Fondo Europeo de Desarrollo Regional (FEDER) funds; Ministerio de Econom\'\i{}a y Competitividad; Xunta de Galicia; European Community 7th Framework Program Grant No.\ FP7-PEOPLE-2012-IEF-328826; USA -- Department of Energy, Contracts No.\ DE-AC02-07CH11359, No.\ DE-FR02-04ER41300, No.\ DE-FG02-99ER41107 and No.\ DE-SC0011689; National Science Foundation, Grant No.\ 0450696; The Grainger Foundation; Marie Curie-IRSES/EPLANET; European Particle Physics Latin American Network; European Union 7th Framework Program, Grant No.\ PIRSES-2009-GA-246806; European Union's Horizon 2020 research and innovation programme (Grant No.\ 646623); and UNESCO.
\end{sloppypar}

% \paragraph{Note added.} This is also a good position for notes added
% after the paper has been written.


\begin{thebibliography}{00}
\bibitem{Auger} A. Aab et al. (Pierre Auger Collaboration), {\it The 
Pierre Auger Cosmic Ray Observatory}, 
\href{http://dx.doi.org/10.1016/j.nima.2015.06.058}{Nucl. Instrum. Meth. A {\bf 
798} (2015) 172--213}
\bibitem{LOPES}H. Falcke et al. (LOPES Collaboration), {\it Detection and 
imaging of atmospheric radio flashes from cosmic ray air showers}, 
\href{http://dx.doi.org/10.1038/nature03614}{Nature {\bf 435} (2005) 313--316}
\bibitem{Codalema}D. Ardouin et al. (CODALEMA Collaboration), 
{\it Radioelectric field features of extensive air showers observed with 
CODALEMA}, 
\href{{https://doi.org/10.1016/j.astropartphys.2006.07.002}}{Astropart. Phys. 
{\bf 26} (2006) 341--350}
\bibitem{AERA}J. Schulz for the Pierre Auger Collaboration, {\it Status 
and prospects of the Auger Engineering Radio Array}, Proc. of the 34th 
ICRC, The Hague, The Netherlands, 2015, 
\href{https://pos.sissa.it/contribution?id=PoS(ICRC2015)615}{[PoS(ICRC2015)615]}
\bibitem{TRex}P.A. Bezyazeekov et al. (Tunka-Rex Collaboration), {\it 
Measurement of cosmic-ray air showers with the Tunka Radio Extension
(Tunka-Rex)}, \href{https://doi.org/10.1016/j.nima.2015.08.061} {Nucl. Instrum. 
Meth. A {\bf 802} (2015) 89--96}
\bibitem{LOFAR}J.R. H\"orandel, {\it Radio detection of cosmic rays with LOFAR},
Proc. of the 34th ICRC, The Hague, The Netherlands, 2015, 
\href{https://pos.sissa.it/contribution?id=PoS(ICRC2015)033} 
{[PoS(ICRC2015)033]}
\bibitem{Tim}T. Huege, {\it Radio detection of cosmic ray air showers in the 
digital era}, \href{https://doi.org/10.1016/j.physrep.2016.02.001}{Phys. Rep. 
{\bf 620} (2016) 1--52}
\bibitem{Frank}F.G. Schr\"oder, {\it Radio detection of cosmic-ray air showers 
and high-energy neutrinos}, 
\href{https://doi.org/10.1016/j.ppnp.2016.12.002}{Prog. in Part. and Nucl. Phys. 
{\bf 93} (2017) 1--68}
\bibitem{Geo-KahnLerche}F.D. Kahn, I. Lerche, {\it Radiation from cosmic ray air 
showers}, \href{https://doi.org/10.1098/rspa.1966.0007}{Proc. R. Soc. Lond. A 
{\bf 289} (1966) 206--213}
\bibitem{Geo-Allan}H.R. Allan, {\it Radio emission from extensive air showers},
Prog. Element. Part. Cosmic Ray Phys. {\bf X} (1971) 169--302
\bibitem{Geo-HuegeFalcke}T. Huege, H. Falcke, {\it Radio emission from cosmic 
ray air showers: coherent geosynchrotron radiation}, 
\href{https://dx.doi.org/10.1051/0004-6361:20031422}{Astron. Astrophys. {\bf 
412} (2003) 19--34}
\bibitem{Geo-Codalema}D. Ardouin et al. (CODALEMA Collaboration), {\it 
Geomagnetic origin of the radio emission from cosmic ray induced air showers 
observed by CODALEMA} \href{}{Astropart. Phys. {\bf 31} (2009) 192--200}
\bibitem{Geo-Auger}P. Abreu et al. (Pierre Auger Collaboration), {\it 
Results of a self-triggered prototype system for radio-detection of extensive 
air showers at the Pierre Auger Observatory}, 
\href{https://doi.org/10.1088/1748-0221/7/11/P11023}{J. Instrum. {\bf 7} (2012) 
P11023}
\bibitem{CE-Ask}G. Askaryan, {\it Excess negative charge of an electron-photon 
shower and its coherent radio emission}, Soviet Phys. JETP Lett. (USSR) {\bf 
14} (1962) 441
\bibitem{CE}J.R. Prescott, J.H. Hough, and J.K. Pidcock, {\it Mechanism of radio 
emission from extensive air showers}, Nature Phys. Sci. {\bf 223} (1971) 
109--110
\bibitem{CE-Codalema}A. Bellétoile et al., {\it Evidence for the 
charge-excess contribution in air shower radio emission observed
by the CODALEMA experiment}, 
\href{http://dx.doi.org/10.1016/j.astropartphys.2015.03.007}{Astropart. Phys. 
{\bf 69} (2015) 50--60}
\bibitem{CE-Lofar}P. Schellart et al. (LOFAR Collaboration), {\it 
Polarized radio emission from extensive air showers measured with LOFAR}, 
\href{https://doi.org/10.1088/1475-7516/2014/10/014}{J. Cosmol. Astropart. Phys.
{\bf 10} (2014) 014}
\bibitem{CE-AERA}A. Aab et al. (Pierre Auger Collaboration), {\it Probing 
the radio emission from air showers with polarization measurements},
\href{https://doi.org/10.1103/PhysRevD.89.052002}{Phys. Rev. D {\bf 89} (2014) 
052002}
\bibitem{CORSIKA}D. Heck, G. Schatz, T. Thouw, J. Knapp and J. Capdevielle, 
{\it CORSIKA: a Monte Carlo code to simulate extensive air showers}, Report 
FZKA 6019, (1998)
\bibitem{ZHAires}J. Alvarez-Muñiz, W.R. Carvalho and E. Zas, {\it Monte Carlo 
simulations of radio pulses in atmospheric showers using ZHAireS}, 
\href{https://doi.org/10.1016/j.astropartphys.2011.10.005}{Astropart. Phys. {\bf 
35} (2012) 325--341}
\bibitem{CoREAS}T. Huege, M. Ludwig and C.W. James, {\it Simulating radio 
emission from air showers with CoREAS}, 
\href{http://dx.doi.org/10.1063/1.4807534}{Proc. of the AIP Conf. {\bf 1535} 
(2013) 128--132}
\bibitem{Radio}C. Glaser, M. Erdmann, J.R. H\"orandel, T. Huege and J. Schulz, 
{\it Simulation of radiation energy release in air showers}, 
\href{https://doi.org/10.1088/1475-7516/2016/09/024}{J. Cosmol. Astropart. Phys. 
{\bf 
09} (2016) 024}
\bibitem{LOPES-Energy}W.D. Apel et al. (LOPES Collaboration), {\it 
Reconstruction of the energy and depth of maximum of cosmic-ray air-showers 
from LOPES radio measurements}, 
\href{https://doi.org/10.1103/PhysRevD.90.062001}{Phys. Rev. D {\bf 90} (2014) 
062001}
\bibitem{LOFAR-Energy}A. Nelles et al., {\it The radio emission pattern 
of air showers as measured with LOFAR — a tool for the reconstruction of the 
energy and the shower maximum}, 
\href{https://doi.org/10.1088/1475-7516/2015/05/018}{J. Cosmol. Astropart. Phys.
{\bf 5} (2015) 18}
\bibitem{TREX-Energy}P.A. Bezyazeekov et al. (Tunka-Rex Collaboration), 
{\it Radio measurements of the energy and the depth of the shower maximum of 
cosmic-ray air showers by Tunka-Rex}, 
\href{https://doi.org/10.1088/1475-7516/2016/01/052}{J. Cosmol. Astropart. Phys. 
{\bf 01} (2016) 052}
\bibitem{AERA-Energy-PRL}A. Aab et al. (Pierre Auger Collaboration), {\it 
Measurement of the radiation energy in the radio signal of extensive air 
showers as a universal estimator of cosmic-ray energy}, 
\href{http://dx.doi.org/10.1103/PhysRevLett.116.241101}{Phys. Rev. Lett. {\bf 
116} (2016) 241101}
\bibitem{AERA-Energy-PRD}A. Aab et al. (Pierre Auger Collaboration), {\it 
Energy estimation of cosmic rays with the engineering
radio array of the Pierre Auger Observatory}, 
\href{https://doi.org/10.1103/PhysRevD.93.122005}{Phys. Rev. D {\bf 93} (2016) 
122005}
\bibitem{LOPES-Calib}K. Link for the LOPES Collaboration, {\it Revised 
absolute amplitude calibration of the LOPES experiment}, Proc. of the 
34th ICRC, The Hague, The Netherlands, 2015, 
\href{https://pos.sissa.it/contribution?id=PoS(ICRC2015)311}{[PoS(ICRC2015)311]}
\bibitem{TRex-Calib}R. Hiller for the Tunka-Rex Collaboration, {\it 
Calibration of the absolute amplitude scale of the Tunka Radio Extension 
(Tunka-Rex)}, Proc. of the 34th ICRC, The Hague, The Netherlands, 2015, 
\href{https://pos.sissa.it/contribution?id=PoS(ICRC2015)573}{[PoS(ICRC2015)573]}
\bibitem{LOFAR-Calib}A. Nelles et al., {\it Calibrating the absolute 
amplitude scale for air showers measured at LOFAR}, 
\href{https://doi.org/10.1088/1748-0221/10/11/P11005}{J. Instrum. {\bf 10} 
(2015) P11005}
\bibitem{KlausPhD}K. Weidenhaupt, {\it Antenna calibration and energy 
measurement of ultra-high energy cosmic rays with
the Auger Engineering Radio Array}, 
PhD Thesis, RWTH Aachen University, (2014)
\bibitem{AERA-Antennas}P. Abreu et al. (Pierre Auger Collaboration), {\it 
Antennas for the detection of radio emission pulses from
cosmic-ray}, \href{http://dx.doi.org/10.1088/1748-0221/7/10/P10011}{J. Instrum. 
{\bf 7} (2012) P10011}
\bibitem{Friis}H.T. Friis (Bell Telephone Laboratories), {\it A note on a simple 
transmission formula}, \href{https://doi.org/10.1109/JRPROC.1946.234568}{Proc. 
of the IRE. Volume {\bf 34}, Issue 5 (1946) 254--256}
\bibitem{NEC2}G. Burke and A. Poggio, {\it Numerical Electromagnetics Code (NEC) 
method of moments}, tech. rep.
Lawrence Livermore National Laboratory, NEC-1 (1977), NEC-2 (1981), NEC-3 (1983), NEC-4 (1992)
\bibitem{RG}coaxial cable specifications,
\url{https://www.bundesnetzagentur.de}
\bibitem{LNA}M. Stephan for the Pierre Auger Collaboration, {\it Antennas, 
filters and preamplifiers designed for the radio
detection of ultra-high-energy cosmic rays}, 
Proc. of the Asia-Pacific-Microwave Conference 2010, Yokohama, Japan, 2010 
1455–1458
\bibitem{OctoXL}Octocopter XL from MikroKopter,
\url{https://www.mikrocontroller.com/index.php?main_page=index&cPath=114}
\bibitem{FDCalib}J. Bäuml for the Pierre Auger Collaboration, {\it 
Measurement of the optical properties of the Auger
Fluorescence Telescopes},
Proc. of the 33rd ICRC, Rio de Janeiro, Brazil, 2013
\bibitem{CromeCalib}R. Smida et al., {\it First experimental 
characterization of microwave emission from cosmic ray air
showers}, \href{http://dx.doi.org/10.1103/PhysRevLett.113.221101}{Phys. Rev. 
Lett. {\bf 113} (2014) 221101}
\bibitem{Camera}Canon Ixus 132,
\url{http://www.canon.de/for_home/product_finder/cameras/digital_camera/ixus/ixus_132/}
\bibitem{OpticalMeth}F. Briechle, {\it Octocopter position reconstruction for 
calibrating the Auger Engineering Radio Array}, 
Master Thesis, RWTH Aachen University, (2015)
\bibitem{OpticalMethProc}F. Briechle for the Pierre Auger Collaboration, 
{\it In-situ absolute calibration of electric-field amplitude
measurements with the radio detector stations of the Pierre Auger Observatory},
Proc. of the 7th ARENA Conf., Groningen, The Netherlands, 2016
\bibitem{DGPS}Hiper V,
\url{https://www.topconpositioning.com/gnss/integrated-gnss-receivers/hiper-v}
\bibitem{RSG1000}RSG1000 Signal Generator,
\url{http://www.teseq.de/produkte/RSG-1000.php}
\bibitem{FSH4}Rohde\&Schwarz FSH4 Spectrum and Network Analyzer,
\url{https://www.rohde-schwarz.com/de/produkt/fsh-produkt-startseite_63493-8180.html}
\bibitem{Agilent}Agilent N9030A ESA Spectrum Analyzer,
\url{http://literature.cdn.keysight.com/litweb/pdf/5990-3952EN.pdf?id=1759326}
\bibitem{Schwarzbeck}Schwarzbeck Mess-Elektronik, Ultra Light BBOC 9217 Biconical Antenna with UBAA 9114 4:1 Balun,
\url{http://schwarzbeck.de/Datenblatt/91149217.pdf}
\bibitem{PHD}~R. Krause, {\it Antenna development and calibration for 
measurements of radio emission from extensive
air showers at the Pierre Auger Observatory}, 
PhD Thesis, RWTH Aachen University, (in preparation)
\bibitem{BarometricFormula}MikroKopter, air pressure sensor as altitute meter,
\url{http://wiki.mikrokopter.de/en/heightsensor}
\bibitem{HillerPHD}R. Hiller, {\it Radio measurements for determining the energy 
scale of cosmic rays}, PhD Thesis, Karlsruhe Institute of Technology (KIT), 
(2016)
\bibitem{Conductivity}W.G. Fano and V. Trainotti, {\it Dielectric properties of 
soils}, \href{https://doi.org/10.1109/CEIDP.2001.963492}{
IEEE Ann. Rept. Conf. Electr. Insul. Dielectr. Phenom. {\bf 2001} (2011) 75}
\bibitem{Offline}P. Abreu et al. (Pierre Auger Collaboration), {\it 
Advanced functionality for radio analysis in the Offline
software framework of the Pierre Auger Observatory}, 
\href{https://doi.org/10.1016/j.nima.2011.01.049}{
Nucl. Instrum. Meth. A {\bf 635} (2011) 92--102}

\end{thebibliography}
\end{document}